\documentclass[twocolumn]{aastex62}

\usepackage[caption=false]{subfig}

\shorttitle{Supervised Machine Learning for Brown Dwarfs}
\shortauthors{Oreshenko et al.}

\newif\ifdraft
\drafttrue

\definecolor{orange}{rgb}{1,0.5,0}

\ifdraft
 \newcommand{\PMN}[1]{{\color{orange}{\bf PMN: #1}}}
 
\else
 \newcommand{\PMN}[1]{{\color{red}{}}}
 
\fi

\begin{document}

\title{Supervised Machine Learning for Inter-comparison of Model Grids of Brown Dwarfs: \\ Application to GJ 570D and the Epsilon Indi B Binary System}

\correspondingauthor{Maria Oreshenko, Kevin Heng}
\email{maria.oreshenko@csh.unibe.ch, kevin.heng@csh.unibe.ch}

\author{Maria Oreshenko}
\affil{University of Bern, Center for Space and Habitability, Gesellschaftsstrasse 6, CH-3012 Bern, Switzerland}

\author{Daniel Kitzmann}
\affil{University of Bern, Center for Space and Habitability, Gesellschaftsstrasse 6, CH-3012 Bern, Switzerland} 

\author{Pablo M\'{a}rquez-Neila}
\affil{University of Bern, Center for Space and Habitability, Gesellschaftsstrasse 6, CH-3012 Bern, Switzerland}
\affil{University of Bern, ARTORG Center for Biomedical Engineering, CH-3008 Bern, Switzerland}

\author{Matej Malik}
\affil{Department of Astronomy, University of Maryland, College Park, MD 20742, USA}

\author{Brendan P. Bowler}
\affil{McDonald Observatory and the University of Texas at Austin \\
Department of Astronomy, 2515 Speedway, Stop C1400, Austin, TX 78712, USA}

\author{Adam J. Burgasser}
\affil{Center for Astrophysics and Space Science, University of California San Diego, La Jolla, CA 92093, USA}

\author{Raphael Sznitman}
\affil{University of Bern, ARTORG Center for Biomedical Engineering, CH-3008 Bern, Switzerland}
      
\author{Chloe E. Fisher}
\affil{University of Bern, Center for Space and Habitability, Gesellschaftsstrasse 6, CH-3012 Bern, Switzerland}

\author{Kevin Heng}
\affil{University of Bern, Center for Space and Habitability, Gesellschaftsstrasse 6, CH-3012 Bern, Switzerland}

\begin{abstract}
Self-consistent model grids of brown dwarfs involve complex physics and chemistry, and are often computed using proprietary computer codes, making it challenging to identify the reasons for discrepancies between model and data as well as between the models produced by different research groups.  In the current study, we demonstrate a novel method for analyzing brown dwarf spectra, which combines the use of the \texttt{Sonora}, \texttt{AMES-cond} and \texttt{HELIOS} model grids with the supervised machine learning method of the random forest.  Besides performing atmospheric retrieval, the random forest enables information content analysis of the three model grids as a natural outcome of the method, both individually on each grid and by comparing the grids against one another, via computing large suites of mock retrievals.  Our analysis reveals that the different choices made in modeling the alkali line shapes hinder the use of the alkali lines as gravity indicators. Nevertheless, the spectrum longward of 1.2 $\mu$m encodes enough information on the surface gravity to allow its inference from retrieval.  Temperature may be accurately and precisely inferred independent of the choice of model grid, but not the surface gravity.  We apply random forest retrieval to three objects: the benchmark T7.5 brown dwarf GJ 570D; and $\epsilon$ Indi Ba (T1.5 brown dwarf) and Bb (T6 brown dwarf), which are part of a binary system and have measured dynamical masses.  For GJ 570D, the inferred effective temperature and surface gravity are consistent with previous studies.  For $\epsilon$ Indi Ba and Bb, the inferred surface gravities are broadly consistent with the values informed by the dynamical masses.
\end{abstract}


\section{Introduction}

\subsection{Traditional use of model grids for interpreting brown dwarf spectra}

The study of atmospheres has a longer history and richer literature for brown dwarfs than for exoplanets.  The availability of high-quality, low- and high-resolution spectra for brown dwarfs serves as a prelude to spectra of exoplanetary atmospheres that we aspire to measure with the James Webb Space Telescope.

Traditionally, there are several ways to analyze brown dwarf data.  One may compare the dynamical mass and observed luminosity with grids of evolutionary models in order to derive the model-dependent age and radius, from which the gravity and effective temperature may be derived (e.g., \citealt{2003A&A...402..701B}, \citealt{2010ApJ...711.1087K}, \citealt{2011ApJ...736...47B}, \citealt{2017ApJS..231...15D}). 


A complementary approach is the development of radiative transfer models of brown dwarf atmospheres, which predict synthetic spectra (see \citealt{Helling2014A&ARv..22...80H} or \citealt{2015ARA&A..53..279M} for recent reviews). A large grid of forward models is then compared with the spectroscopic data (e.g., \citealt{1993ApJ...406..158B}, \citealt{1997ApJ...491..856B}, \citealt{2007ApJ...657..494B}, \citealt{2008ASPC..384..111C}, \citealt{2009ApJ...702..154S}).
 One may also interpolate the model spectra of the grid to fit the measured spectrum (e.g., \citealt{2010ApJS..186...63R}), or use the grid as the basis for a Markov Chain Monte Carlo calculation (\citealt{2014ApJ...793...33L}). See \cite{2017ApJ...848...83L} for a review.
 
 The model grids often involve complex physics and chemistry, and the models are computed using proprietary computer codes.  It is often challenging or infeasible to diagnose the strengths and weaknesses of model grids produced by different research groups, even if one has access to these codes.  When the match between model and data is discrepant, it is not easy to identify the reason.  These challenges motivate the development of a method to analyze the information content of model grids of brown dwarfs \textit{even without having access to the computer codes used to produce them}.  It provides the practitioner with an extra set of tools, which may either pinpoint the source of model discrepancies or allow follow-up questions to be asked. The main focus of the current study is to elucidate such a method using supervised machine learning. Such a method allows models to be confronted by data more decisively, which will motivate the development of improved models. The ultimate goal of resolving and elucidating all of the discrepancies between brown dwarf model grids is neither the intention, nor within the scope of, the current study. 
 

\subsection{Combining the use of model grids with supervised machine learning}

Given a measured spectrum, atmospheric retrieval solves the inverse problem of inferring the properties of the object (see \citealt{2018haex.bookE.104M} for a recent review).  Traditionally, atmospheric retrieval is performed using a simple forward model that is computed on the fly and used in tandem with a Bayesian method such as nested sampling or Markov Chain Monte Carlo, e.g., \cite{Line2015ApJ...807..183L}.  The Bayesian method allows for the computation of posterior distributions of properties such as the surface gravity.

In the current study, we wish to demonstrate an alternative method of performing retrieval that combines the use of pre-computed model grids with a supervised method of machine learning known as the ``random forest" \citep{marquez18}.  Random forest retrieval was previously used to perform retrieval on low-resolution transmission spectra of exoplanets \citep{marquez18}, but it has never been used to analyze brown dwarf spectra.  It offers several distinct advantages over traditional retrieval methods:
\begin{itemize}
    
    \item The forward model, which computes synthetic spectra given a set of assumptions, need not be computed on the fly.  It not only shifts the computational burden offline, but allows retrieval modelers to harness the collective effort of the community by using longstanding or classical model grids such as \texttt{BT-Settl} \citep{2014IAUS..299..271A}, \texttt{AMES-cond} \citep{allard00}, the \cite{1997ApJ...491..856B} grid, evolutionary models from \cite{2008ApJ...689.1327S}, etc.  The model grid may be of arbitrary physical (and chemical) sophistication.  It resolves the tension between computational feasibility and physical realism.
    
    \item The explicit need to provide a grid of (forward) models as the training set of the random forest means that the models need to be stored.  In traditional retrieval, models computed on the fly may, in principle, be stored, but this does not happen in practice.  This encourages reproducibility of the computed models.
    
    \item Large suites of $\gtrsim 10^2$ retrievals may be performed to quantify the predictive power of the models with respect to each parameter of the model.  For example, we demonstrate in the current study that temperature may be both accurately and precisely inferred, but this is not the case for the surface gravity (which is inferred accurately but not precisely).  In the random forest method, this is known as the ``real versus predicted (parameter values)" analysis.
    
    \item The relative importance of each data point in the synthetic spectrum towards determining the value of each parameter of the model may be quantified.  In traditional retrieval, this may be performed as an additional step by computing the Jacobians of the model, but such a step is expensive and is thus restricted to models that have a small number of parameters.  In random forest retrieval, this ``feature importance" step is a natural outcome of the method with no added computational burden and is not restricted by the complexity of the model.
    
\end{itemize}

In the current study, we apply random forest retrieval to interpret the spectra of the benchmark brown dwarf GJ 570D \citep{2000ApJ...531L..57B}  and two brown dwarfs that are part of a binary system ($\epsilon$ Indi Ba \& Bb; \citealt{2003A&A...398L..29S, 2004A&A...413.1029M}).  For GJ 570D, we demonstrate that we obtain values for the temperature and gravity that are broadly consistent between the three model grids used.  For $\epsilon$ Indi Ba \& Bb, we obtain values for the gravities that are broadly consistent with those derived from the dynamical masses.

\subsection{Layout of study}

In Section \ref{sec::brown_dwarf_data}, we describe the archival data used for the current study.  In Section \ref{sec:methods}, we describe our methodology.  Outcomes from a battery of tests as well as retrievals performed on measured spectra are presented in Section \ref{sec:results}.  Section \ref{sec:discussion} presents a summary of the study, its implications and opportunities for future work.

\section{Archival data of brown dwarfs}
\label{sec::brown_dwarf_data}

\begin{figure}
\begin{center}
\includegraphics[width=\columnwidth]{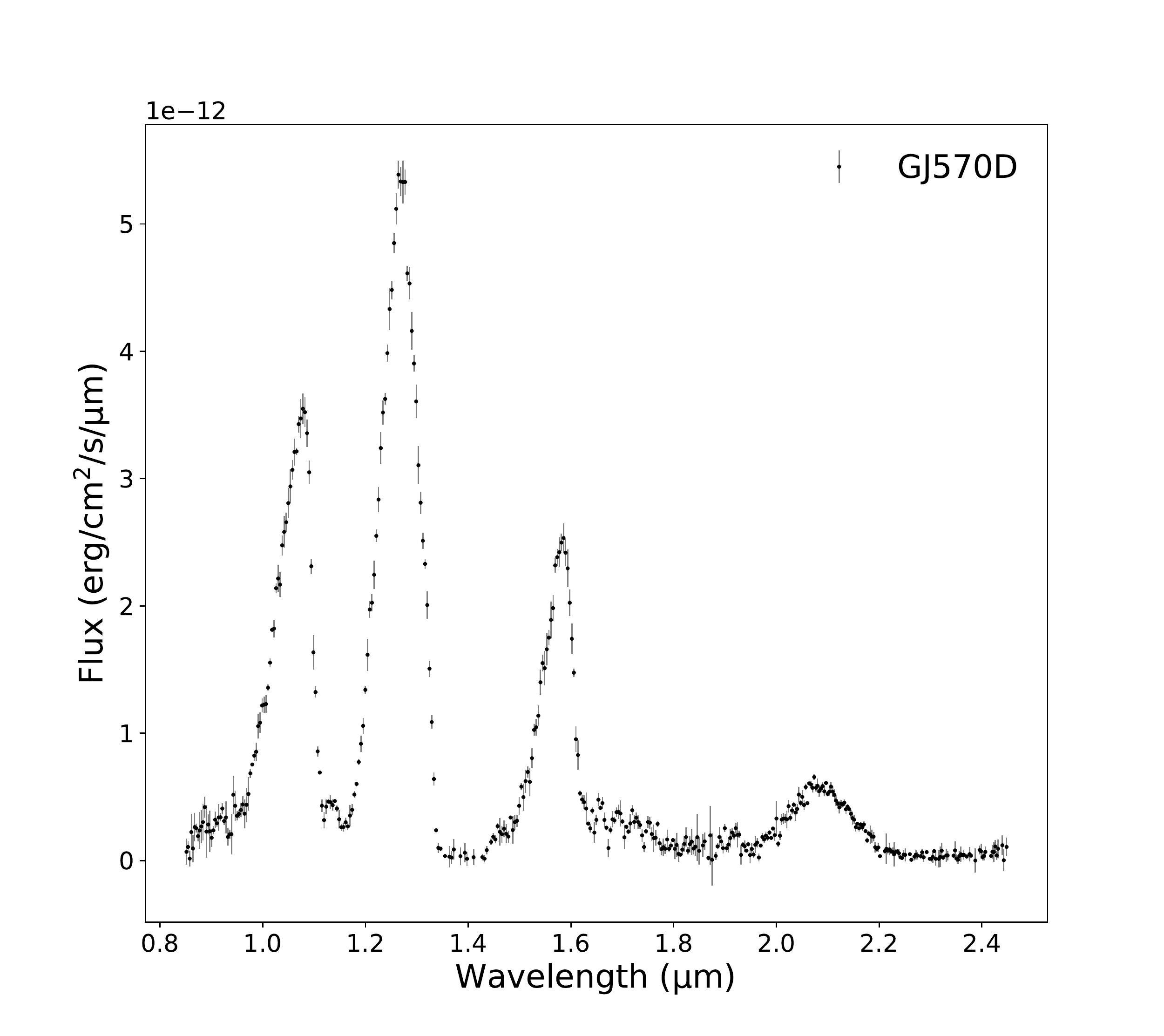}
\includegraphics[width=\columnwidth]{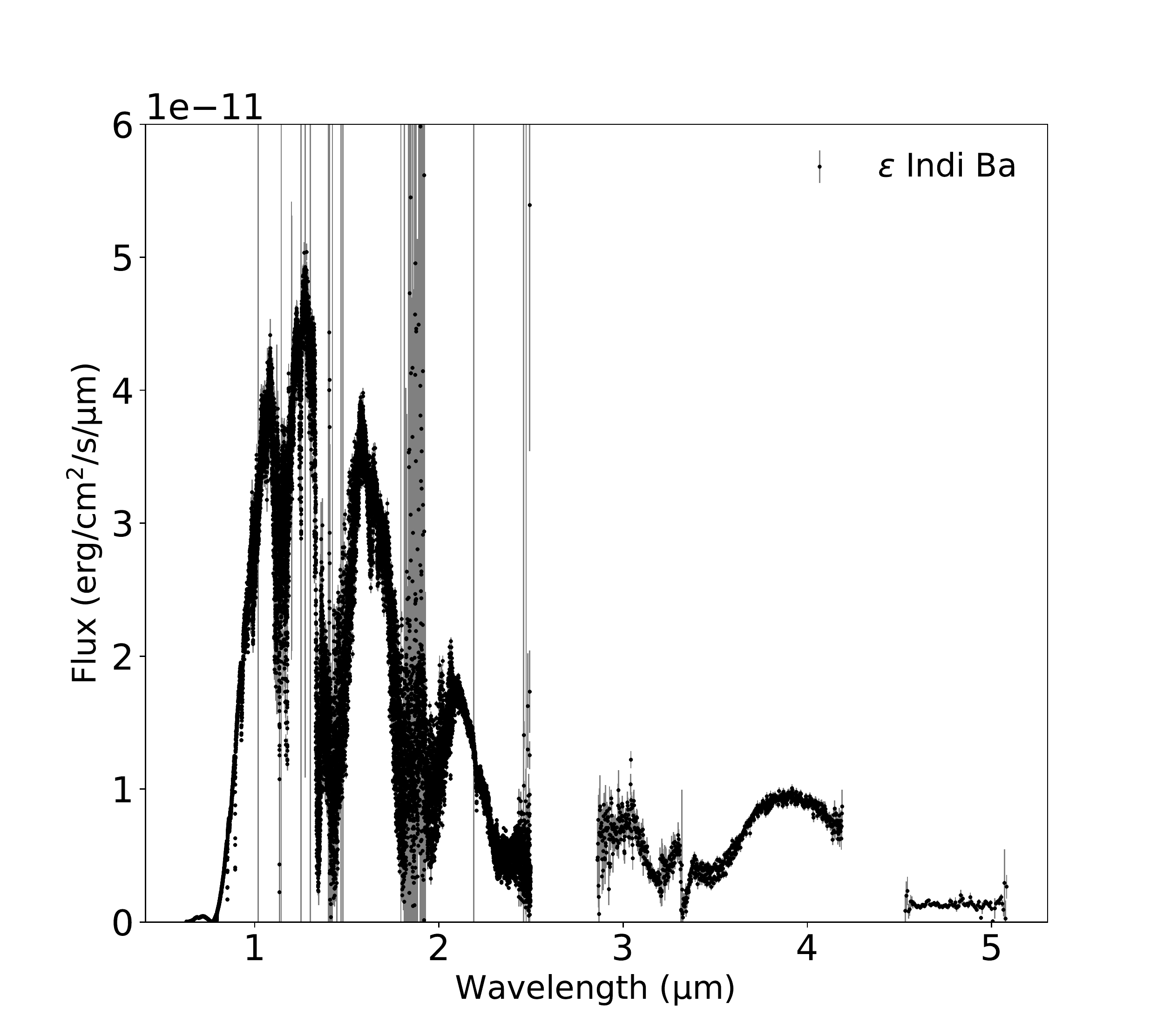}
\includegraphics[width=\columnwidth]{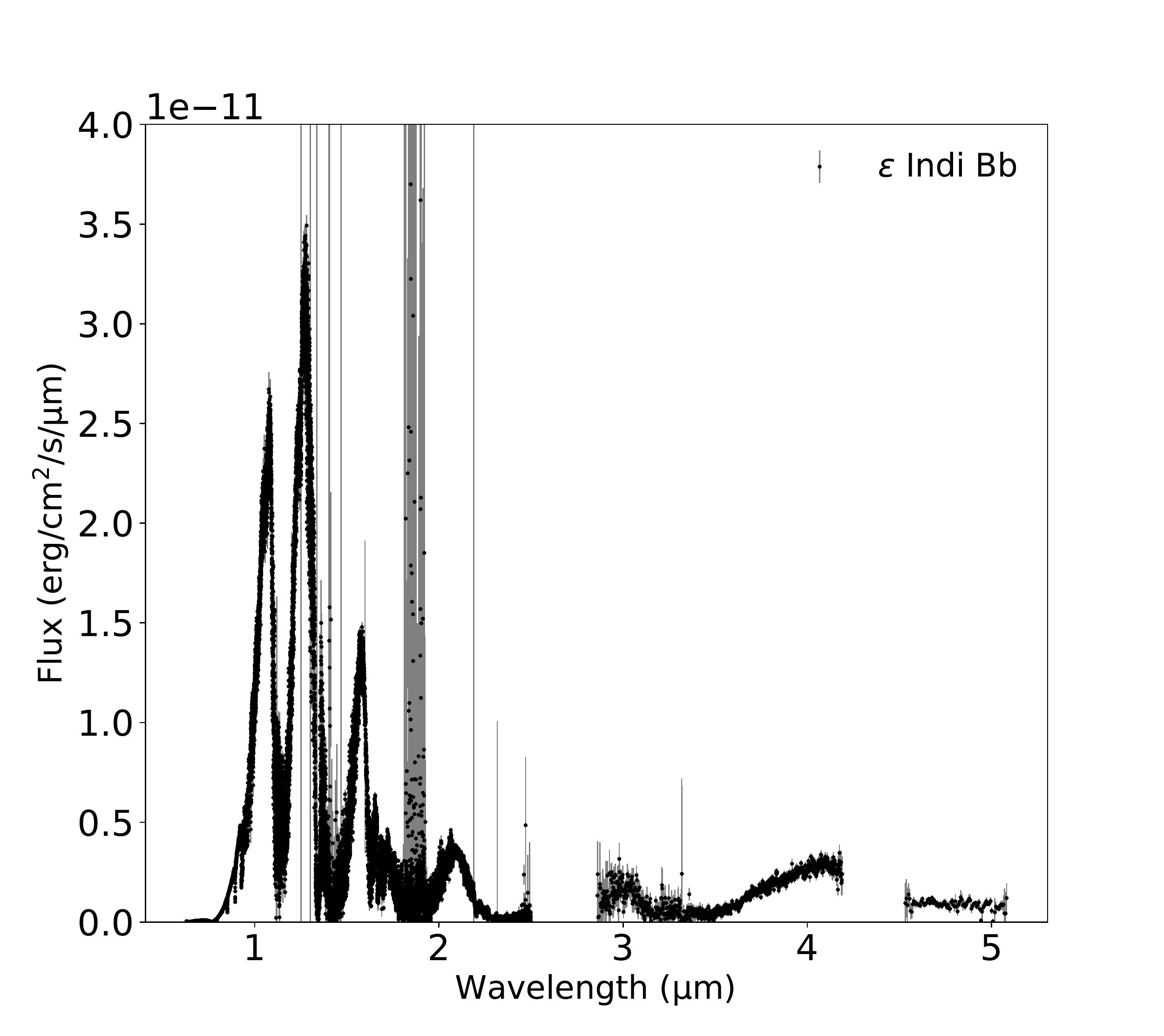}
\end{center}
\caption{Spectra of the three brown dwarfs used in this study: GJ 570D (top panel), $\epsilon$ Indi Ba (middle panel), and $\epsilon$ Indi Bb (bottom panel). For the $\epsilon$ Indi brown dwarfs, the measurements by \cite{king10} are used, while the data for GJ 570D are based on \cite{burgasser04}. }
\label{fig:data_spectra}
\end{figure}

In this study, we focus on three different brown dwarfs: the late T-dwarf GJ 570D, as well as members of a binary $\epsilon$ Indi Ba \& Bb (T1.5 and T6 class objects, respectively) \citep{2004A&A...413.1029M}. More specifically, we use a spectrum of GJ 570D from \cite{burgasser04} and two spectra of the brown dwarfs in the  $\epsilon$ Indi system published by \cite{king10}. All three measurements are shown in Fig.\ \ref{fig:data_spectra}.

The spectrum of GJ 570D \citep{burgasser04} has been taken by the SpeX instrument at the NASA Infrared Telescope Facility and provides around 400 data points from about 0.8 $\mu$m to 2.4 $\mu$m. The spectral resolution varies between about 35 and 200 throughout the spectrum. The SpeX prism spectrum of GJ 570D is flux-calibrated using 2MASS
photometry.  The multiplicative factor that scales the spectrum to
match the measured 2MASS photometry is computed separately for the $J$
(15.32 $\pm$ 0.05 mag), $H$ (15.27 $\pm$ 0.09 mag), and $K_S$ (15.24
$\pm$ 0.16 mag) bandpasses following the approach described in
\cite{Cushing:2005ed}.  Uncertainties in the scale factor take into
account spectral measurement errors and photometric uncertainties.  We
adopt the weighted mean of these three values for our final flux
calibration scale factor for GJ 570D.

The data for the two brown brown dwarfs in the $\epsilon$ Indi system each feature more than 20,000 data points from 0.63 $\mu$m to 5.1 $\mu$m. Both spectra consist of different measurements taken by the Very Large Telescope (VLT), 
using the FORS2 instrument in the optical wavelength range and the ISAAC spectrograph in the near-infrared and infrared. Compared to the GJ 570D data, the $\epsilon$ Indi brown dwarf spectra offer a much higher resolution but also have spectral regions with elevated noise levels. Especially the data points at about 1.4 $\mu$m and 1.8 $\mu$m show a very wide spread in the flux values, with large error bars, which might have been caused by imperfect telluric correction.

\section{Methodology}
\label{sec:methods}

\subsection{Atmosphere model grids}
\label{sec:model_grids}

In this study, we use pre-computed grids from three different atmosphere models: \texttt{HELIOS} \citep{malik17, malik19}, \texttt{AMES-cond} \citep{allard00}, and the \texttt{Sonora} model (Marley et al.\ in prep). Unlike the simplified forward models in a standard retrieval approach \citep[e.g.][]{Line2015ApJ...807..183L}, each of the models employed here is a self-consistent atmosphere model, i.e.\ the only free parameters are the effective temperature $T$, the surface gravity $\log(g)$, and the elemental abundances. The atmospheric structure problem is described in detail in \cite{Helling2014A&ARv..22...80H} and \cite{2015ARA&A..53..279M}. In principle, the abundances of the chemical elements (O/H, C/H, N/H, etc.\ ) could be varied independently. However, for simplification, a common approach is to keep their \textit{ratios} at their solar value and scale all abundances by a common factor with respect to hydrogen. This factor, or more precisely its logarithm, is usually referred to as the metallicity. In the following, we briefly summarize the main features of each model.

\paragraph{\textbf{Sonora}} The \texttt{Sonora} spectral model set\footnote{https://zenodo.org/record/1309035} is derived by first computing radiative-convective equilibrium atmosphere structures for a specified set of effective temperatures and gravities. The \texttt{Sonora} models employ a layer-by-layer convective adjustment method permitting solutions for detached convective zones. Chemistry is computed using the rainout method in which condensed species are removed from the atmosphere and not permitted to further react with gaseous species at lower temperatures. This choice plays an important role in the alkali chemistry in particular and is a principal difference among the model sets employed here. The model grid used for this study neglects cloud opacity. Once an atmospheric thermal model is converged a final emergent spectra at high spectral resolution ($R\sim25,000$) is computed given the computed abundances and the opacities described in \cite{2008ApJS..174..504F} and \cite{2014ApJS..214...25F}. More details can be found in Marley et al. (in prep).  


\paragraph{\textbf{HELIOS}} The open-source radiative transfer code \texttt{HELIOS} \citep{malik17, malik19}\footnote{https://github.com/exoclime/HELIOS} utilizes an improved hemispheric two-stream method \citep{2017ApJS..232...20H, heng18} with convective adjustment to obtain the converged atmospheric solution in radiative-convective equilibrium. The included opacity sources and the corresponding line lists are given in Table 1 of \cite{malik19}. Most opacities are calculated with \texttt{HELIOS-K} \citep{grimm15} at a resolution of 10$^{-2}$ cm$^{-1}$, using a Voigt profile with a wing cut-off at 100 cm$^{-1}$. Pressure broadening is included as provided by default in the ExoMol and Hitran online databases. The Na and K treatment is based on \cite{burrows00} and \cite{burrows03}, described in detail in Appendix A of \cite{malik19}.The equilibrium gas-phase chemistry is calculated with \texttt{FastChem} \citep{stock18} based on the elemental abundances given in Table 1 of \cite{asplund09}. Removal of species due to condensation is included for H$_2$O, TiO, VO, SiO, N and K, and described in Appendix B of \cite{malik19}. \texttt{HELIOS} employs the $\kappa$-distribution method with correlated-$\kappa$ approximation, using 300 wavelength bins with 20 Gaussian quadrature points over 0.33 $\mu$m - 10 cm. The spectra are post-processed at a resolution of 3000.

\begin{figure}

\begin{center}
\includegraphics[width=\columnwidth]{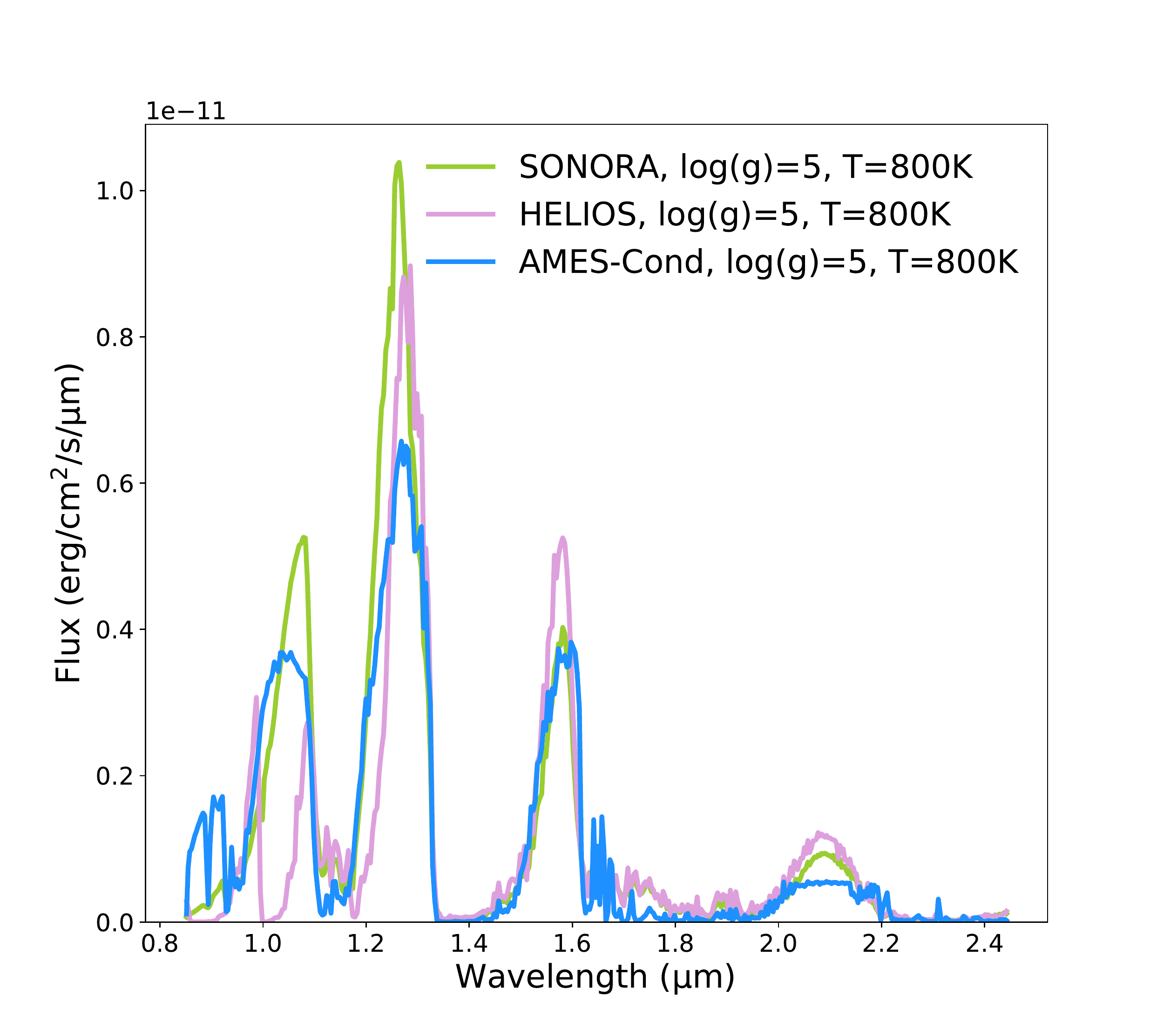}
\end{center}
\caption{Examples of spectra from the different model grids for a typical late T-dwarf. The spectra are shown for an effective temperature of 800 K, a $\log(g)$ of 5, and solar metallicity. For presentational reasons, the original high-resolution spectra provided by each model are binned down to a lower resolution in this figure.}
\label{fig:spectra}
\end{figure}

\paragraph{\textbf{AMES-cond}} 
The brown dwarf model grid \texttt{AMES-cond} \citep{allard00}\footnote{http://perso.ens-lyon.fr/france.allard} is based on the well-known \texttt{PHOENIX} stellar atmosphere code \citep{Hauschildt1992JQSRT..47..433H, Hauschildt1997ApJ...483..390H}. The \texttt{PHOENIX} model solves the radiative transfer equation by using an accelerated lambda iteration approach in combination with the opacity sampling technique. 
The \texttt{AMES-cond} model builds upon the previously published atmospheric grid \texttt{NEXTGEN} \citep{Allard1997ARA&A..35..137A}, but includes further improvements with respect to the dust chemistry and opacities required to describe the cool atmospheres of brown dwarfs. Absorption coefficients for H$_2$O and TiO have been replaced by \cite{2000ApJ...540.1005A} in the \texttt{PHOENIX} model. Condensation of dust is included by assuming continuous chemical equilibrium between the gas and all condensed phases. Similar to \texttt{Sonora}, the resulting dust opacity is neglected. 
\\

Each of the three models differ in terms of model physics complexity, chemistry, or opacities. To illustrate the differences, Fig.\ \ref{fig:spectra} shows an exemplary spectrum from each grid for solar metallicity, an effective temperature of 800 K and a $\log(g)$ value of 5\footnote{Unless stated otherwise, we express values of $g$ and $\log(g)$ in cgs units throughout this work.}. These parameters resemble a typical late T-dwarf with a cloud-free photosphere. The figure clearly suggests that even for the same model parameters, the spectra can show significant differences. This is especially true for the wavelength range below 1.2 $\mu$m. This region is dominated by the line wings of the alkali resonance lines. These lines are known to have a strong non-Lorentzian far-wing line profile, for which various approximations have been developed in the past \citep[e.g.][]{Tsuji1999ApJ...520L.119T, burrows00, burrows03, Allard2012A&A...543A.159A, Allard2016A&A...589A..21A}. Depending on what approach is used for these line profiles in an atmospheric model, the resulting spectra can exhibit large discrepancies. Moreover, \texttt{AMES-cond} models use equilibrium chemistry, not rain-out chemistry. This will result in significant differences at around 800K.

Additionally, \texttt{HELIOS} shows a pronounced feature near 1 $\mu$m due to CrH absorption that neither \texttt{Sonora} nor \texttt{AMES-cond} possess. Furthermore, all three models also partially disagree in regions of opacity minima, at the spectral peaks in J, H, and K bands. At 1.3 $\mu$m, for example, the flux provided by \texttt{AMES-cond} is by a factor of about 1.7 smaller than the one predicted by the \texttt{Sonora} model.

\begin{deluxetable}{lccc}
\tablecaption{Properties of the three different atmosphere model grids used in this study. \label{tab:model_grids}}
\tablecolumns{4}
\tablewidth{0pt}
\tablehead{
\colhead{} &
\colhead{\texttt{Sonora}} &
\colhead{\texttt{AMES-cond}} &
\colhead{\texttt{HELIOS}}
}
\startdata
Total \# of models& 400 & 150 & 700 \\
Range of $T$ (K) & 200 - 2400 & 300 - 2400 & 200 - 3000 \\
Range of $\log(g)$ & 3.25 - 5.5 & 3.0 - 6.0  & 1.4 - 6.0 \\ 
\enddata
\end{deluxetable}

A grid of self-consistent brown dwarf atmospheres has been generated by each of the models. The grids provide tabulated photospheric spectra of brown dwarfs as a function of effective temperature and surface gravity. They are, however, restricted to solar metallicity. The grid sizes (in terms of temperature range or $\log(g)$ values), as well as the grid step sizes differ greatly between the three models. Table \ref{tab:model_grids} gives a summary of the three different grids. \texttt{AMES-cond} offers the smallest grid, with only 150 models. The number of models in each grid ranges from 150 to 700, with \texttt{HELIOS} offering the largest grid, but there is a good overlap in the phase space explored in the range 300 to 2400K and $\log{(g)}$ 3.25 to 5.5.

\subsection{Atmospheric retrieval using the random forest method}

For the retrieval calculations in this study, we employ the random forest technique. In particular, we use our open-source code \texttt{HELA}\footnote{https://github.com/exoclime/HELA} that has previously been applied for analyzing WFC3 data of exoplanet atmospheres \citep{marquez18}. A detailed description of \texttt{HELA} can be found in \cite{marquez18}. The code implements the random forest algorithm \citep{Ho98, breiman01} applied to a pre-computed grid of forward models for brown dwarf atmospheres. 

We use 3000~regression trees in our forests. We do not impose a maximum depth of the trees during training. Instead, each tree grows by splitting the space of models until the decrease in variance of further splits is smaller than~$0.01$. 

To check how well the random forest is performing, it is necessary to use a subset of the grid for a testing procedure. This procedure tests models with known parameters on a trained random forest and compares the outcome with the actual, true parameters of the injected testing set. Each model grid is therefore divided into training and testing parts, with 20 percent of the grid models being used for testing.

For a proper performance of the random forest's training algorithm, a large grid of spectra is usually required. However, even the largest grid used in this study only provides about 700 unique models distributed throughout the parameter space. Especially the $\log(g)$ parameter space is rather restricted compared to the temperature space in all three model grids. Training the random forest with such few models per grid would lead to a non-convergence of the random forest algorithm. We therefore need to artificially increase the grid sizes by creating new spectra via interpolation within the grids.

By running a number of training tests (not shown), we estimate that increasing the grid size in the $\log(g)$ space by a factor of ten is sufficient to properly train the random forest. The temperature spacing in the grids is usually already sampled densely enough, such that adding new temperature points to the grids proved to be unnecessary. New spectra are therefore generated by interpolating linearly between $\log(g)$ values. 

It should be noted that because the original grids already have different sizes and sampling steps, the final grids used for the random forest also differ in parameter range, total grid size, and parameter step sizes. We perform the retrieval using each of the grids separately in order to compare the results from different forward models, but it is in principle also possible to use all three grids at once to train the random forest algorithm. 

\subsubsection{Retrieval parameters}

The two main retrieval parameters are the effective temperature $T$ and the surface gravity $\log(g)$. Since the grids are restricted to solar elemental abundances, neither the metallicity nor the C/O ratio can be retrieved, even though the actual brown dwarf atmospheres are not expected to all have a solar-like elemental composition. The priors for $T$ and $\log(g)$ are given by the tabulated parameter ranges of each grid (see Table \ref{tab:model_grids}).

For the retrieval of actual brown dwarf data (see Section \ref{sec:retrieval_spectra}), we also need to take into account the geometric dilution of the photospheric spectrum, depending on the stellar radius and the distance of the star to the observer. For the distances, (more or less accurate) parallax measurements are usually available. Radii, on the other hand, cannot be directly measured unless the brown dwarf is transiting a host star. Here, we need to choose values based on, for example, evolutionary track calculations of brown dwarfs (see Section~\ref{sec::brown_dwarf_data} for details).

In addition to the effective temperature and gravity, we add a flux calibration factor $f$ as a third parameter to our retrieval. It is used to scale the radius-distance relation for the flux $F_\nu$ of the brown dwarf as measured by the observer:
\begin{equation}
F_\nu = F_{\nu,*} f \left( \frac{\mathcal{R}}{d}\right)^2 \ ,
\label{eq:distance_relation}
\end{equation}
where $F_{\nu,*}$ is the photospheric flux of the brown dwarf, $\mathcal{R}$ the brown dwarf radius, and $d$ the distance. The calibration factor accounts for uncertainties in measured distances and inferred radii but also the impact of the photometric calibration or inadequacies of the atmospheric models. As prior, we use values between 0.5 and 2 for $f$ in the following.

\section{Results}
\label{sec:results}

\subsection{Model dependence of alkali lines as gravity diagnostics}
\label{sec:spectrum_cut}

One of the natural outcomes of the random forest's training procedure are the so-called \textit{feature importance} plots \citep{marquez18}. These plots describe the contribution of certain wavelengths to learning a specific model parameter. They are inherently useful to obtain estimates on the information content of certain wavelength regions, which can be directly exploited for e.g.~planning future observations.
The feature importance plots can also be used to compare the different grids and to verify that the random forest algorithm is not dominated by adapting to the noise. Therefore, one expects that the data points with low signal-to-noise level do not have a large feature importance value for the parameter estimation (e.g., in the plots corresponding to the GJ 570D data one can see that the region around 1.4 $\mu$m is ``empty'' in all feature importance plots).

\begin{figure}[h!]
\includegraphics[width=0.9\columnwidth]{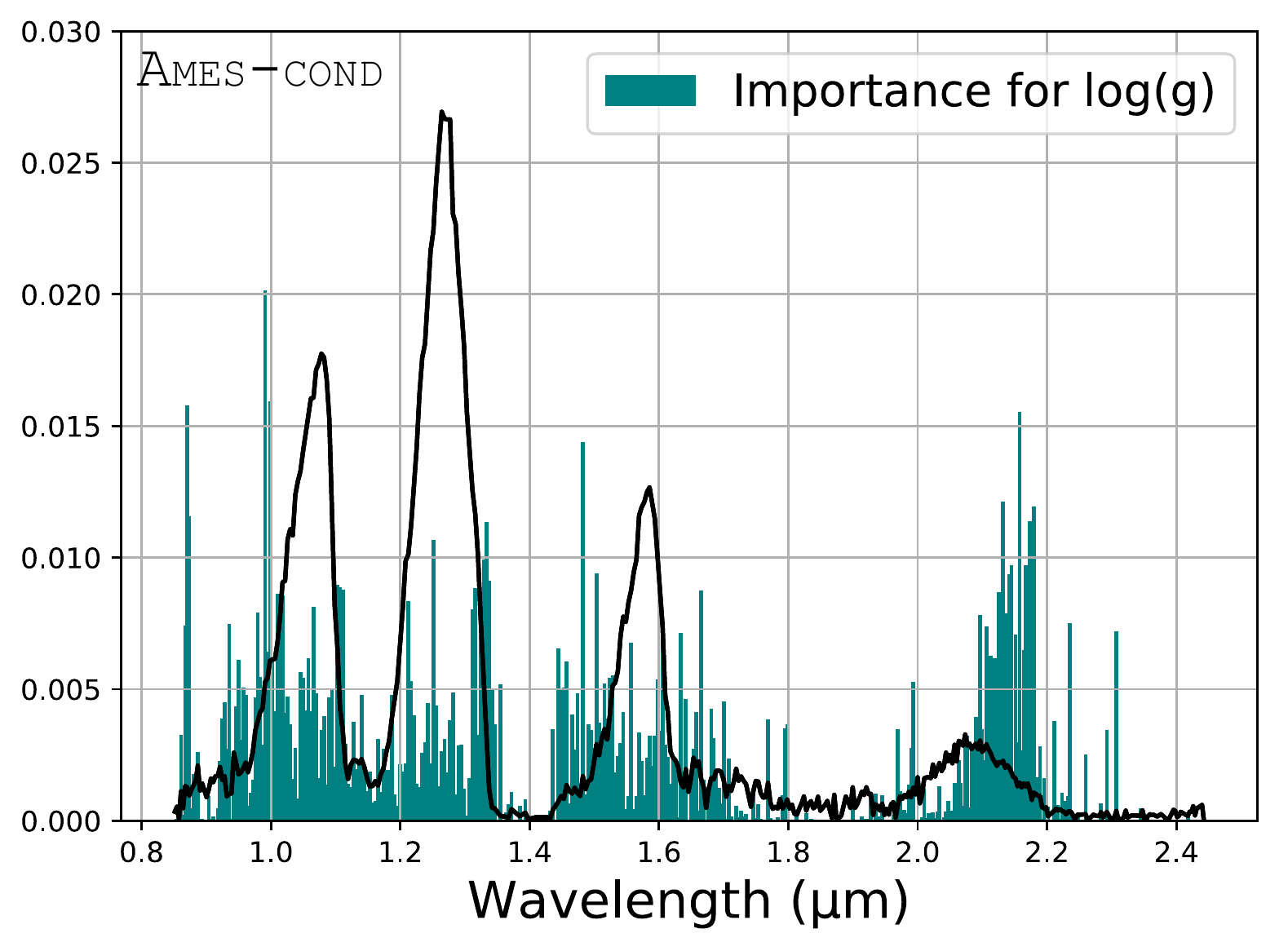}\\
\includegraphics[width=0.9\columnwidth]{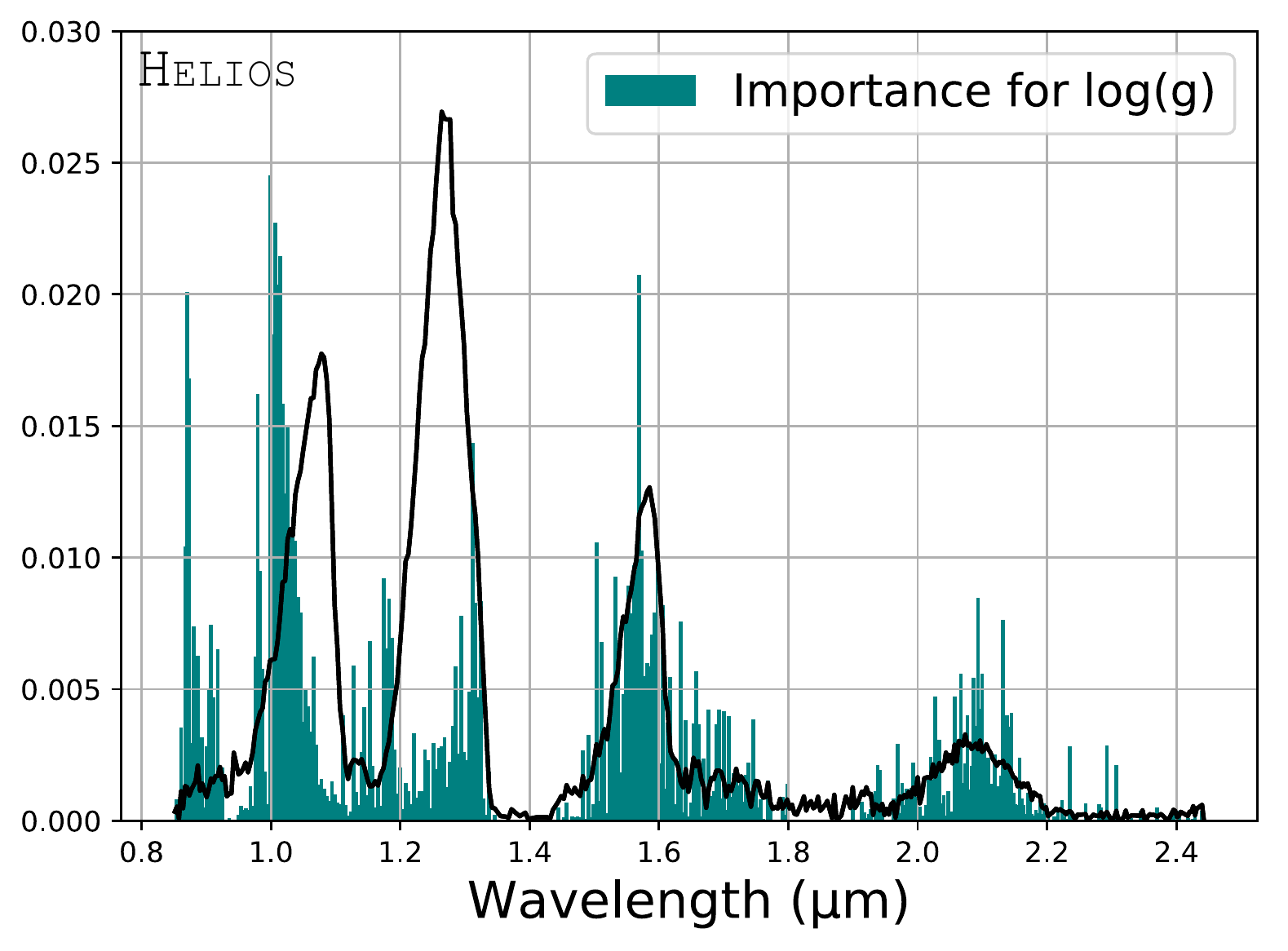}\\
\includegraphics[width=0.9\columnwidth]{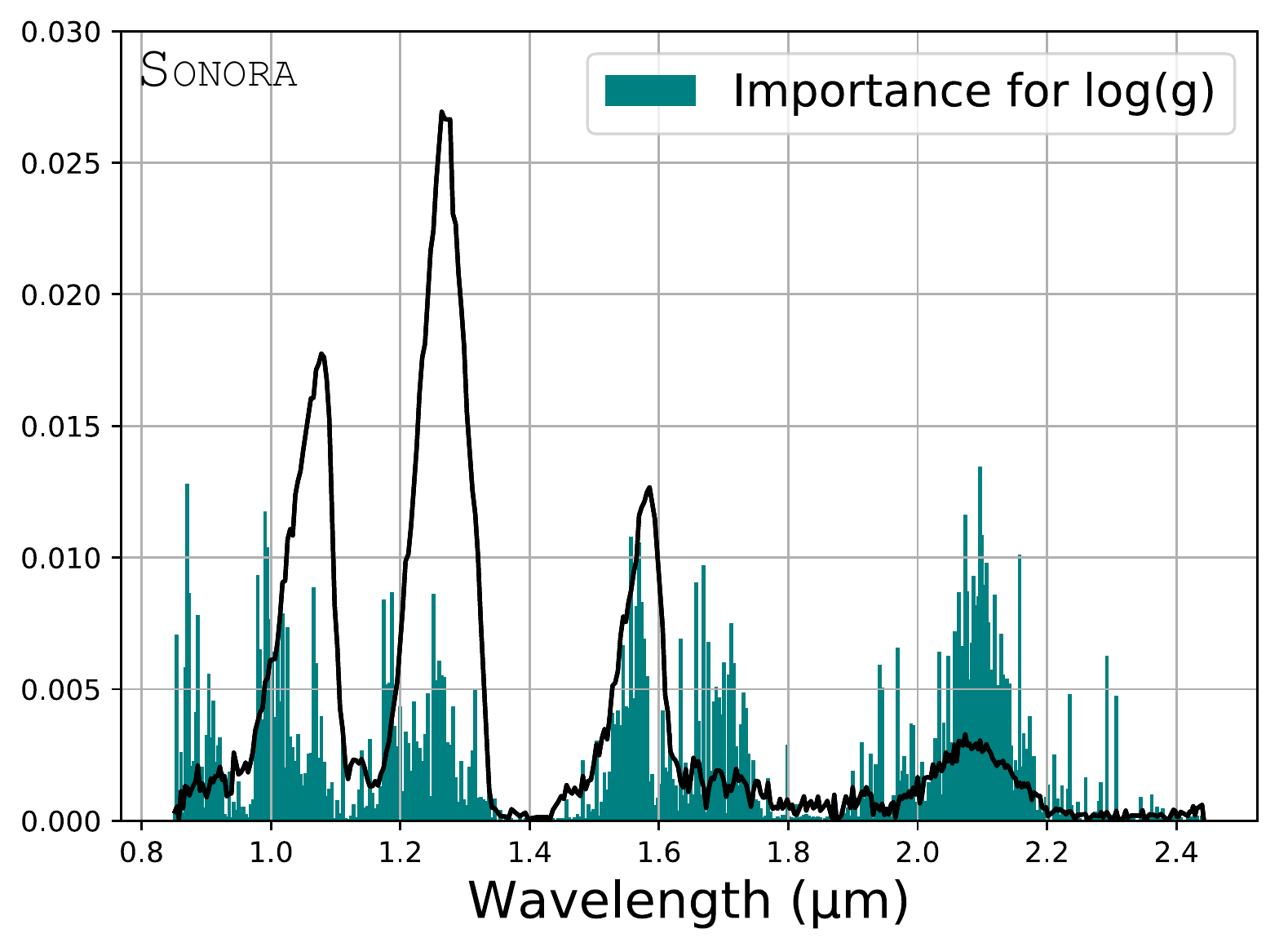}
\caption{Feature importance plots for gravity: \texttt{AMES-cond}, \texttt{HELIOS}, \texttt{Sonora}.  The x-axis is the wavelength in $\mu$m and the y-axis is the \emph{importance}. The importance of a feature is the normalized reduction in variance brought by that feature during training. More specifically, the importance is the sum of the reduction of variance achieved each time a feature is used for a split in a tree of the forest. The scaled spectrum of GJ 570D (solid, black lines) is shown for reference. }
\label{fig:feature_imp_g}
\end{figure}

Figure \ref{fig:feature_imp_g} shows the feature importance plots for the surface gravity. For reference, we also add the scaled spectrum of GJ 570D to the feature importance plot to visualize the connection between the spectral features and the feature importance values as a function of wavelength.

The results presented in Figure \ref{fig:feature_imp_g} clearly suggest that wavelengths shorter than about 1 $\mu$m are highly important for the prediction of the $\log(g)$ values from observed spectra. This confirms the outcome of the parameter sensitivity analysis by \cite{Line2015ApJ...807..183L} who obtained similar results for predicting the surface gravity.

On the other hand, as mentioned in Section \ref{sec:model_grids} and depicted in Figure \ref{fig:spectra}, this is also the wavelength region where the three model grids show the largest discrepancies. Owing to the strong alkali resonance line wing opacity and the various approaches to describe it, the results in this part of the spectrum are highly model-dependent. For consistency in applying the three different model grids, we therefore neglect the wavelength region below 1.2 $\mu$m in the following. Our predictions for $\log(g)$, thus, focus on the other important regions, most notably the ones at about 2.1 $\mu$m and 1.6 $\mu$m which seems to provide constraints on the surface gravity by all three grids. Those regions have been empirically demonstrated to be log(g)-sensitive, see for example \cite{2006ApJ...639.1095B}.

\subsection{Comparing predictions of model grids} 
\label{sec:model_comparison}

\begin{figure*}
\gridline{\fig{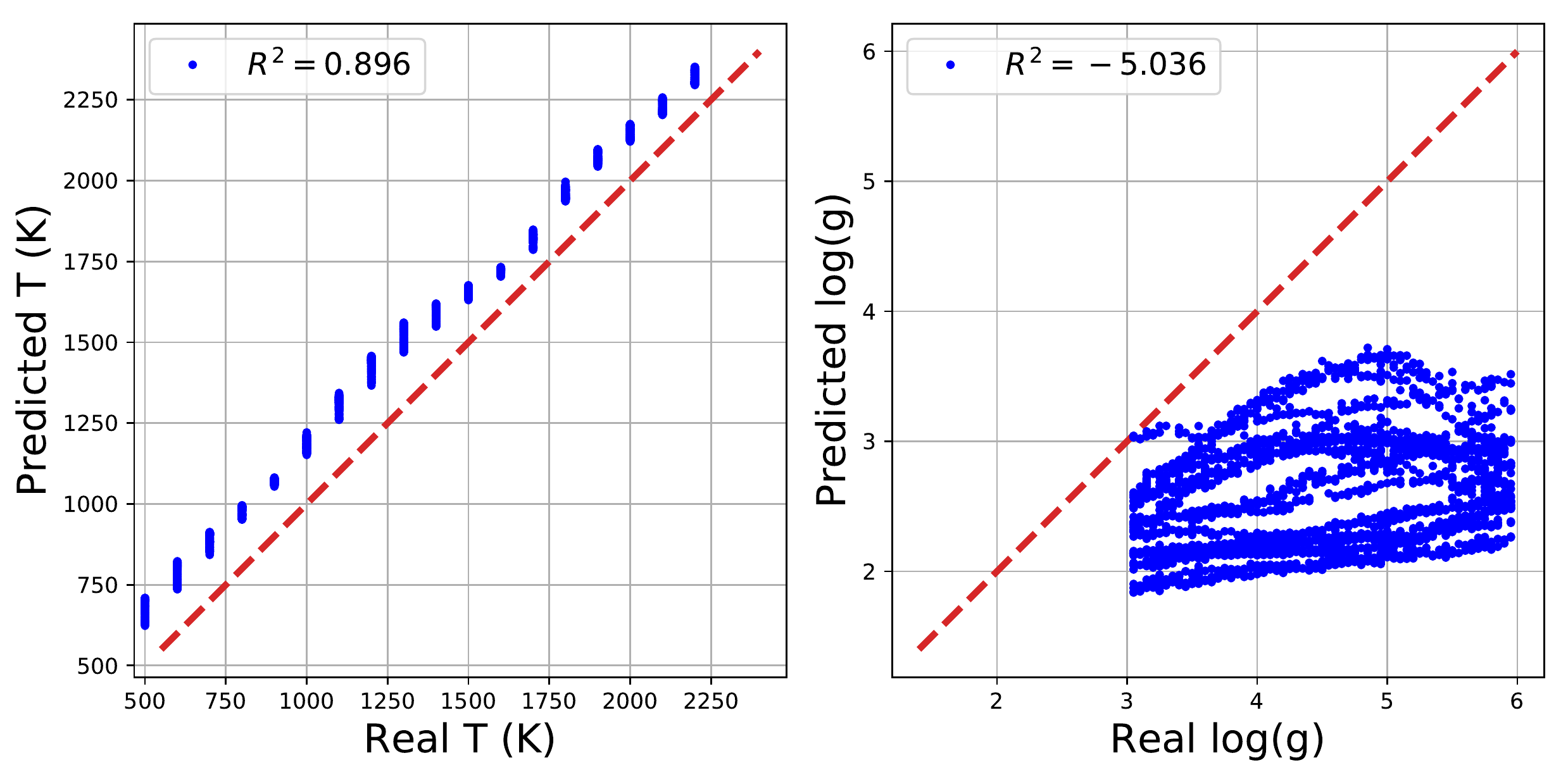}{0.49\textwidth}{(a) Tested on \texttt{AMES-cond}, full spectral range}
          \fig{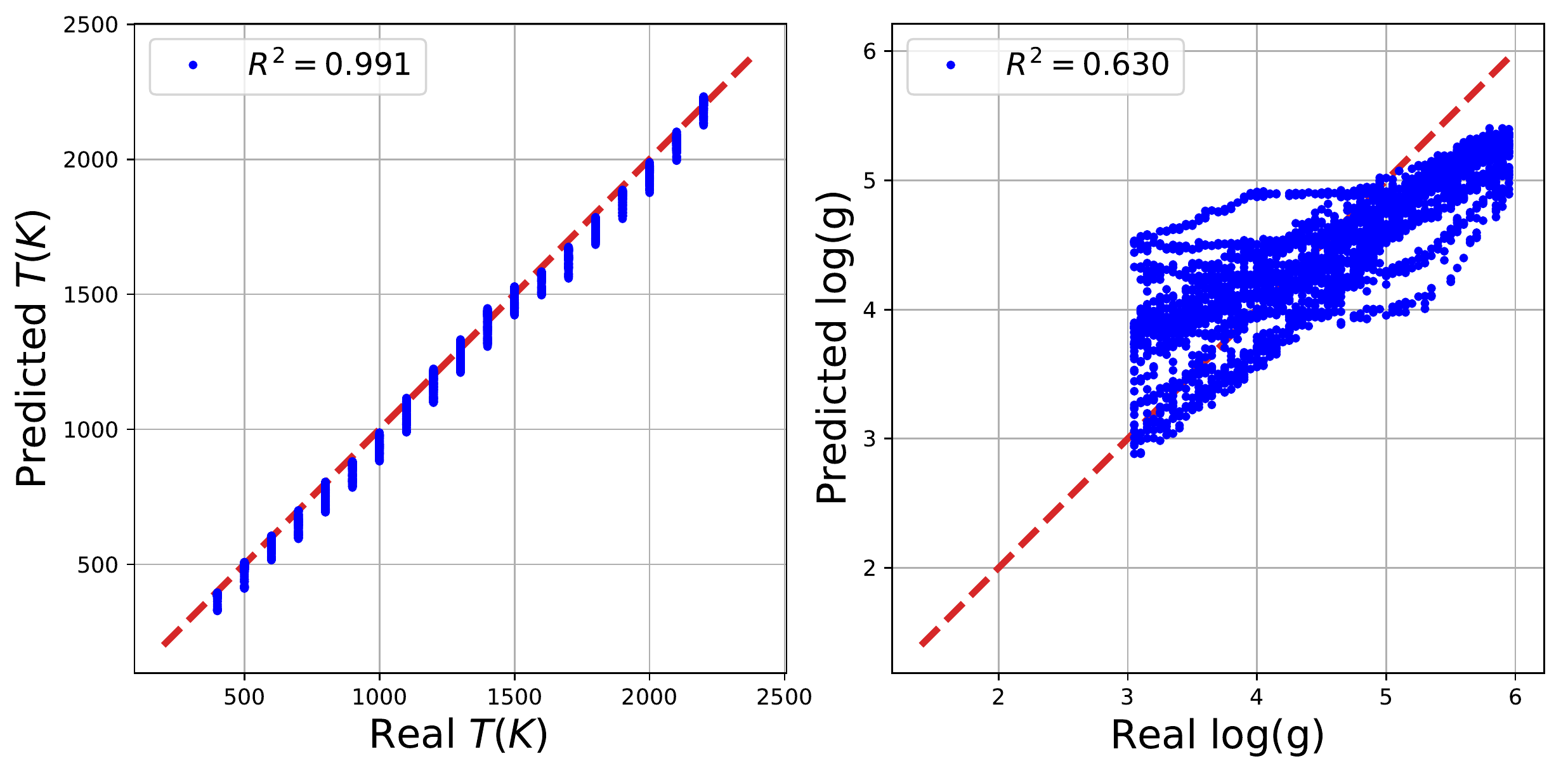}{0.49\textwidth}{(b) Tested on \texttt{AMES-cond}, spectrum cut below 1.2 $\mu$m}
          }
\gridline{\fig{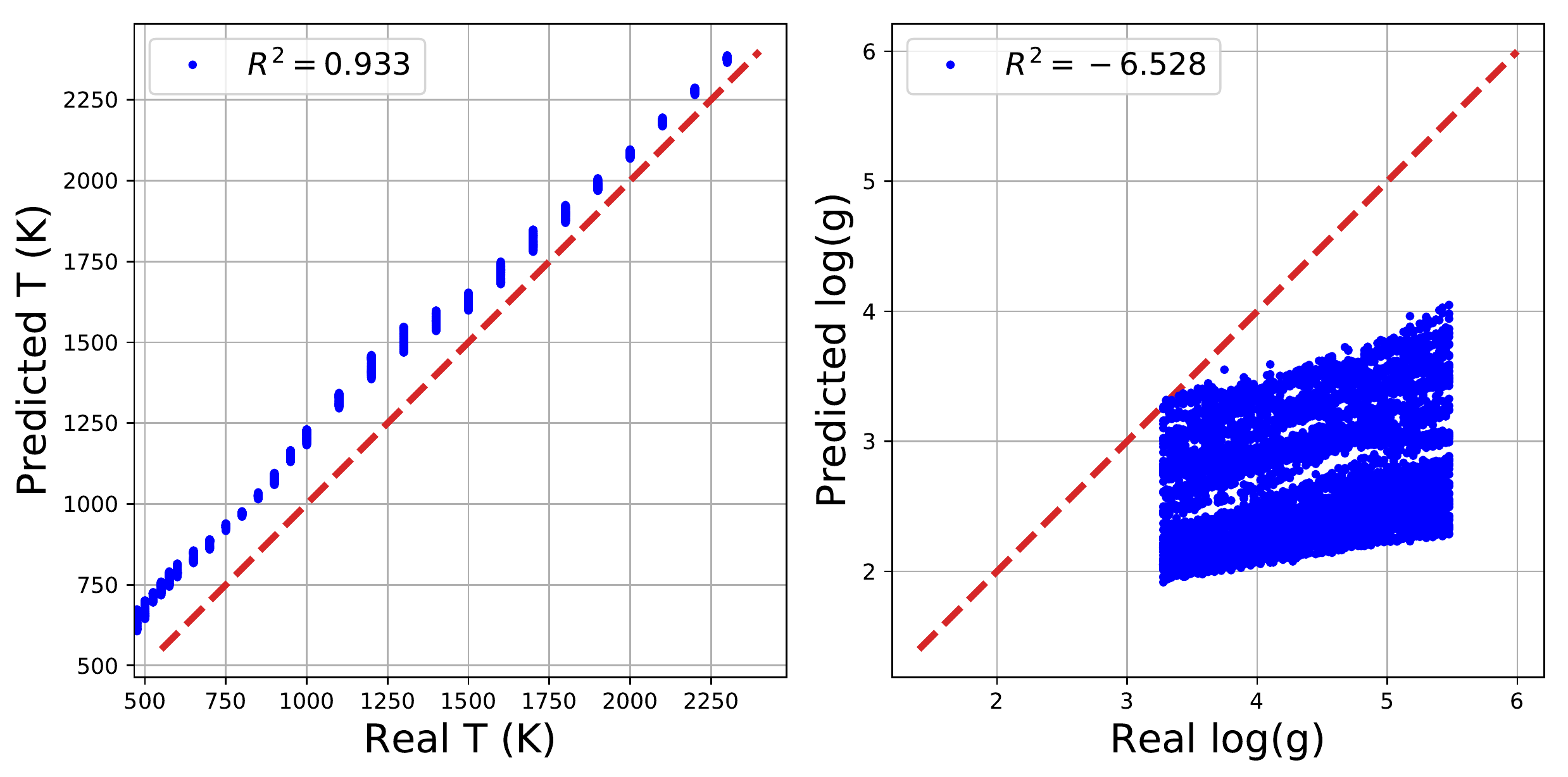}{0.49\textwidth}{(c) Tested on \texttt{Sonora}, full spectral range}
          \fig{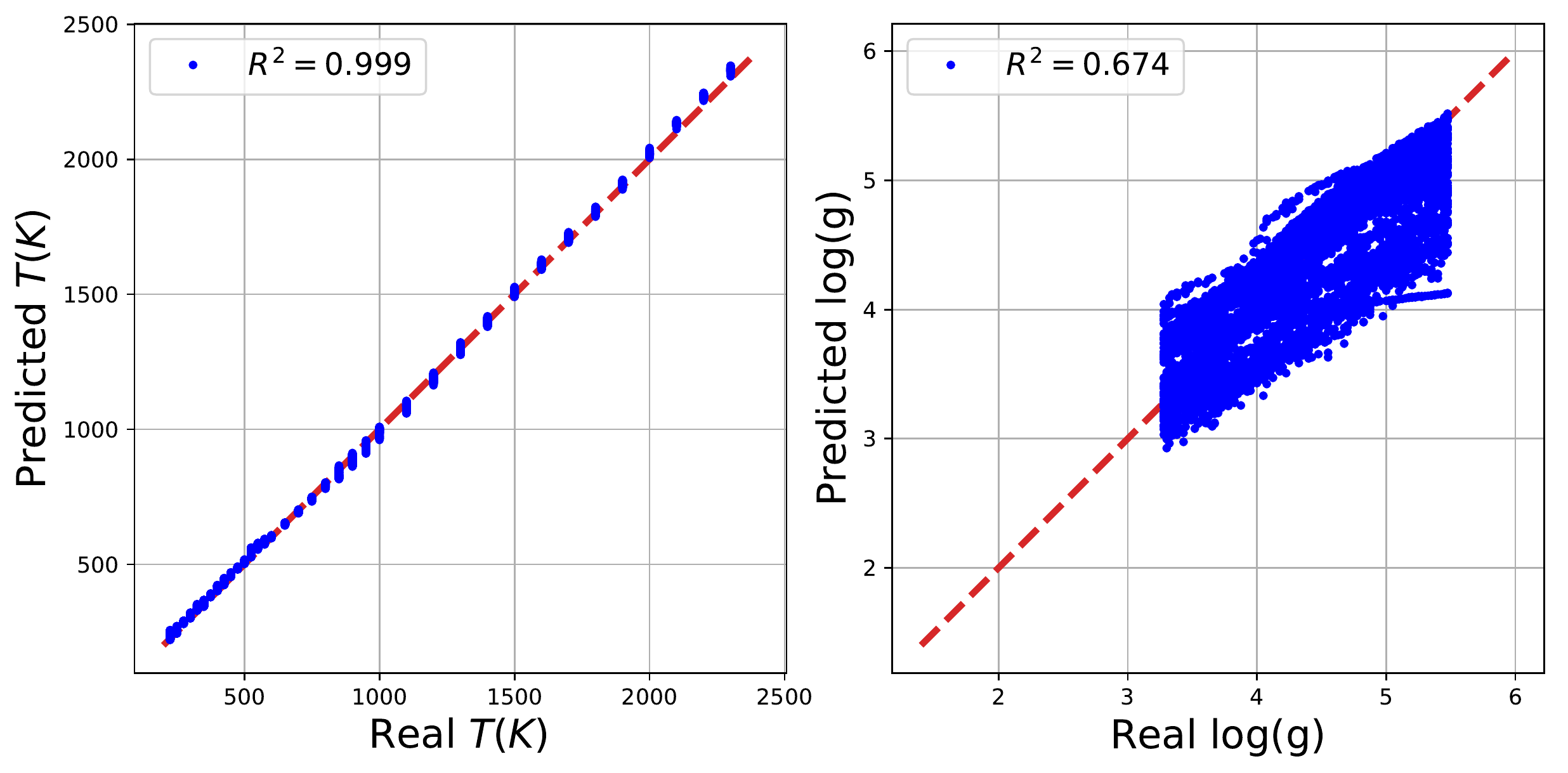}{0.49\textwidth}{(d) Tested on \texttt{Sonora}, spectrum cut below 1.2 $\mu$m}
          }
\caption{Comparison of the random forest training and testing procedure using different grids. The random forest is trained on the full \texttt{HELIOS} grid and then tested on \texttt{AMES-cond} (upper panel) and \texttt{Sonora} (lower panel) spectra. Note that both \texttt{Sonora} and \texttt{AMES-cond} cover a smaller parameter range than \texttt{HELIOS}. The left panels use the full spectral range from 0.8 $\mu$m to 2.4 $\mu$m, while the spectra corresponding to the right-hand side panels are cut below 1.2 $\mu$m (see text for details). }
\label{fig:compare}
\end{figure*}

One of the advantages of the random forest framework is the ability to quantify the differences between model grids. We therefore start by performing a grid comparison before we proceed to the retrieval itself.
The comparison of forward models grids by means of training and testing the random forest algorithm on the subsets from different grids is a novel way to analyze the differences between the radiative transfer models, opacity treatments, chemistry calculations, or numerical implementations used to create the different grids. 

The comparison is done by training the random forest algorithm on one of the grids and then performing the testing phase on a different grid, essentially performing a large suite of mock retrievals. We choose to do the training procedure of the random forest on the \texttt{HELIOS} grid because it offers the widest parameter range in terms of effective temperatures and surface gravities. After the training, we tested the random forest on the \texttt{Sonora} and \texttt{AMES-cond} grids. For the testing, we randomly pick parameter combinations from the testing set and let the random forest predict the retrieved parameters based on its training data. The corresponding results are shown in Fig. \ref{fig:compare}. We perform the training and testing on both, the full wavelength range from 0.8 $\mu$m to 2.4 $\mu$m as well as the one cut at 1.2 $\mu$m (see Sect. \ref{sec:spectrum_cut}). Note that we do not perform the testing for the calibration factor $f$ since it is not part of the original atmosphere grids and, therefore, model-independent.

For a perfectly trained random forest, the predicted, retrieved values would correspond to the parameters from the injected testing set, i.e. all values should lie on the red lines in Fig. \ref{fig:compare}, with $R^2$ values equal to unity. The $R^2$ values essentially provide a metric for describing the similarities in the spectral features that constrain parameters like the effective temperature or surface gravity.
Obviously, as suggested by the results shown in Fig. \ref{fig:compare}, this is not the case. In general, all grids are able to predict the effective temperatures more or less consistently. The $R^2$ values for these cases are mostly larger than 0.9, especially for the cases where the wavelength range is cut at 1.2 $\mu$m.

Predicting the gravity, on the other hand, seems to be more challenging. The predictions when testing with \texttt{Sonora} and \texttt{AMES-cond} and using the full wavelength range provide $\log(g)$ values that are much lower than those of the training set, resulting in negative $R^2$ values. Based on the large differences of the actual spectra in Fig. \ref{fig:spectra} and our discussion about the model discrepancies below 1.2 $\mu$m due to the alkali line wing prescriptions and CrH feature in \texttt{HELIOS} spectra, this result is not surprising. This outcome further corroborates our approach of neglecting this part of the spectrum for the actual retrieval of brown dwarf spectra.

When using the smaller wavelength range (right panel in Figure \ref{fig:compare}), the predicted values for the surface gravities correspond much better to the actual values drawn from the testing set; the $R^2$ values improve quite significantly after the data below 1.2 $\mu$m are discarded. Compared to the quite tight predictions of the effective temperatures, the testing on the $\log(g)$ values reveals a wider spread that is due to a combination of model degeneracies and different modeling choices.

Thus, what the random forest recognizes as ``gravity features" in the spectra seems to be more model specific than spectral features that constrain the effective temperature. 
Reasons for this outcome may be based on different model treatments of e.g. chemistry, opacities, or numerical implementations, all of which are not reflected in our limited set of retrieved parameters and, thus, will be ``hidden" in what the random forest model understands as ``gravity features". This emphasizes the fact that retrieving surface gravities directly from spectra does often not yield very precise results and is quite model dependent. 

In addition to training and testing on different grids, we also performed this procedure using each grid independently. The results are shown and discussed in Appendix \ref{sec:appendix_comparison}.

\subsection{Information content analysis of spectra from different model grids}

As mentioned in Section \ref{sec:spectrum_cut}, the feature importance plots describe the contribution of certain data points to learning the impact of a given parameter on the spectrum. These plots are a natural outcome of the training phase.

\begin{figure*}[p]
\begin{center}
\subfloat[\texttt{AMES-cond}]{\includegraphics[width=0.82\textwidth]{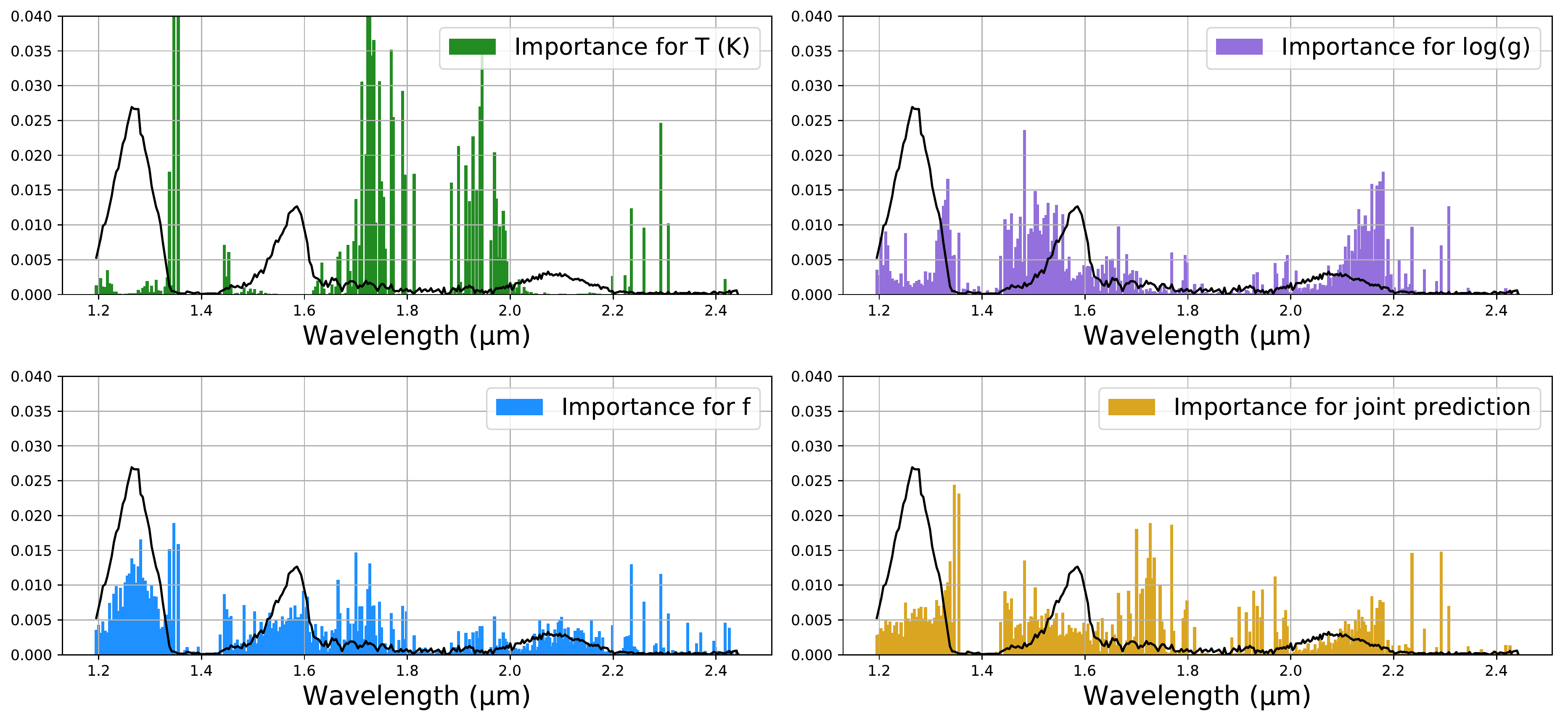}}

\subfloat[\texttt{HELIOS}]{\includegraphics[width=0.82\textwidth]{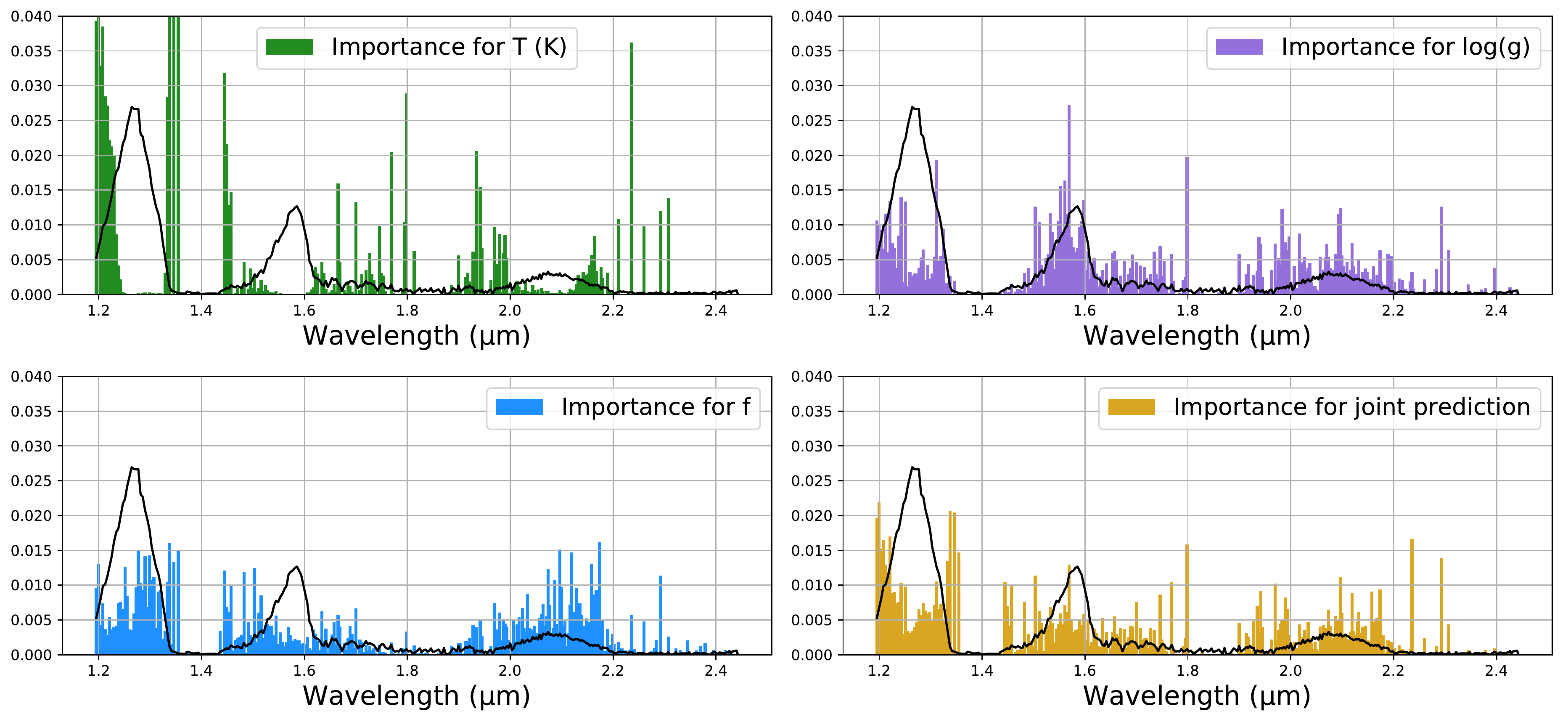}}

\subfloat[\texttt{Sonora}]{\includegraphics[width=0.82\textwidth]{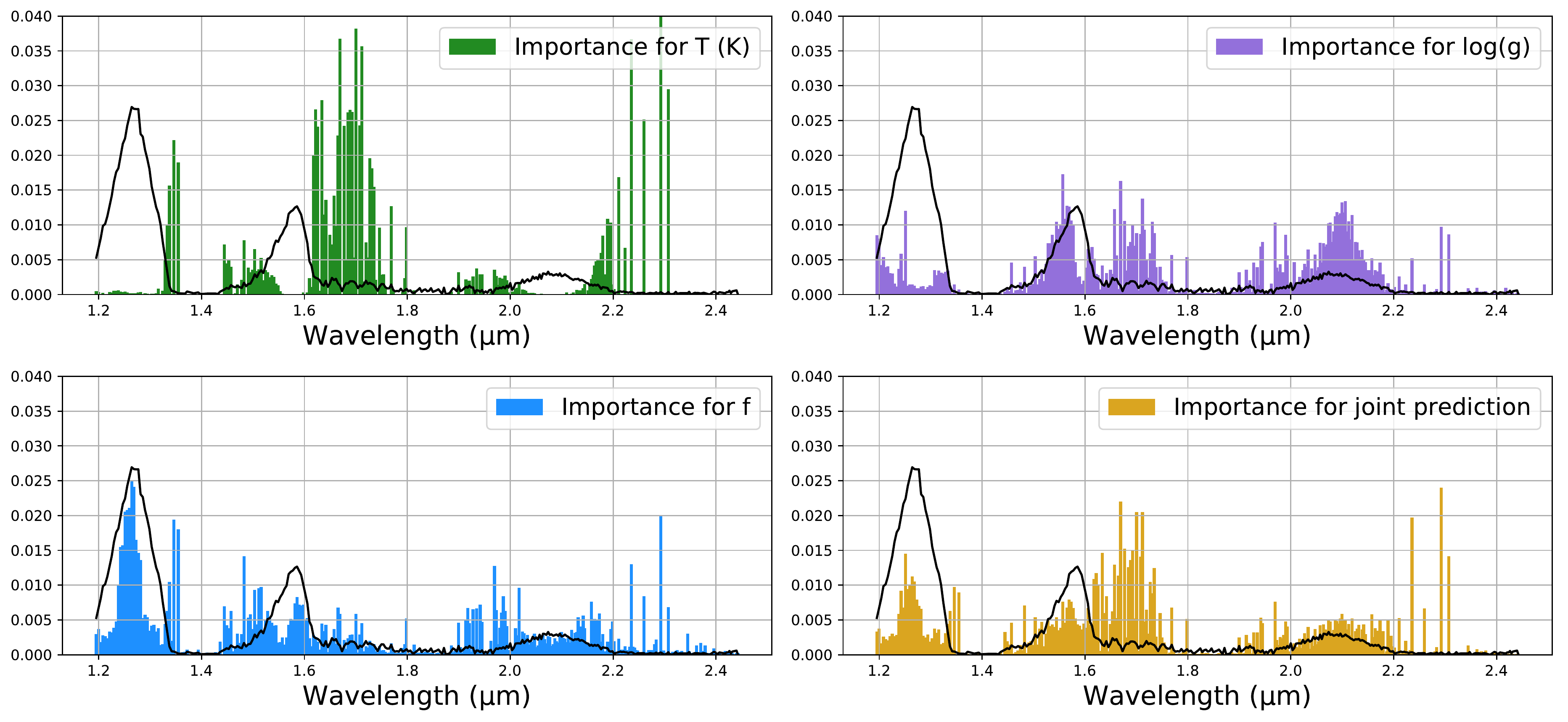}}
\end{center}
\caption{Feature importance plots for the retrieval of the GJ 570D spectrum. The  feature importance plots are shown for the temperature $T$ (green), the surface gravity $\log(g)$ (purple), the calibration factor $f$ (blue), and their joint retrieval (yellow). The results are provided for each of the grids in three separate panels: (a) \texttt{AMES-cond}, (b) \texttt{HELIOS}, and (c) \texttt{Sonora}. For comparison, the scaled, measured spectrum of GJ 570D (solid, black line) has been added to each plot.}
\label{fig:fi_gj}
\end{figure*}

The feature importance plots for the effective temperature $T$, the surface gravity $\log(g)$, and the calibration factor are shown in Figure \ref{fig:fi_gj} for all three model grids. The grids have been trained on the wavelength range of the SpeX measurement of GJ 570D, excluding the wavelengths below 1.2 $\mu$m. For comparison, we also add a scaled spectrum of GJ 570D to each plot. It is important to note here that feature importance plots are obtained based on the whole grid, and since the parameter ranges differ slightly for three model grids, this might introduce small differences in the results.

Given the results of the model comparison from the previous subsection, we expect that all three model grids should roughly yield the same feature importance with respect to the effective temperature because this parameter was more or less consistently predicted during the training and testing phase using different model grids. 

Figure \ref{fig:fi_gj} indeed suggests that the feature importance distributions for the temperature are quite similar and seem to be concentrated at the strong molecular absorption bands within the spectrum. The random forest predicts that wavelength regions at about 1.35 $\mu$m, between 1.4 $\mu$m and 1.8 $\mu$m, as well as around 2.2 $\mu$m are important for the inference of the effective temperature, independent of the employed grid. While the relative contributions of the feature importance values differ, the location of the wavelength regions for determining the effective temperatures seem to coincide for all three grids. This is consistent with the outcome of the previous subsection and indicates that the effective temperature is a more or less robust parameter that can be retrieved from brown dwarf spectra, independent of the atmosphere model grid.

For the surface gravity, however, the results are less clear. While overall the gravity features seem to be concentrated at the slopes of the peaks within the spectra, the actual distribution for the three grids seem to differ quite strongly. The \texttt{Sonora} and \texttt{AMES-cond} grids both show a pronounced feature near 2.1 $\mu$m, however, the maximum of the one from \texttt{AMES-cond} is shifted towards larger wavelengths. Using the \texttt{HELIOS} grid, on the other hand, results in a very flat distribution of the feature importance values in this region. The same effect can also be seen at 1.5 $\mu$m, where again all three grids show gravity features, however the one of \texttt{AMES-cond} seems to be shifted to smaller wavelengths compared to \texttt{HELIOS} and \texttt{Sonora}. Additionally, \texttt{Sonora} predicts a very high feature importance at 1.7 $\mu$m that none of the two other grids agrees with. At smaller wavelengths, between 1.2 $\mu$m and 1.35 $\mu$m \texttt{AMES-cond} and \texttt{HELIOS} show a similar distribution of the feature importance values, while \texttt{Sonora} seems to put less weight in this region for constraining $\log(g)$. This again confirms our results of the model comparison from the previous subsection.

As mentioned in the previous section, the calibration factor $f$ is largely model-independent. This is clearly reflected in its corresponding feature importance plots for the three grids. The spectral regions that are important for constraining $f$ are the same for each grid. Furthermore, as expected, the regions with the highest flux values have the strongest impact on retrieving the calibration factor. Since this factor is used in Equation \ref{eq:distance_relation} to scale the stellar flux, spectral points with higher flux values will naturally have a much higher constraining power for this parameter than regions where the flux is almost zero.

\subsection{Atmospheric retrieval of measured brown dwarf spectra}
\label{sec:retrieval_spectra}
After testing and comparing the grids, we perform the retrieval on actual brown dwarf observations. Details on the observational data can be found in Section \ref{sec::brown_dwarf_data}.

For each atmosphere grid, we bin the theoretical model spectra to the same pixel sampling as that of the observational data. Due to the aforementioned problems of the model grids below $1.2$ $\mu$m, originating from the alkali line wings, we discard this wavelength region from the measured spectra for all three retrievals in the following.

In contrast to the standard retrieval techniques where the model is trying to find a fit within the data error bars, the random forest algorithm accounts for the noise in the data in a different way. The spectra used for the training and testing procedures should include the simulated noise at a level comparable to that of the measured spectra, whereas several noise instances per each model spectrum is required to be run through the training. This procedure makes the random forest robust to the presence of noise in real data and it avoids overfitting.

 We therefore calculate the relative noise of the real data at each wavelength bin ( $\mathrm{ \frac{F_{error}}{F}}$) and use this value as $1\sigma$ for calculating the Gaussian noise at a given wavelength for all the models in the grid. As a result, the spectra in the training set will feature the same noise level distribution as the real data (i.e., the measured data point with a large error bar will be reflected in all the models having a data point with large noise added in this bin). For each model spectrum, we add several noise instances (randomly drawn from a Gaussian distribution) to the training set. This procedure prevents the random forest algorithm from adapting to a specific error distribution (``learning the noise").

To perform the retrieval analysis, we also need to provide the distance and the radius in Equation \ref{eq:distance_relation}. We use a value of $5.84\pm0.03$ pc for GJ 570D from Hipparcos parallax measurements \citep{vanLeeuwen2007A&A...474..653V} and $3.6224\pm 0.0037$ pc for $\epsilon$ Indi  \citep{king10}, respectively. Based on Table 6 from \cite{bayliss17}, who measured the transit radii of 12 brown dwarfs, we adopt a value of 1 Jupiter radius for $R$ (see also Figure 1 in \cite{1993RvMP...65..301B}, that shows the predicted radii for brown dwarfs being around 1 Jupiter radius, independently of the object's mass). We note, however, that any error in either the distance or the assumed radius will be included in our retrieved calibration factor $f$.

All three brown dwarfs have been subjects of previous characterization studies. For the $\epsilon$ Indi objects, dynamic masses have been reported by \cite{dieterich18}.  A SpeX spectrum of GJ 570D has previously been analyzed by \cite{Line2015ApJ...807..183L} using a classical MCMC approach. Other surface gravity estimates based on theoretical stellar evolution models have been published by \cite{geballe09} and \cite{saumon06}. We use their reported retrieval parameters for comparison. 
A summary of our retrieval results and a comparison with previous studies is given in Table \ref{tab:retrieval_results}.

\begin{deluxetable*}{llcccc}[t]
\tablecaption{Results of the random forest retrieval for the three different brown dwarfs and comparison to previously published values. \label{tab:retrieval_results}}
\tablehead{
\colhead{} &
\colhead{Parameter} &
\multicolumn{3}{c}{This work} &
\colhead{Previous work} \\
\colhead{} &
\colhead{} &
\colhead{\texttt{Sonora}} &
\colhead{\texttt{AMES-cond}} &
\colhead{\texttt{HELIOS}} &
\colhead{}
}
\startdata
\textbf{GJ 570D} &  $T$ (K) & $808^{+43}_{-27}$ & $878^{+23}_{-78}$ & $800^{+14}_{-100}$ & $800 - 820$ \citep{geballe09} \\ 
                & & & &  & $780-820$ \citep{2006ApJ...639.1095B} \\
              & & & &  & $948^{+53}_{-53}$ \citep{delBurgo2009} \\
              & & & &  & $900$ \citep{Testi2009} \\              
                & & & &  & $714^{+20}_{-23}$ \citep{Line2015ApJ...807..183L} \\
                   & & & &  & $759^{+63}_{-63}$ \citep{2015ApJ...810..158F}  \\
 & $\log(g)$ & $4.93^{+0.38}_{-0.55}$ & $5.27^{+0.43}_{-0.67}$ & $5.08^{+0.62}_{-0.68}$ & $5.09 - 5.23$ \citep{geballe09}\\
  & & & &  & $5.1$ \citep{2006ApJ...639.1095B} \\
    & & & &  & $4.5^{+0.5}_{-0.5}$ \citep{delBurgo2009} \\
  & & & &  & $5.0$ \citep{Testi2009} \\
 & & & &  & $4.76^{+0.27}_{-0.28}$ \citep{Line2015ApJ...807..183L} \\
  & & & &  & $4.90^{+0.5}_{-0.5}$ \citep{2015ApJ...810..158F} \\
 & $f$  & $0.618^{+0.156}_{-0.053}$ & $0.633^{+0.147}_{-0.070}$  & $0.686^{+0.592}_{-0.109}$ &\\
 \hline
 \textbf{$\mathbf \epsilon$ Indi Ba} &  $T$ (K)\tablenotemark{b} & $1530^{+173}_{-127}$/$1300^{+100}_{-100}$ & $1600^{+122}_{-100}$/$1300^{+100}_{-100}$ & $1530^{+145}_{-127}$/ $1300^{+102}_{-97}$ & 1300-1340 \citep{king10}, atm. models\\
 & & & &  & $1352-1385$ \citep{king10}, evo. models \\
& & & &  & $1400-1600$ \citep{2003ApJ...599L.107S} \\
& & & &  & $1250$ \citep{2004ApJS..154..418R}\\
& & & &  & $1250-1300$ \citep{2009ApJ...695..788K}\\
 & $\log(g)$\tablenotemark{b} & $5.17^{+0.24}_{-0.52}$/$4.39^{+0.748}_{-0.603}$ & $5.68^{+0.20}_{-0.35}$/$5.5^{+0.434}_{-0.866}$ & $5.54^{+0.22}_{-1.56}$/ $5.62^{+0.269}_{-1.12}$ & 5.25 \citep{king10}\\
                  & & & &  & $5.13$ \citep{2004ApJS..154..418R}\\
 & & & &  & $5.2-5.3$ \citep{2009ApJ...695..788K}\\
                 & & & &  & $5.269\pm 1.055$\tablenotemark{a} \citep{dieterich18}\\
 & $f$  & $0.582^{+0.052}_{-0.030}$/ - & $0.557^{+0.053}_{-0.031}$/ -  & $0.593^{+0.119}_{-0.030}$/ -  &\\
 \hline
 \textbf{$\mathbf \epsilon$ Indi Bb} &  $T$ (K)\tablenotemark{b} & $1130^{+352}_{-157}$/$900^{+76.7}_{-26.6}$ & $1100^{+150}_{-100}$/$930^{+70}_{-30}$ & $1180^{+239}_{-181}$/$900^{+100}_{-100}$ &880-940 \citep{king10}\\
                  & & & &  & $840$ \citep{2004ApJS..154..418R}\\
 & & & &  & $875-925$ \citep{2009ApJ...695..788K}\\
 & $\log(g)$\tablenotemark{b} & $5.19^{+0.199}_{-1.02}$/$5.45^{+0.0176}_{-0.0995}$ & $5.53^{+0.33}_{-0.47}$/$5.73^{+0.224}_{-0.399}$ & $5.32^{+0.47}_{-1.28}$/$5.86^{+0.0482}_{-0.321}$ & 5.50 \citep{king10}\\
                   & & & &  & $4.89$ \citep{2004ApJS..154..418R}\\
                    & & & &  & $4.9-5.1$ \citep{2009ApJ...695..788K}\\
                  & & & &  & $5.240 \pm 1.049$\tablenotemark{a} \citep{dieterich18}\\
 & $f$  & $0.593^{+0.0907}_{-0.0419}$/ -  & $0.585^{+0.099}_{-0.041}$/ -  & $0.610^{+0.234}_{-0.041}$/ - &\\
\enddata
\tablenotetext{a}{Derived parameter, based on the measured dynamical mass and assuming $R=(1\pm 0.1)R_J$}
\tablenotetext{b}{The values based on three- and two-parameter retrieval models are given in the left and right columns respectively}
\end{deluxetable*}

\subsubsection{Gliese 570D}

\begin{figure*}[h]
\gridline{\fig{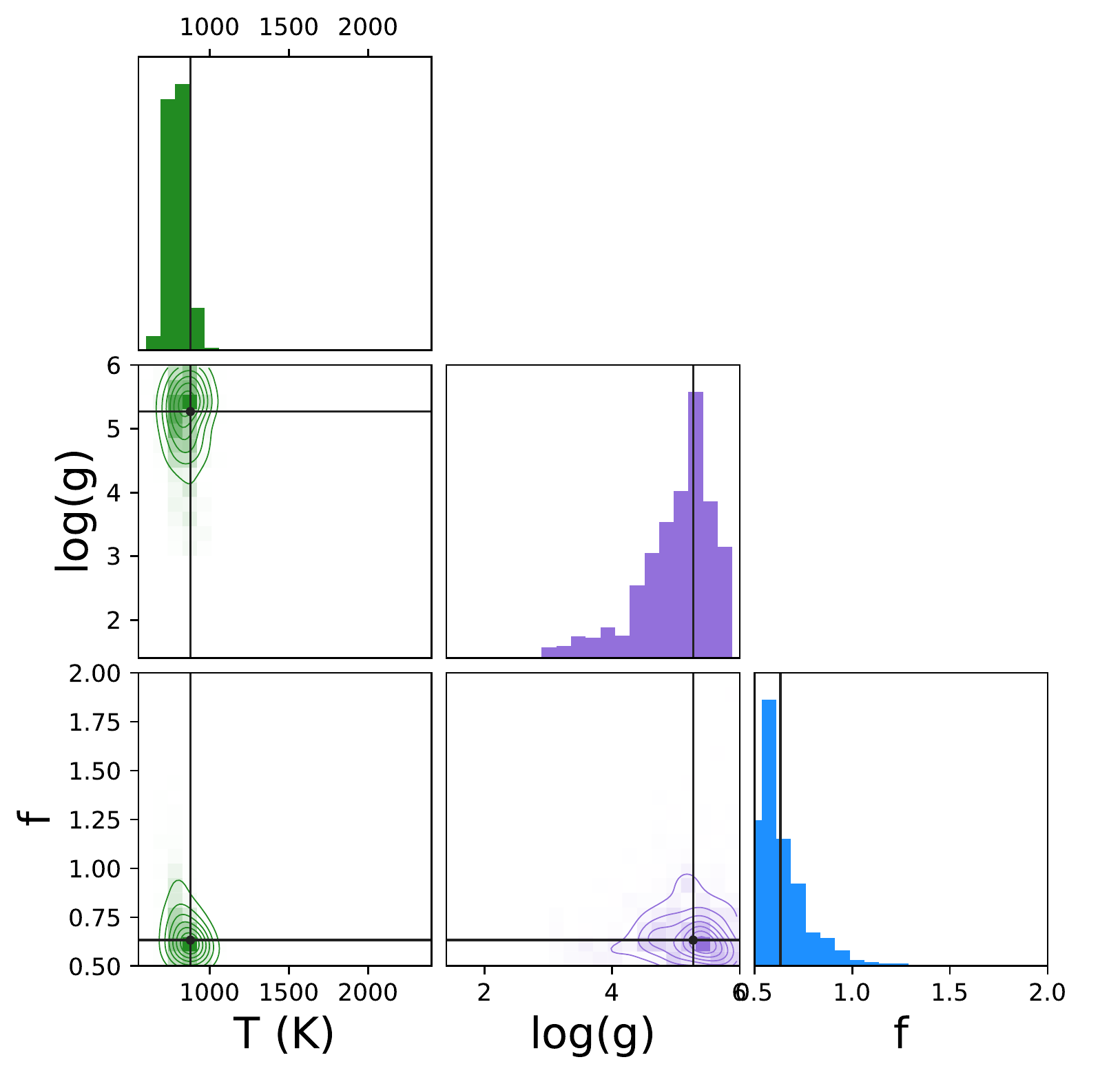}{0.33\textwidth}{}
          \fig{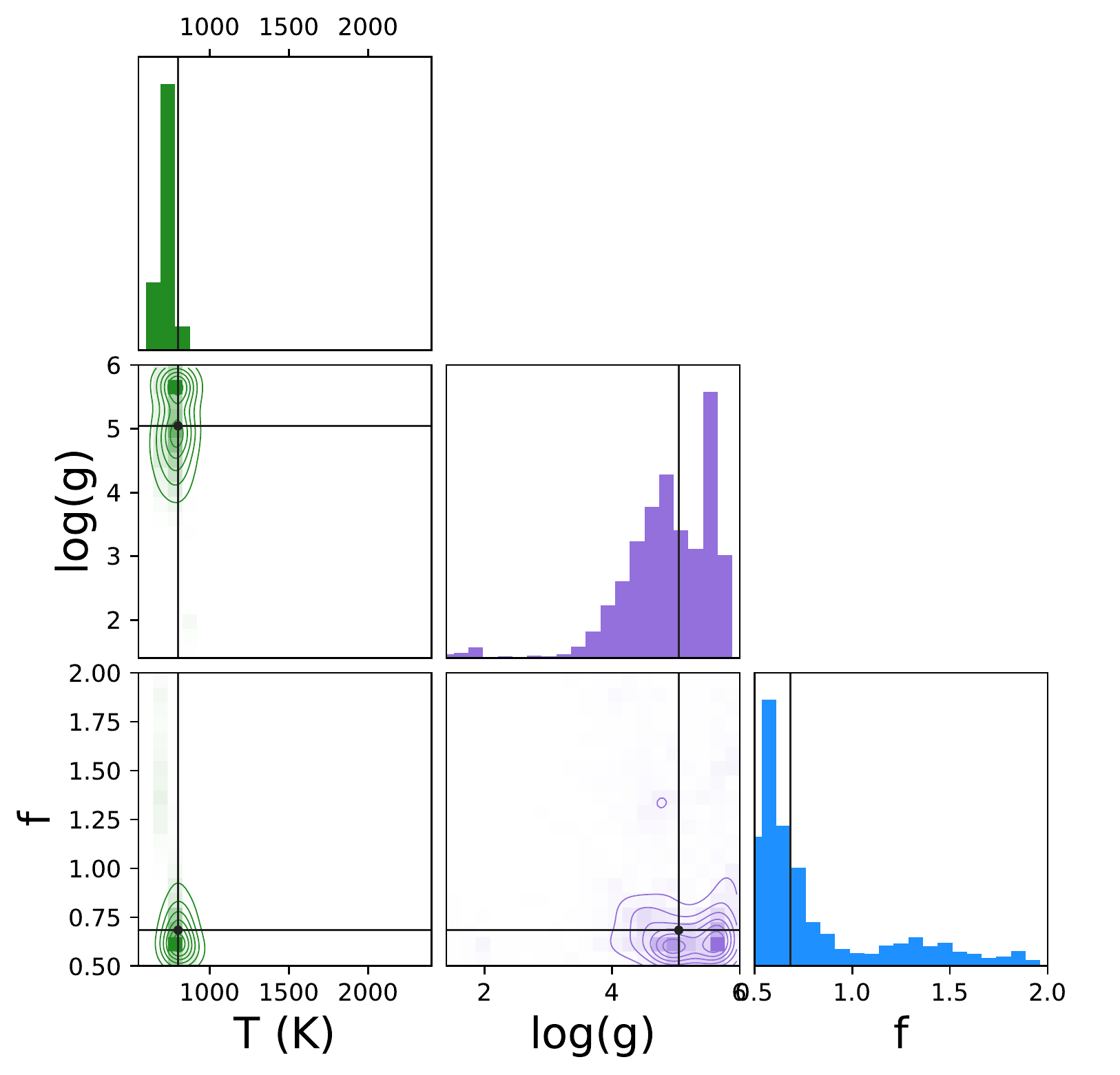}{0.33\textwidth}{}
          \fig{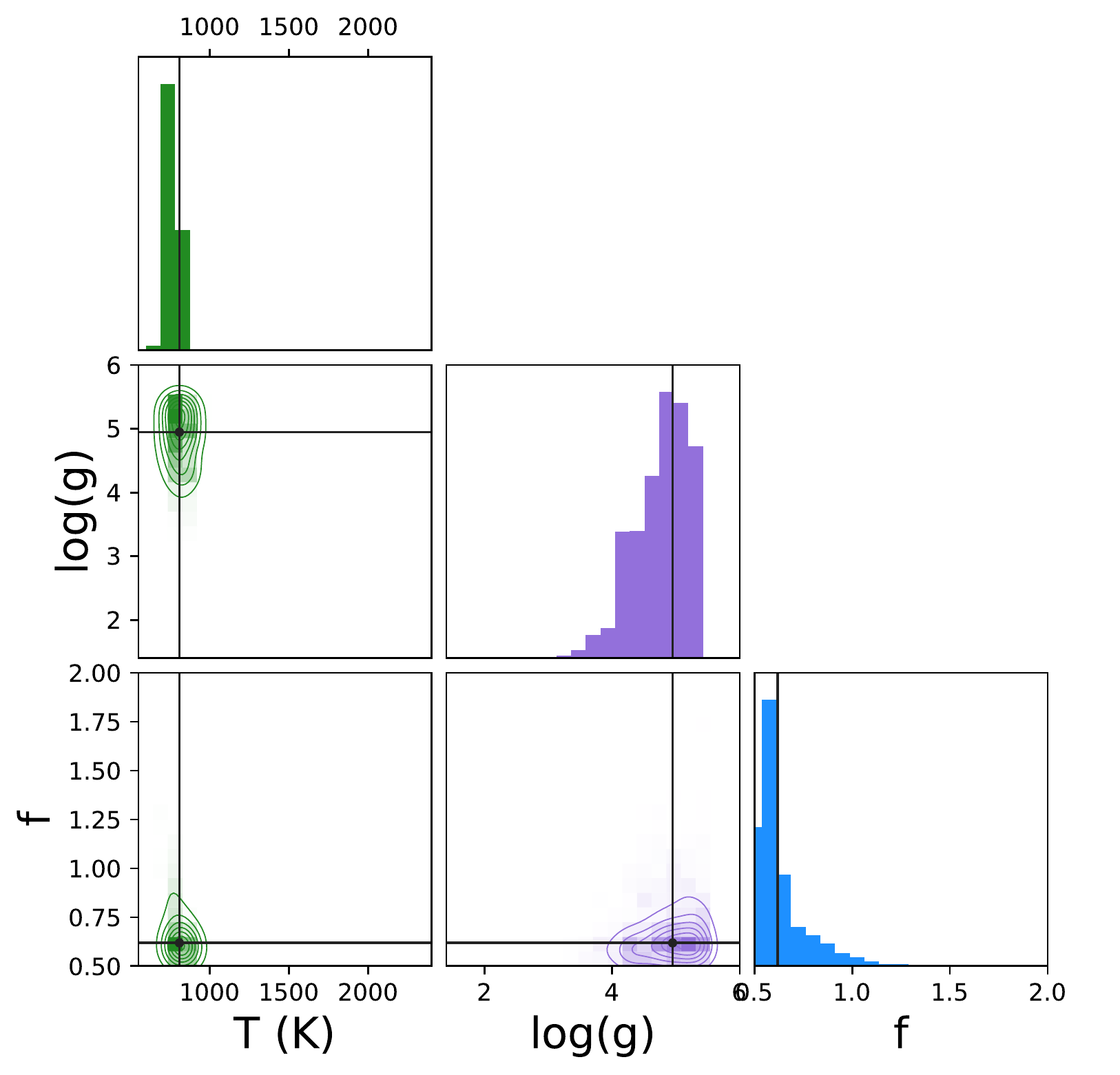}{0.33\textwidth}{}
         }
\vspace{-1.0cm}
\gridline{\fig{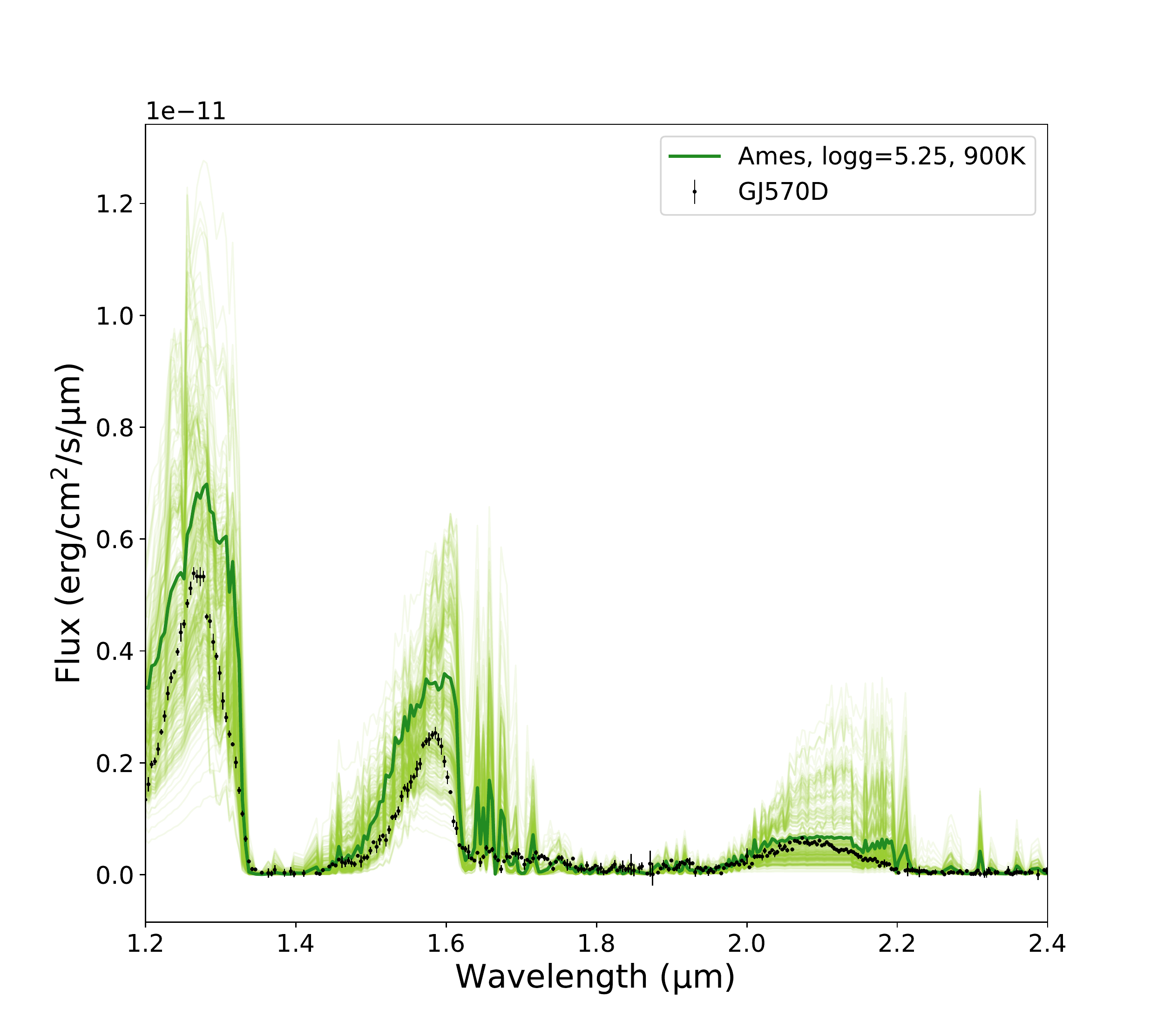}{0.33\textwidth}{(a) \texttt{Ames-cond}}
          \fig{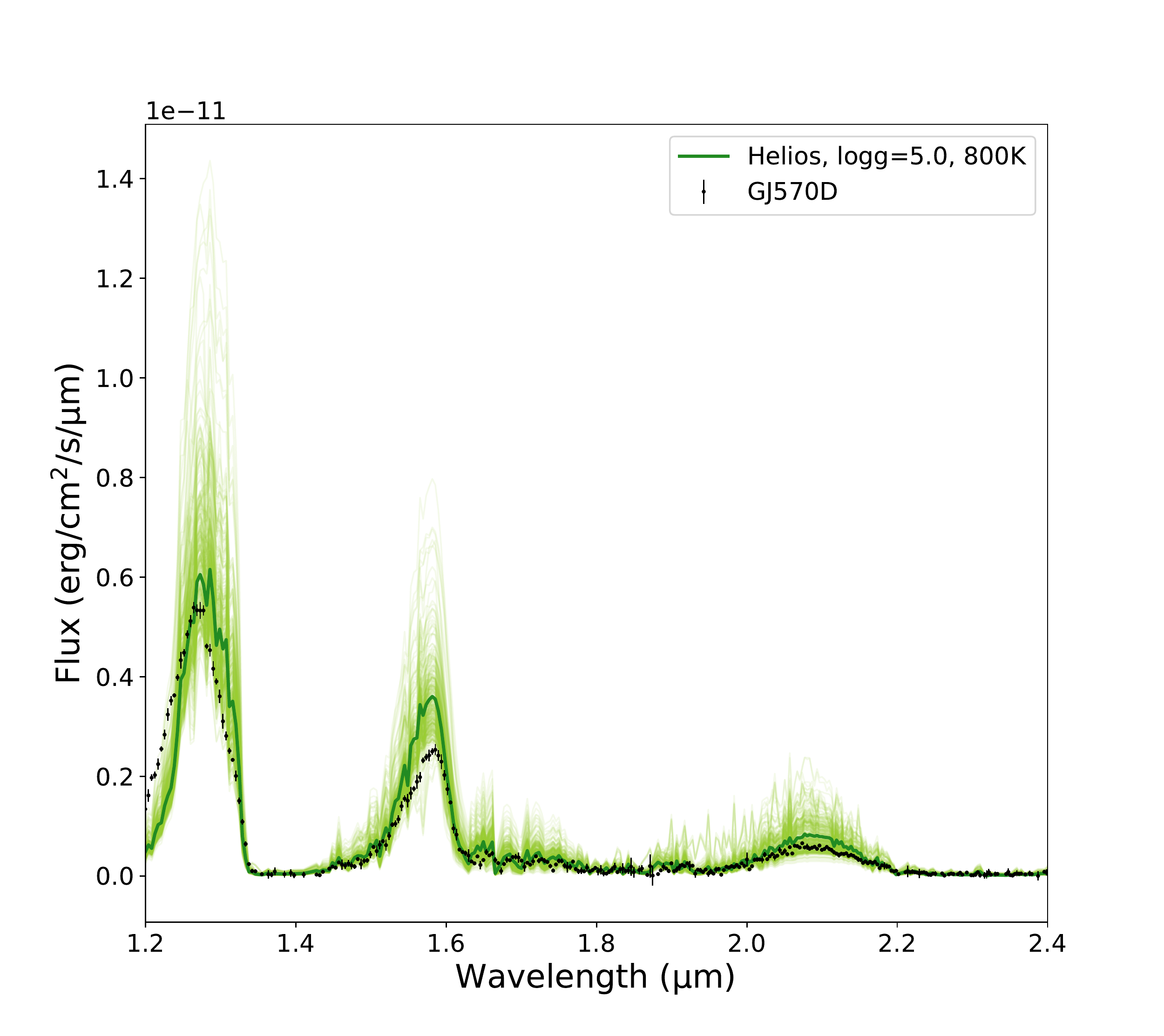}{0.33\textwidth}{(b) \texttt{HELIOS}}
          \fig{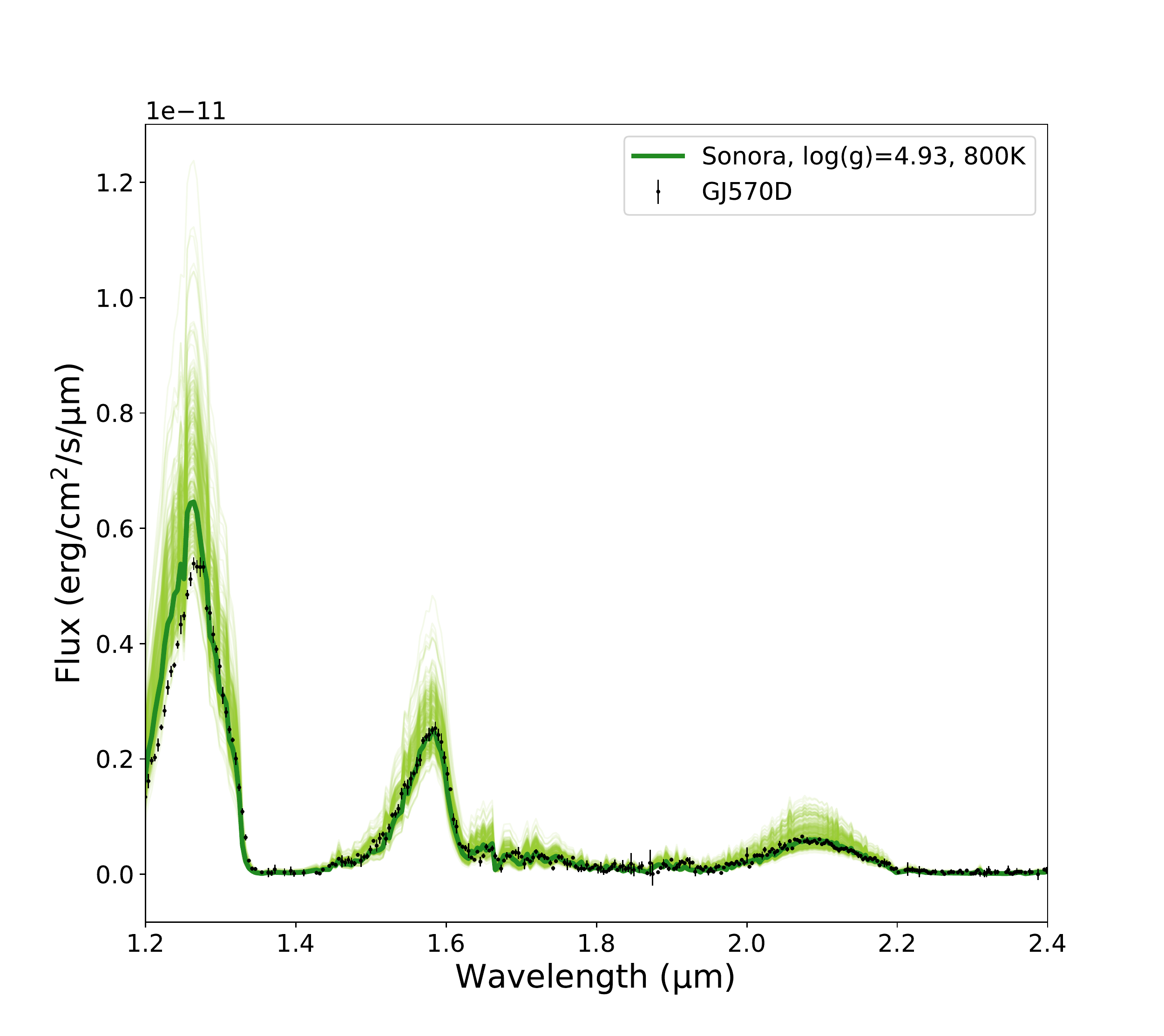}{0.33\textwidth}{(c) \texttt{Sonora}}
         }

\caption{Posteriors distributions for the retrieval of GJ 570D using the three different grids: \texttt{AMES-cond} (a), \texttt{HELIOS} (b), and \texttt{Sonora} (c). 
The resulting posterior distributions for the effective temperature $T$, the surface gravity $\log(g)$, and the calibration factor are shown in the top panel. Solid, black lines mark the median value of each distribution. The corresponding spectra from all posterior samples are depicted in the lower panel. Dark green lines refer to the spectrum corresponding to the median values of $T$, $\log(g)$, and $f$ from the posterior distribution, while all the other spectra are shown in light green. The black data points and error bars denote the measured SpeX spectrum of GJ 570D.}
\label{fig:gj}
\end{figure*}

The resulting posterior distributions for $T$, $\log(g)$, and $f$ of the random forest applied to the spectrum of GJ 570D are shown in Figure \ref{fig:gj}. Median values, confidence intervals, and a comparison with previous studies are summarized in Table \ref{tab:retrieval_results}.  Our deliberate use of the calibration factor $f$ is to facilitate comparison with \cite{Line2015ApJ...807..183L}.  The retrieved value of $f\approx 0.6$ may be interpreted as corresponding to a radius of about $0.8 R_{\rm J}$.

The results for all three grids yield similar estimates: effective temperatures of around 800--900 K and $\log(g)$ values of about 5. Especially the posterior distributions for the temperatures show a very narrow peak for all grids. The surface gravity, on the other hand, is not as well constrained. Together with the fact that the $R^2$ values for gravity are very high (around 0.9) the broadness of the posterior suggests that there are differences between the models and the data.

Overall, the values for the effective temperature and surface gravity of GJ 570D from our random forest retrieval are consistent with previous studies (cf. Table \ref{tab:retrieval_results}).

\subsubsection{The triple star system $\epsilon$ Indi}

Based on the measured dynamical masses \citep{dieterich18} and assuming $R=(1\pm 0.1)R_J$, the values for log(g) are: $5.269\pm 1.055$ for $\epsilon$ Indi Ba and $5.240 \pm 1.049$ for $\epsilon$ Indi Bb.  Since there is (approximate) ``ground truth" for the surface gravities, we explicitly perform pairs of retrievals that include or exclude the calibration factor in order to investigate its effects.

The retrieval results for both objects differ significantly depending on whether or not the calibration factor is added as a third parameter. With only T and log(g) as parameters, the temperature estimates are largely consistent with previous studies \citep{king10},  but the gravity is discrepant from previous estimations and from the calculated values mentioned above.

When the scaling parameter is added, the temperature estimate is less consistent, but gravity is more consistent with the calculations based on dynamical masses. Figures \ref{fig:epsa} and \ref{fig:epsb} show the posteriors and model spectra sampled from posterior distribution.\\

\begin{figure*}
\begin{center}
\gridline{\fig{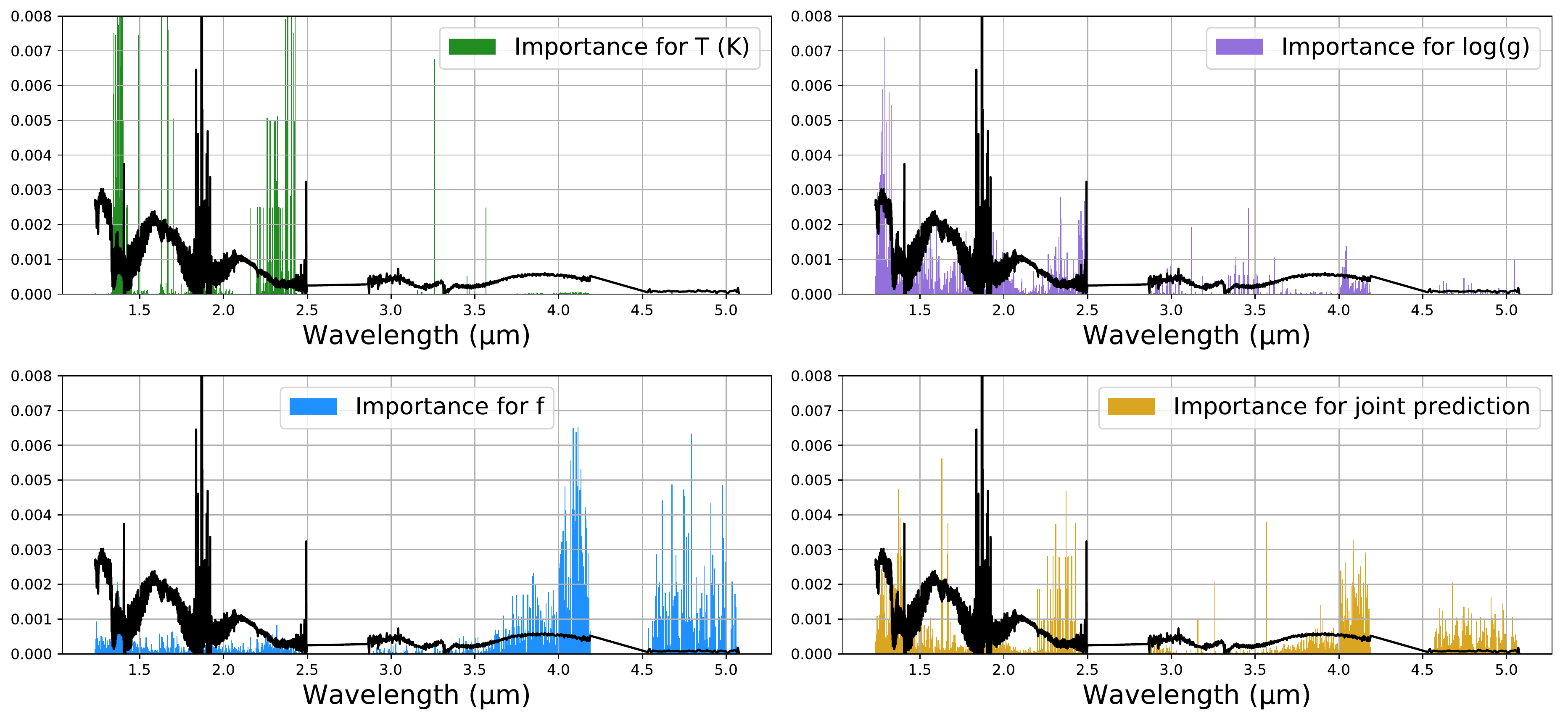}{2\columnwidth}{(a) \texttt{HELIOS}, retrieval including $f$}
         }
\gridline{\fig{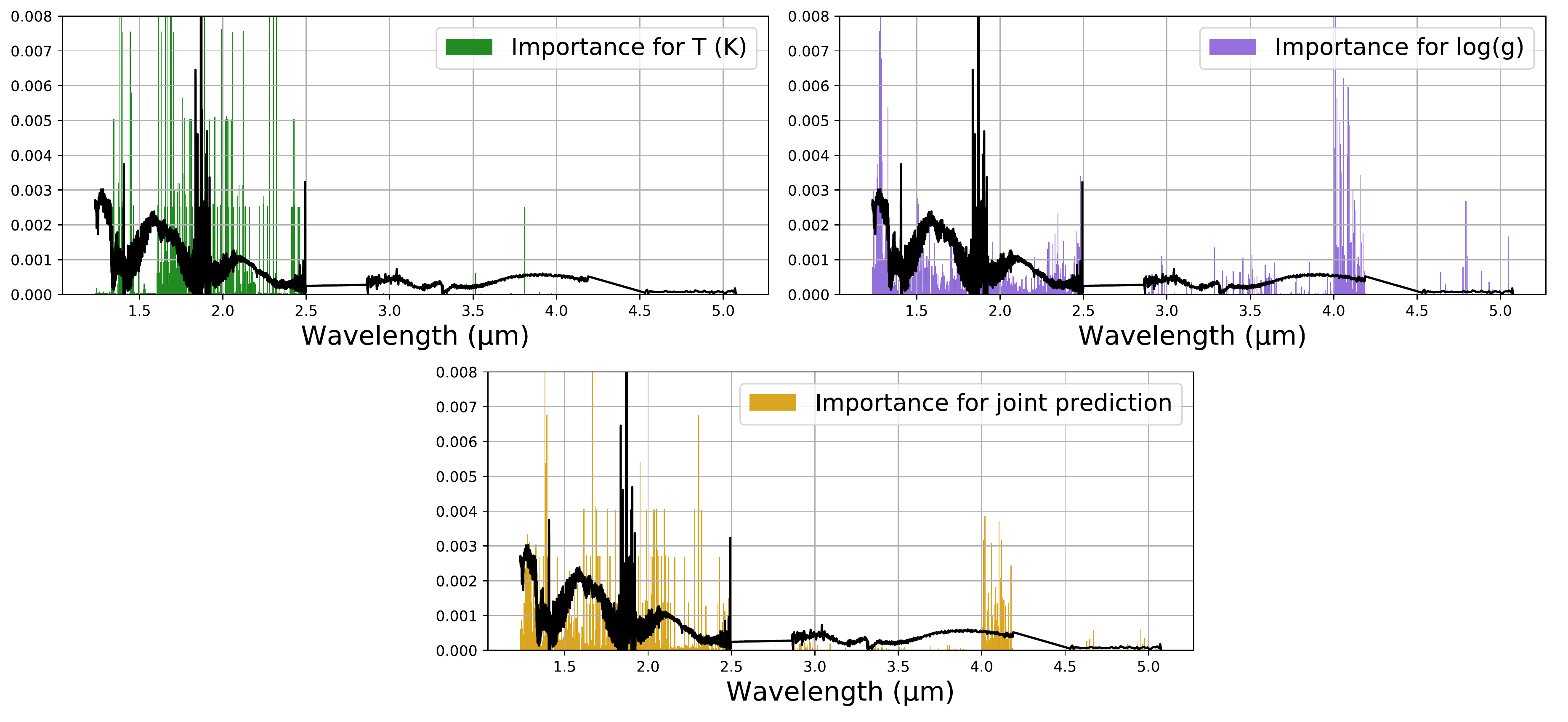}{2\columnwidth}{(b) \texttt{HELIOS}, retrieval excluding $f$}
         }
\end{center}
\caption{Feature importance plots for the retrieval of the $\epsilon$ Indi Ba spectrum using the \texttt{HELIOS} grid. The  feature importance plots are shown for the temperature $T$ (green), the surface gravity $\log(g)$ (purple), the calibration factor $f$ (blue), and their joint retrieval (yellow). The results are provided for two cases: (a) retrieval including the calibration factor $f$, (b) retrieval excluding $f$. For comparison, the scaled, measured spectrum of $\epsilon$ Indi Ba (solid, black line) is shown in each plot.}
\label{fig:fi_epsb}
\end{figure*}

In general, the results for $\epsilon$ Indi  objects are more discrepant both from the previous results and between the different model grids. This is reflected in the posteriors for both temperature and gravity being in general broader for both $\epsilon$ Indi objects than for the retrievals for GJ 570D object, suggesting the differences between the data and the models.


\section{Discussion}
\label{sec:discussion}

\subsection{Summary}

An important difference between the three objects used in our study is that GJ 570D is a T7.5 object and is expected to have a cloudless photosphere. This is not so obvious for the $\epsilon$ Indi  objects, especially for $\epsilon$ Indi Ba which is T1.5 class and does most likely require cloud formation to be included in the models.

Moreover, the grids used in this study are calculated for solar metallicity. This is consistent with our expectation for GJ 570D, based on the metallicity of the primary (\citealt{2000A&A...363..692T,2005A&A...437.1127S,2005ApJS..159..141V}), but again is likely not to hold true for $\epsilon$ Indi  objects. The reported metallicity for the primary, $\epsilon$ Indi A, is sub-solar (see e.g. \citealt{1988A&A...206..100A,2001A&A...379..999S}). Some of the previous studies suggest that the spectra of $\epsilon$ Indi Ba and Bb are best explained by sub-solar metallicity \citep{king10}. This together with the fact that the early-type objects are most likely cloudy may explain the results.

Moreover, it is important to note that much has changed since the development of the \texttt{AMES-cond} grid. In particular, some of the opacities used are outdated. A major difference is the methane line lists, since methane opacity is crucial for modelling T-dwarfs. It is therefore to be expected that the results from \texttt{AMES-cond} grid differ from the estimations obtained with more recent grids, \texttt{HELIOS} and \texttt{Sonora}.

\subsection{Opportunities for future work}

In the current study, we have chosen to consider only temperature and gravity as parameters, with C/O and metallicity being set to the solar value. Future work should include the C/O ratio and metallicity as retrieval parameters by increasing the grid dimensionality. Moreover, it is further possible to add object-specific parameters, such as radius and distance, in the same manner. This will provide the opportunity to study a suite of objects without the necessity to re-train the random forest. 

The unique architecture of the code allows it to be easily modified to perform retrieval studies based on various types of observations and a variety of objects, since the random forest training set is a generic vector of inputs and may include both theoretical and observational parameters. 

There are several interesting objects that might serve as comparisons or benchmarks, such as the directly imaged planet 51 Eri b \citep{2015Sci...350...64M}, or the brown dwarf companion Gl 758B, for which a precise dynamical mass measurement is also available \citep{2018AJ....155..159B}.

\acknowledgments
MO, DK, PM-N, CF and KH acknowledge partial financial support from the Center for Space and Habitability (CSH), the PlanetS National Center of Competence in Research (NCCR), the Swiss National Science Foundation and the Swiss-based MERAC Foundation.  We would like to express gratitude to Robert King (rob@astro.ex.ac.uk) for providing the data for $\epsilon$ Indi Ba and Bb in electronic form.
We thank Mark Marley for providing the \texttt{Sonora} model grid.


\bibliographystyle{aasjournal}
\bibliography{rfNotes}

\appendix

\section{Additional results for the grid comparison}
\label{sec:appendix_comparison}

\begin{figure}[b]
\gridline{\fig{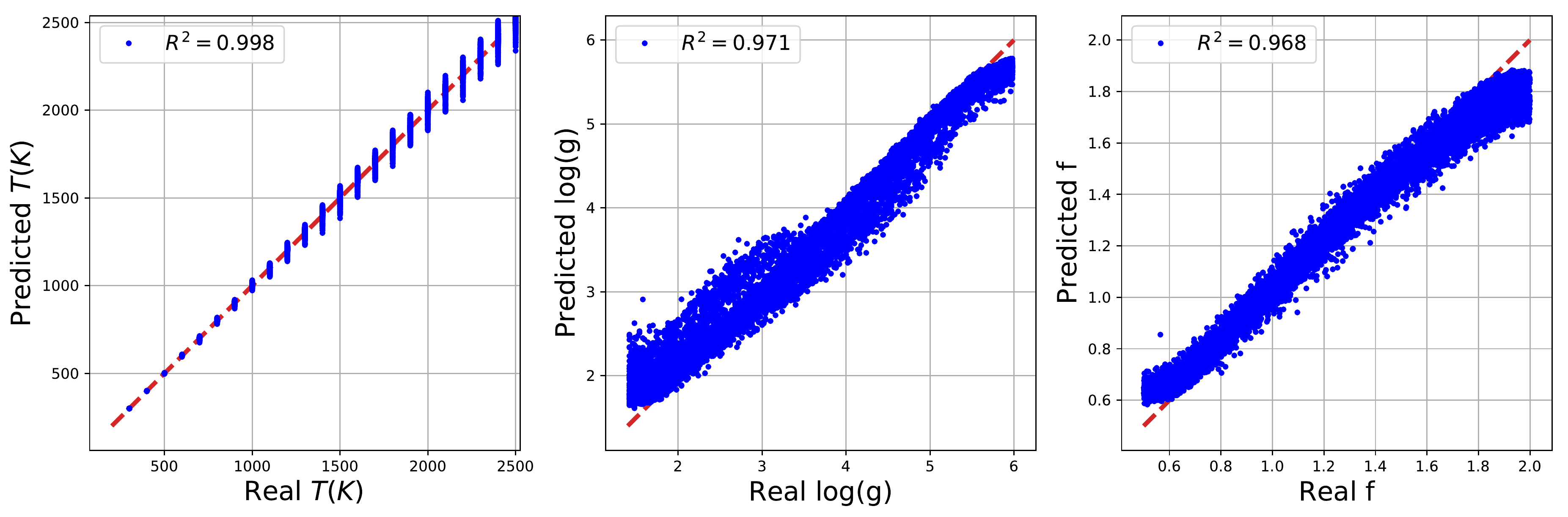}{0.5\columnwidth}{(a) \texttt{HELIOS}, SpeX instrument}
          \fig{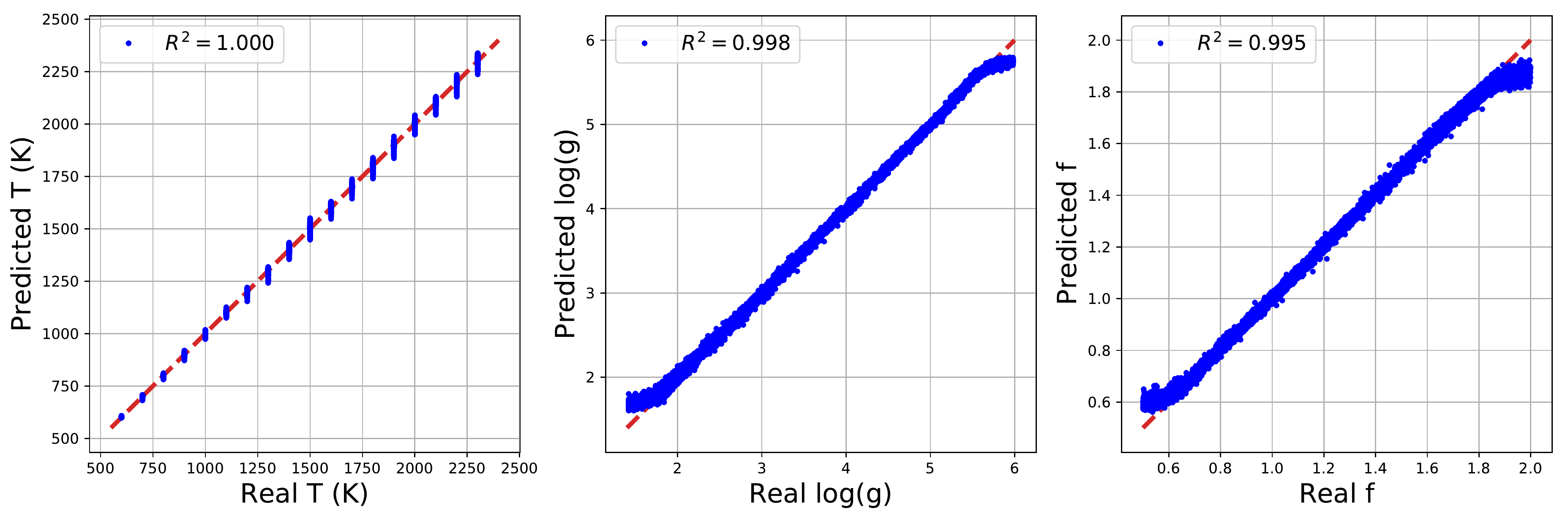}{0.5\columnwidth}{(b) \texttt{HELIOS}, ISAAC instrument}
         }
\gridline{\fig{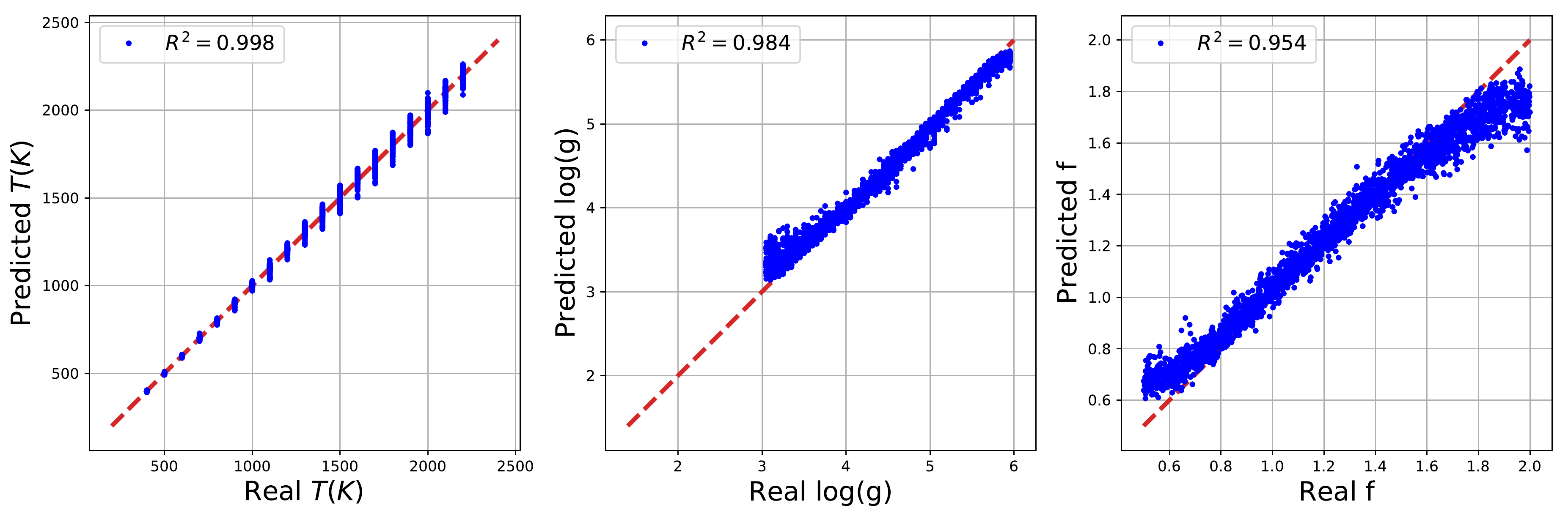}{0.5\columnwidth}{(c) \texttt{AMES-cond}, SpeX instrument}
          \fig{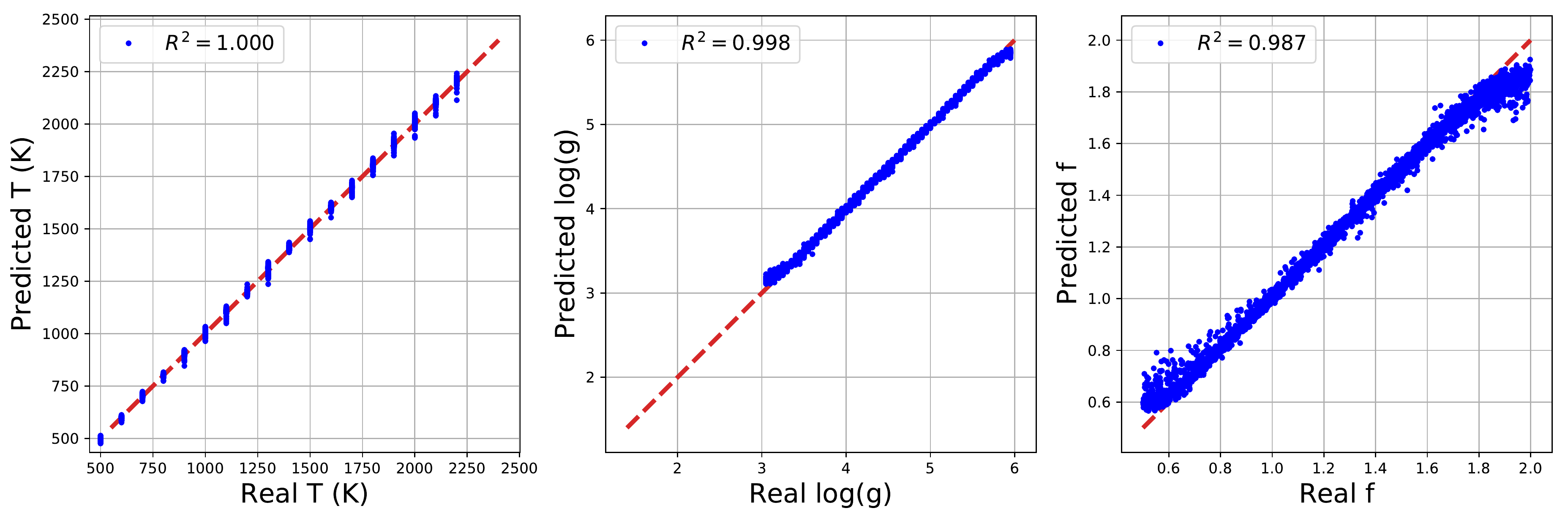}{0.5\columnwidth}{(d) \texttt{AMES-cond}, ISAAC instrument}
         }
\gridline{\fig{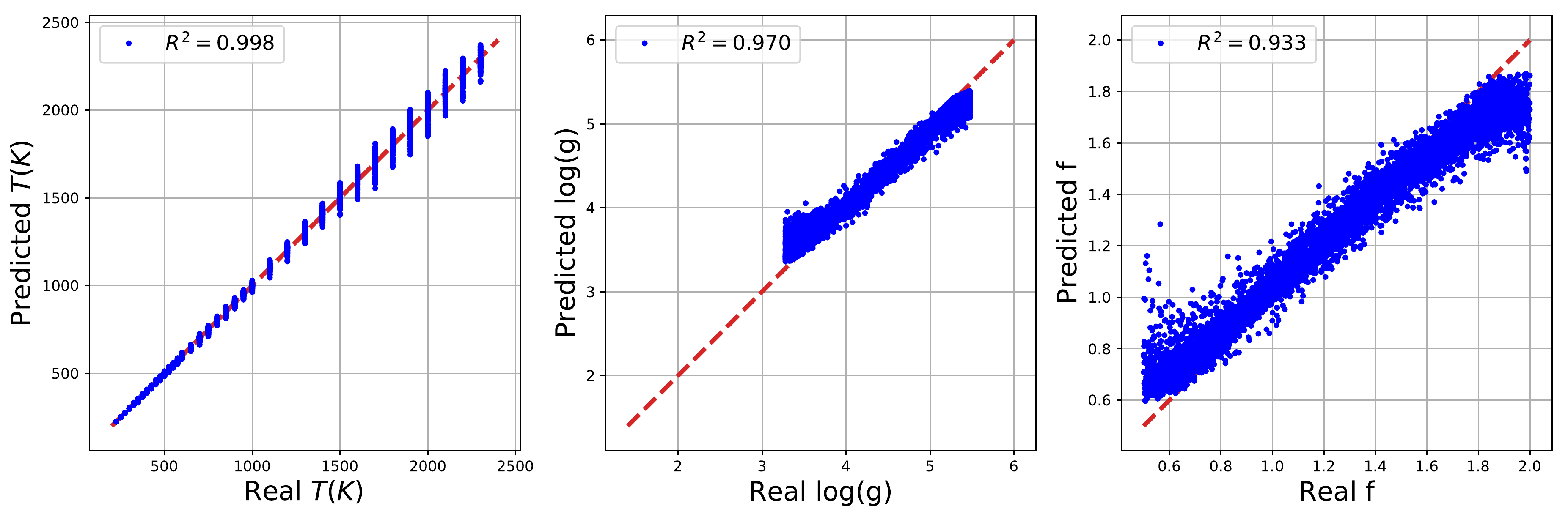}{0.5\columnwidth}{(e) \texttt{Sonora}, SpeX instrument}
          \fig{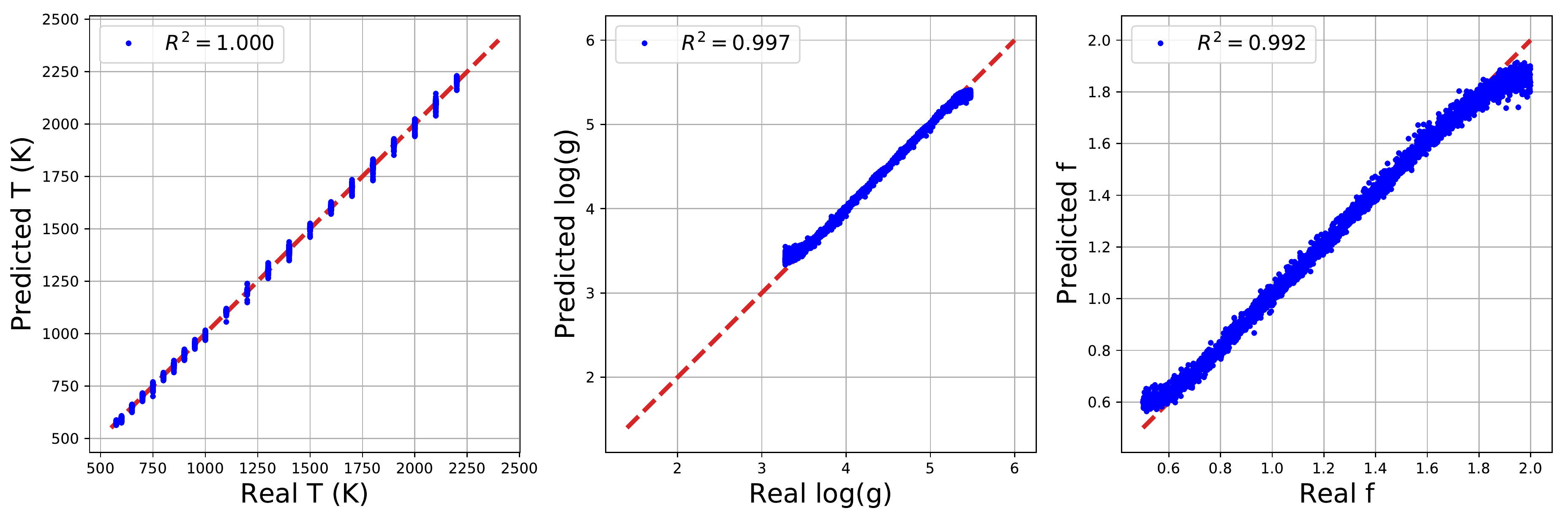}{0.5\columnwidth}{(f) \texttt{Sonora}, ISAAC instrument}
         }
\caption{Comparison of the random forest training and testing procedure using single grids. Upper panel: \texttt{HELIOS}, middle panel: \texttt{AMES-cond}, lower panel: \texttt{Sonora}. The training and testing is performed for each grid individually on either the spectral resolution and wavelength coverage of the GJ 570D spectrum taken with the SpeX prism (left-hand side) or the one from the ISAAC measurement of the brown dwarfs in the $\epsilon$ Indi system (right-hand side). Note that both, \texttt{Sonora} and \texttt{AMES-cond} cover a smaller parameter range than \texttt{HELIOS}. }
\label{fig:helios_r2}
\end{figure}

In Sect. \ref{sec:model_comparison} we present a model comparison by training the random forest on the \texttt{HELIOS} grid and testing it on \texttt{Sonora} and \texttt{Ames-cond}. In addition to that comparison, we show the outcome of training and testing on the same grid in this section. This allows us to evaluate how well the random forest can find and use the spectral features in the spectra of each model to constrain the retrieval parameters. The corresponding results for the three model grids are shown in Fig. \ref{fig:helios_r2} for the spectral wavelength ranges and spectral resolutions of the GJ 570D SpeX measurement and the $\epsilon$ Indi brown dwarfs taken by the ISAAC instrument, respectively. Analogously to Sect. \ref{sec:model_comparison}, all spectra are cut below 1.2 $\mu$m due to the model inconsistencies of describing the alkali line wings.

As expected, when only a single grid is used for testing and training, the predicted and actual values from the testing set match much better than for cases where two different grids are used (cf. Fig. \ref{fig:compare} and Sect. \ref{sec:model_comparison}). It is noteworthy that, as discussed before in Sect. \ref{sec:model_comparison}, the effective temperatures are predicted with a much higher accuracy than the surface gravities. Predicted $\log(g)$ values show a much wider scatter with respect to their real values. The `gravity features" that the random forest tries to locate and use for the prediction of $\log(g)$, thus, seem to be less constrictive than the one it uses to retrieve effective temperatures. 

Overall, the gravity values seem to be better constrained for the spectral resolution of the $\epsilon$ Indi brown dwarfs. This resolution is roughly two orders of magnitude higher than the one provided by the SpeX prism used for the measurement of GJ 570D. On the other hand, the ISAAC measurement of the $\epsilon$ Indi Ba and Bb also offers a larger wavelength coverage towards the infrared. In the feature importance plot for $\epsilon$ Indi Bb shown in Fig. \ref{fig:fi_epsb}, a gravity feature identified by the random forest can be seen at around 4 $\mu$m. These results suggest that constraining the gravity with better precision might require a higher spectral resolution and wavelength coverage than provided by the medium-resolution SpeX spectra.

\begin{figure*}[h!]
\vspace{-0.1in}
\begin{center}
\subfloat[HELIOS, with calibration factor]{\includegraphics[width=0.25\columnwidth,height=0.2\textheight]{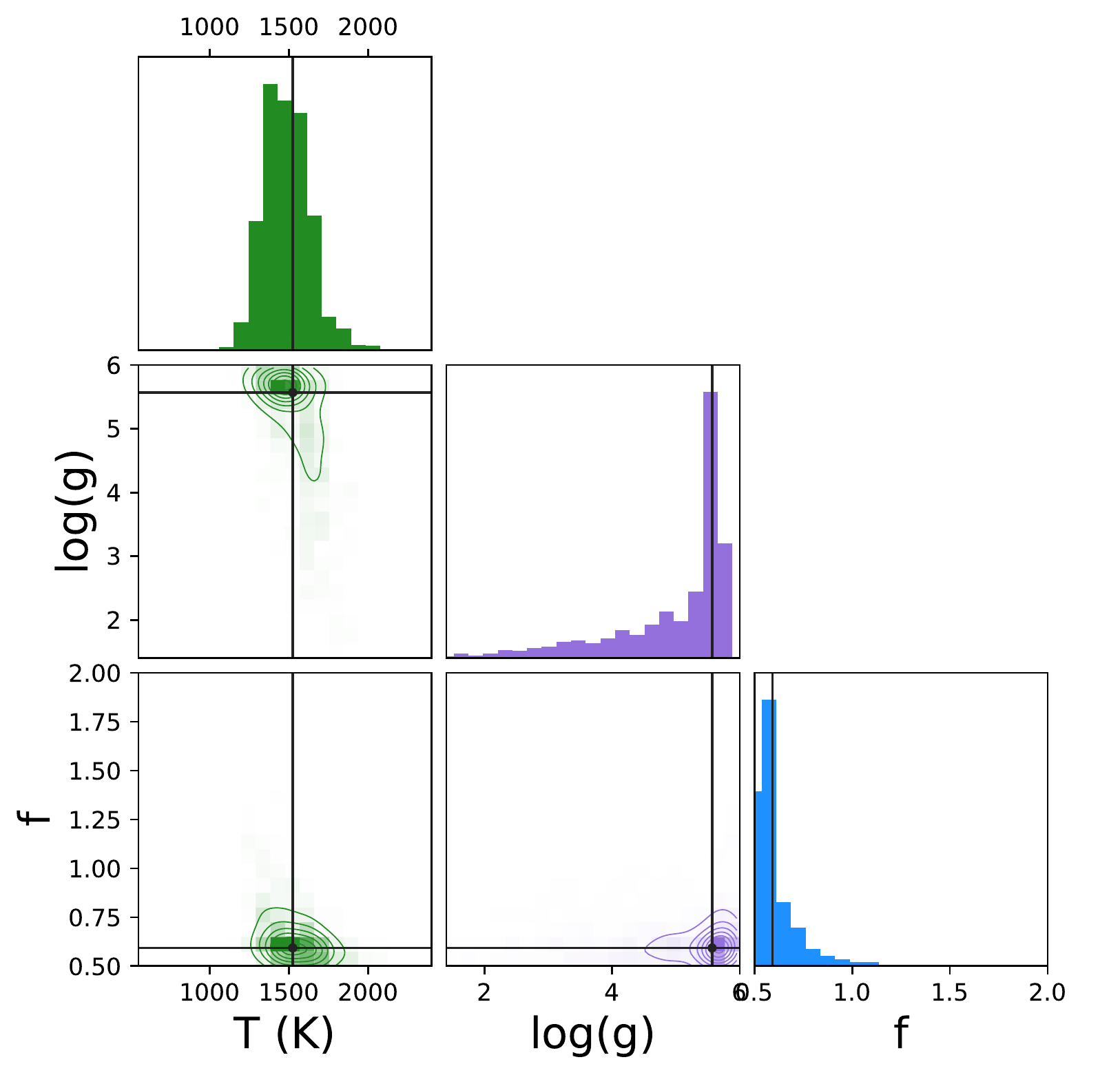}
\includegraphics[width=0.25\columnwidth,height=0.2\textheight]{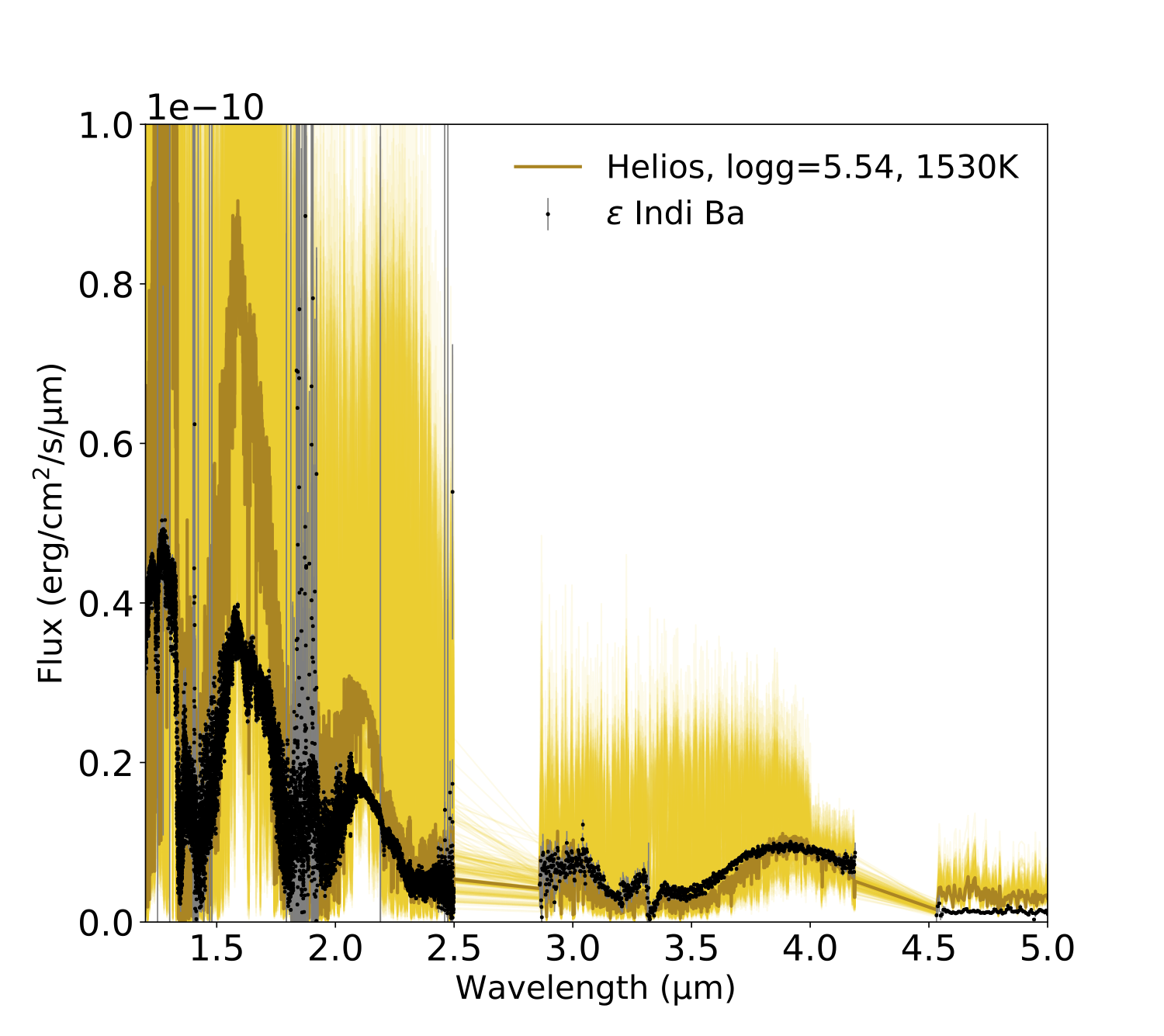}}
\subfloat[HELIOS, no calibration factor]{\includegraphics[width=0.25\columnwidth,height=0.2\textheight]{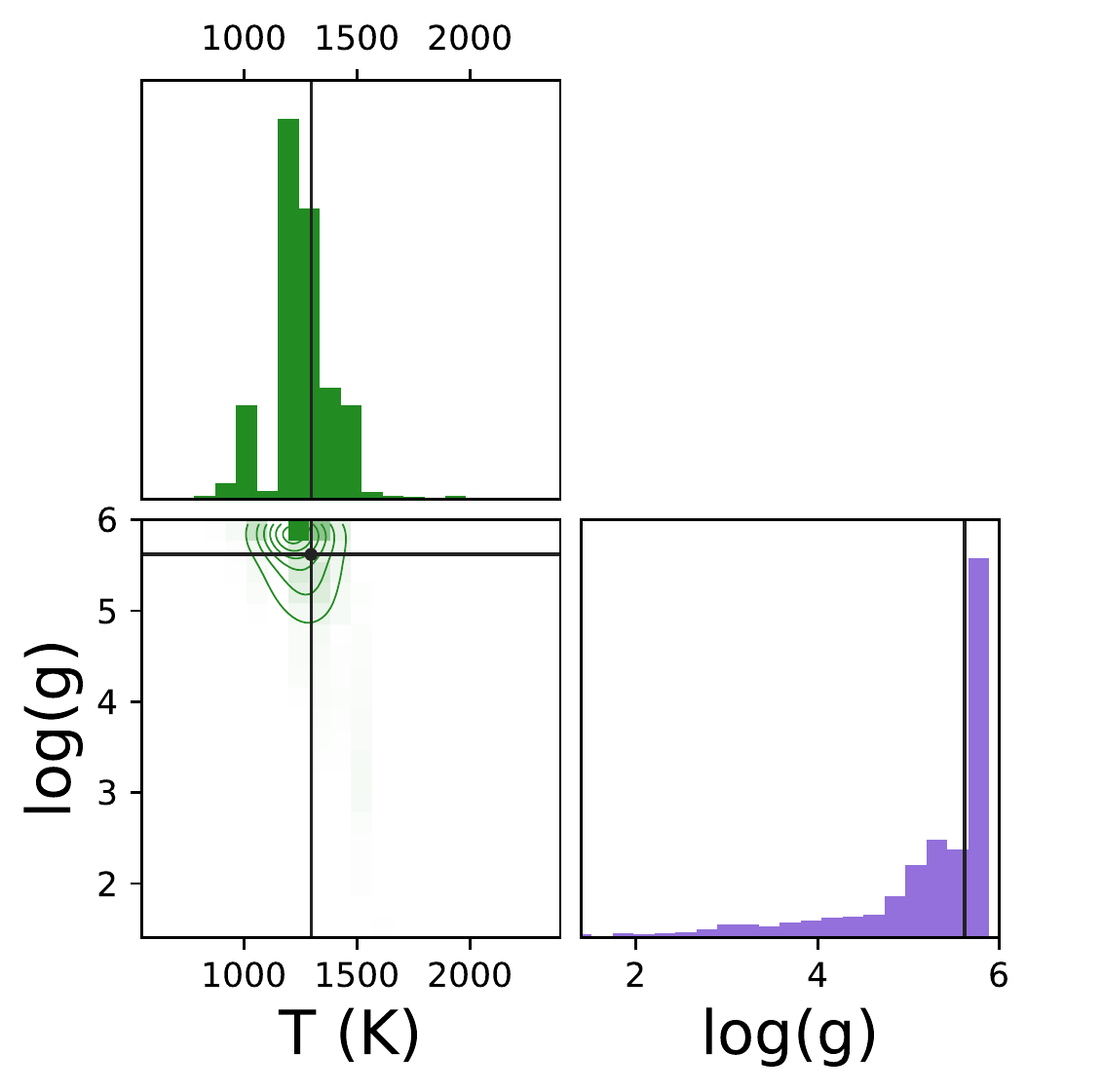}
\includegraphics[width=0.25\columnwidth,height=0.2\textheight]{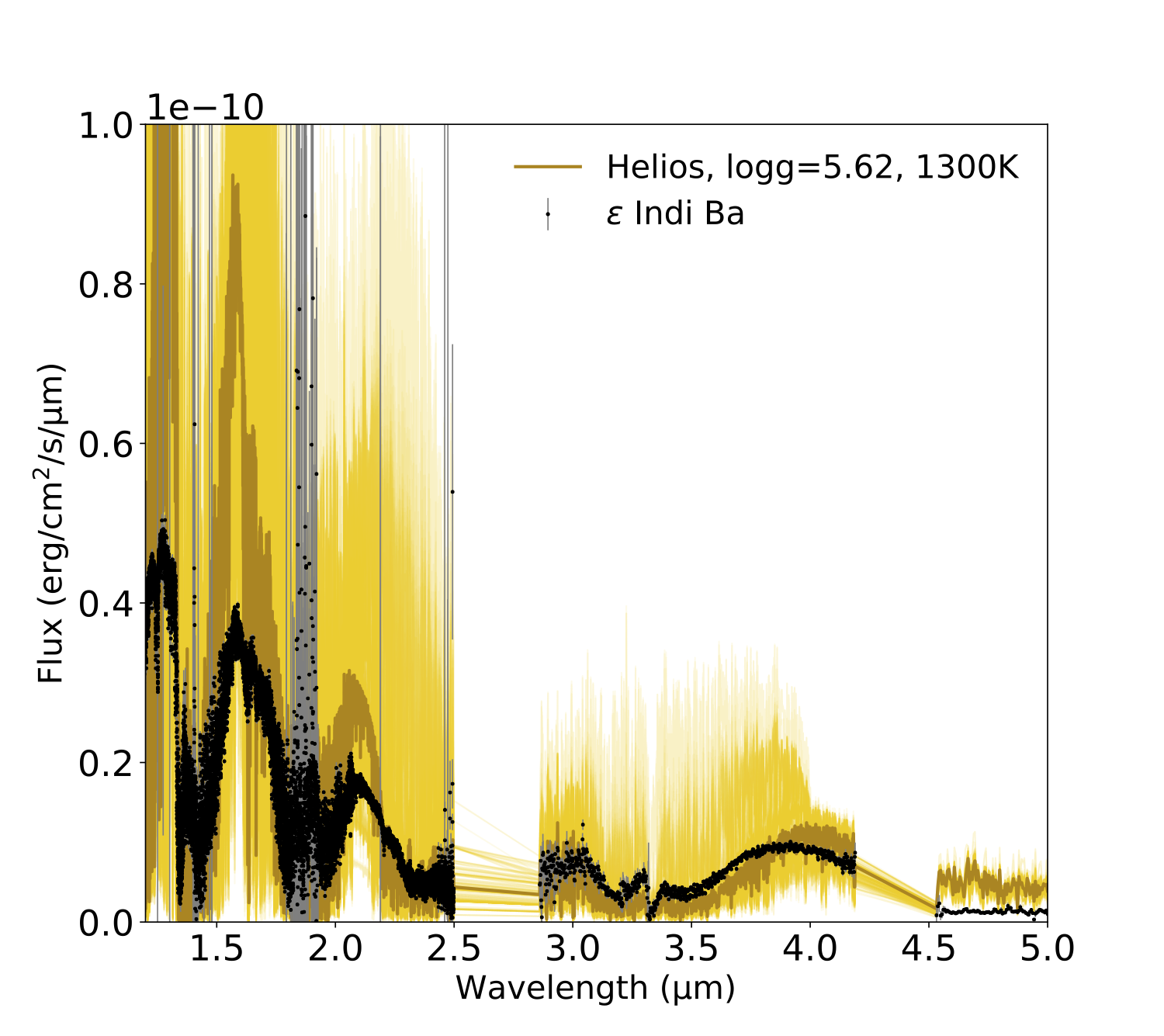}}

\subfloat[Sonora, with calibration factor]{\includegraphics[width=0.25\columnwidth,height=0.2\textheight]{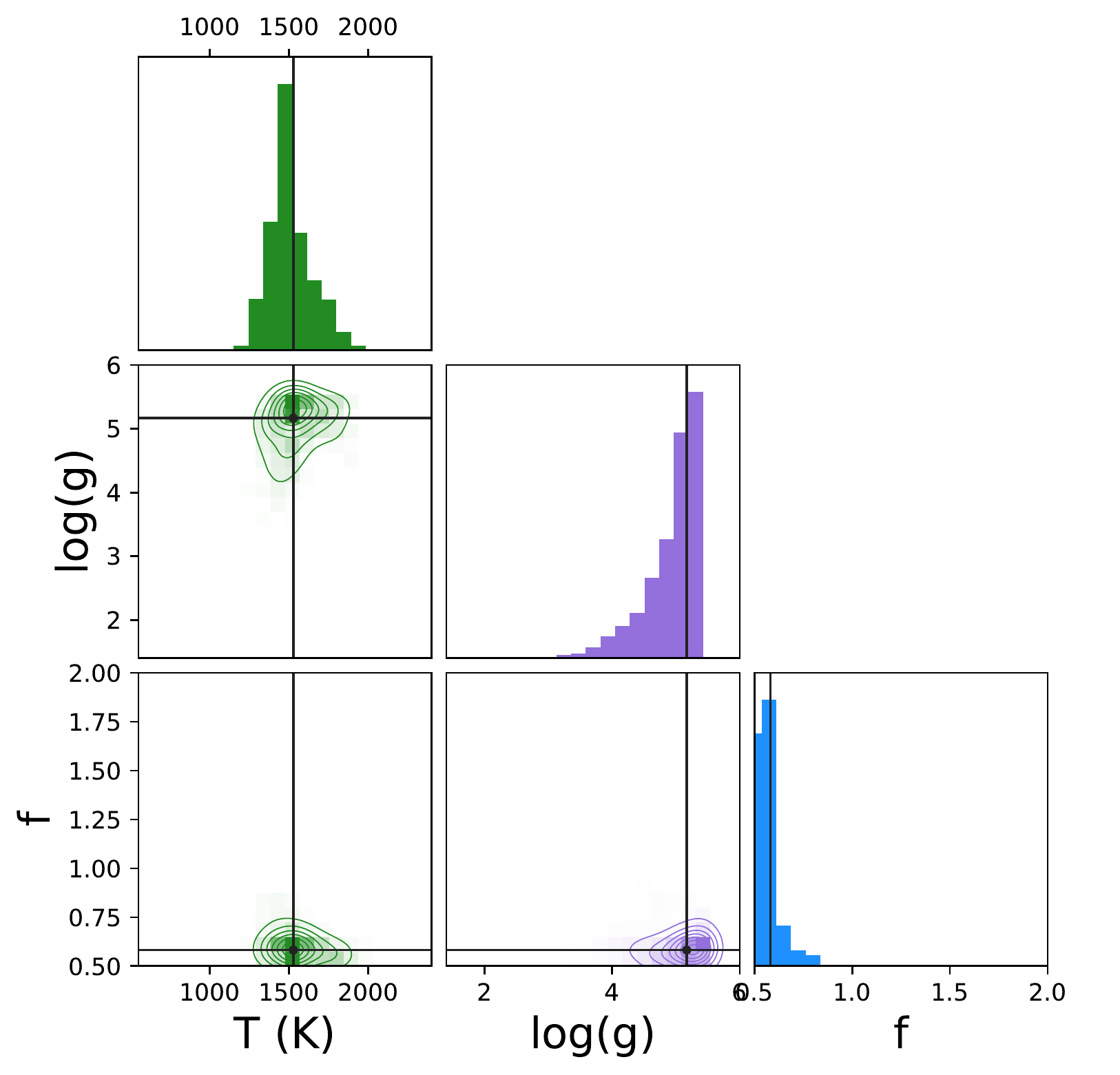}
\includegraphics[width=0.25\columnwidth,height=0.2\textheight]{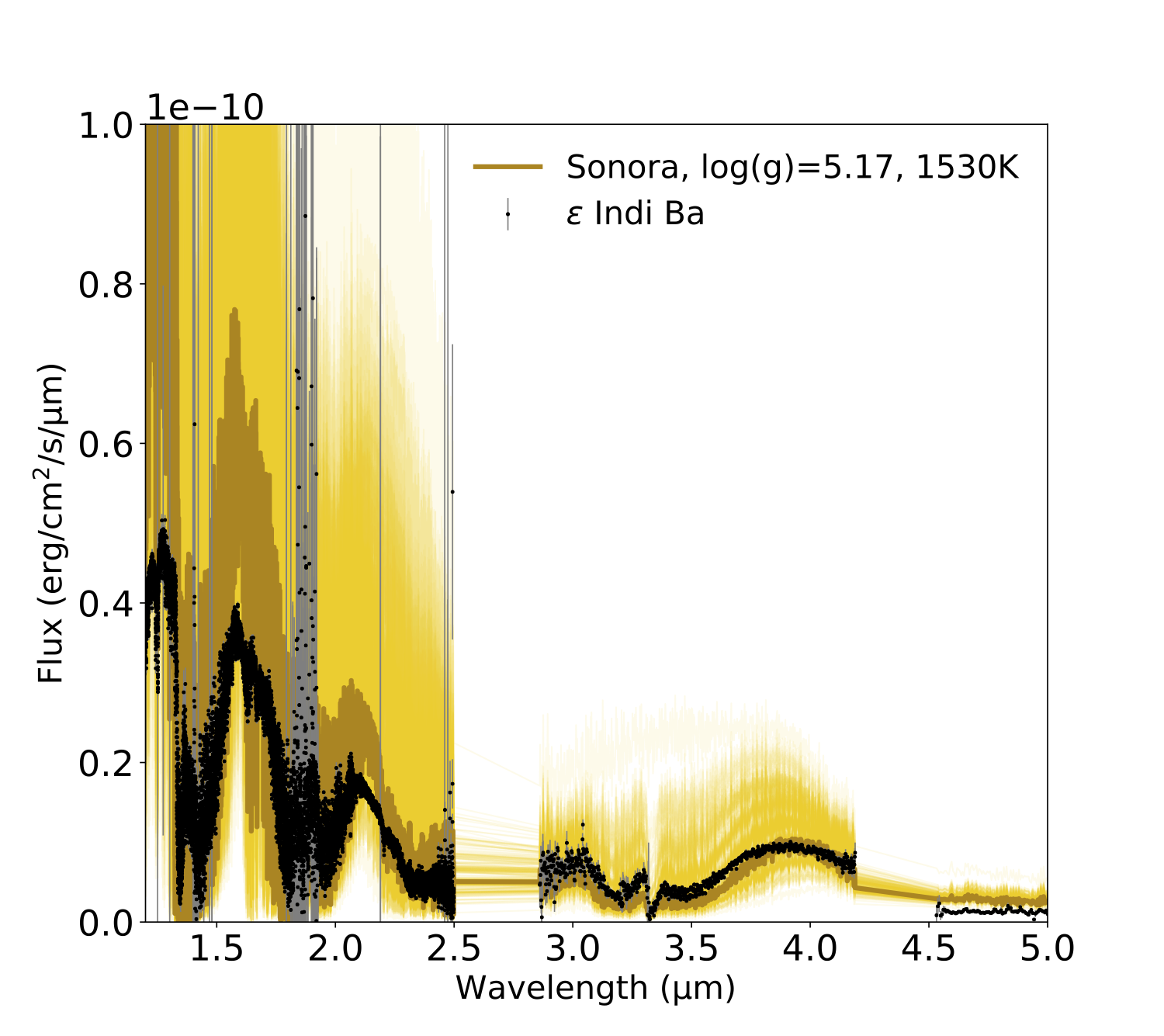}}
\subfloat[Sonora, no calibration factor]{\includegraphics[width=0.25\columnwidth,height=0.2\textheight]{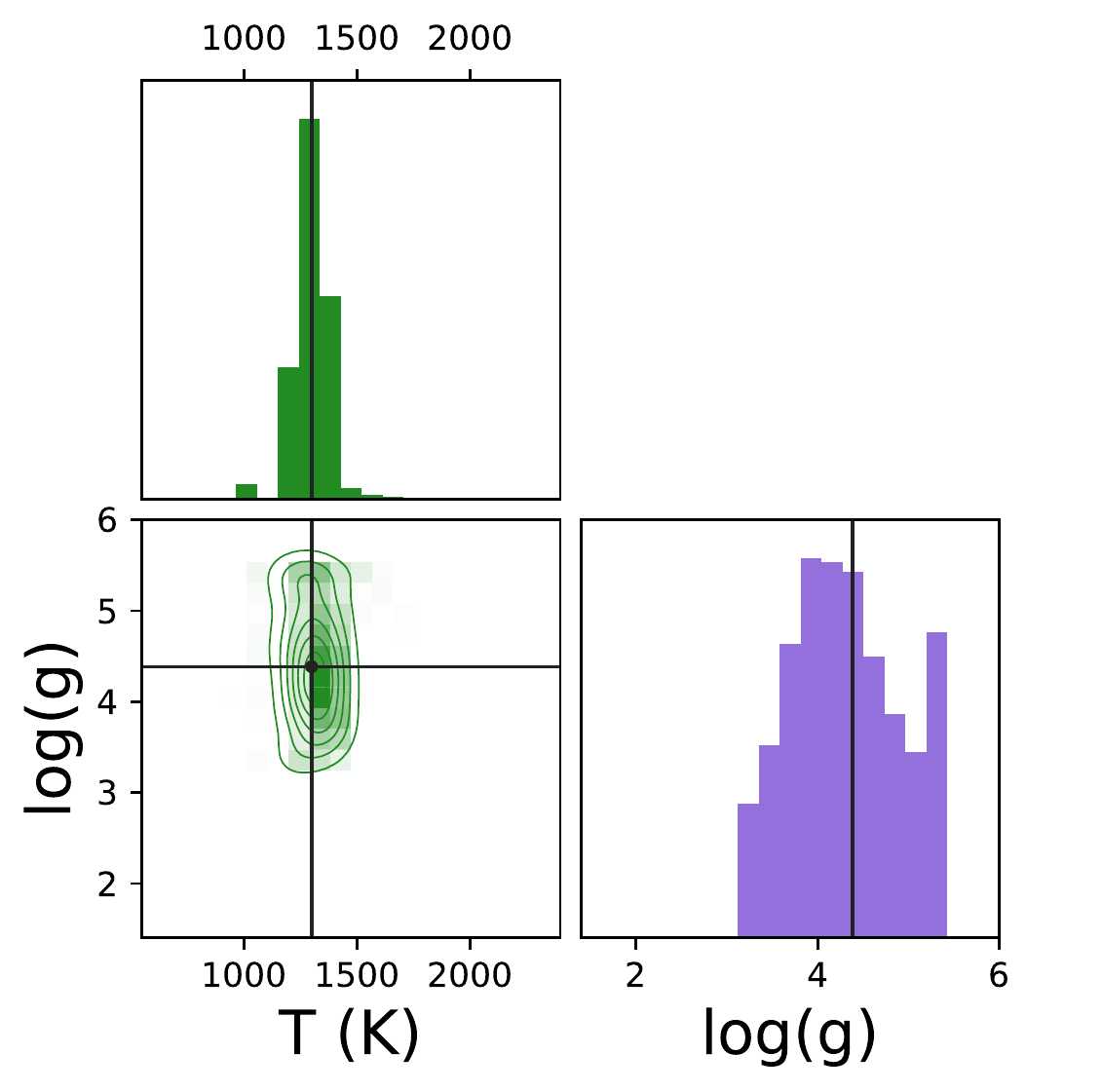}
\includegraphics[width=0.25\columnwidth,height=0.2\textheight]{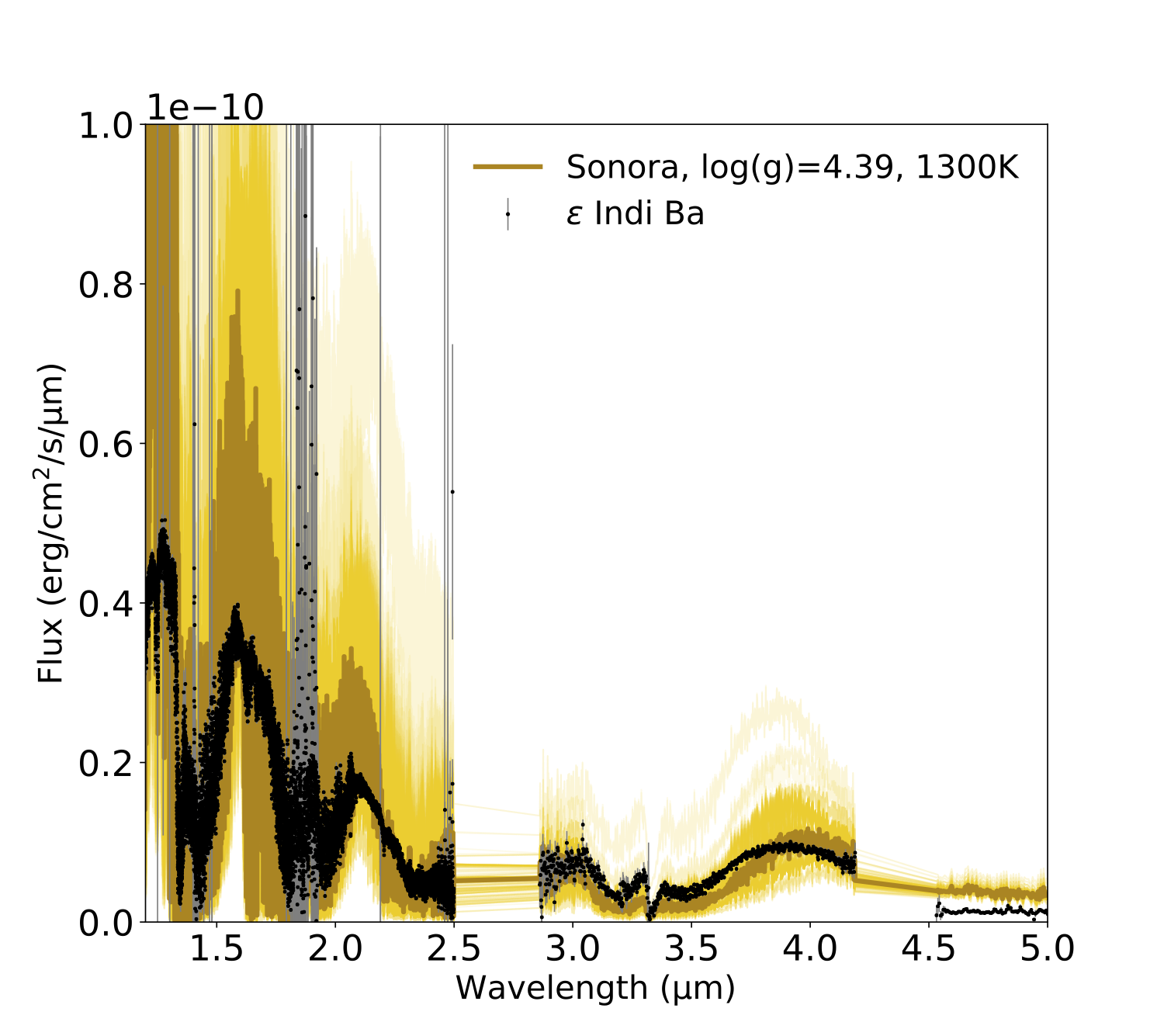}}

\subfloat[AMES-cond, with calibration factor]{\includegraphics[width=0.25\columnwidth,height=0.2\textheight]{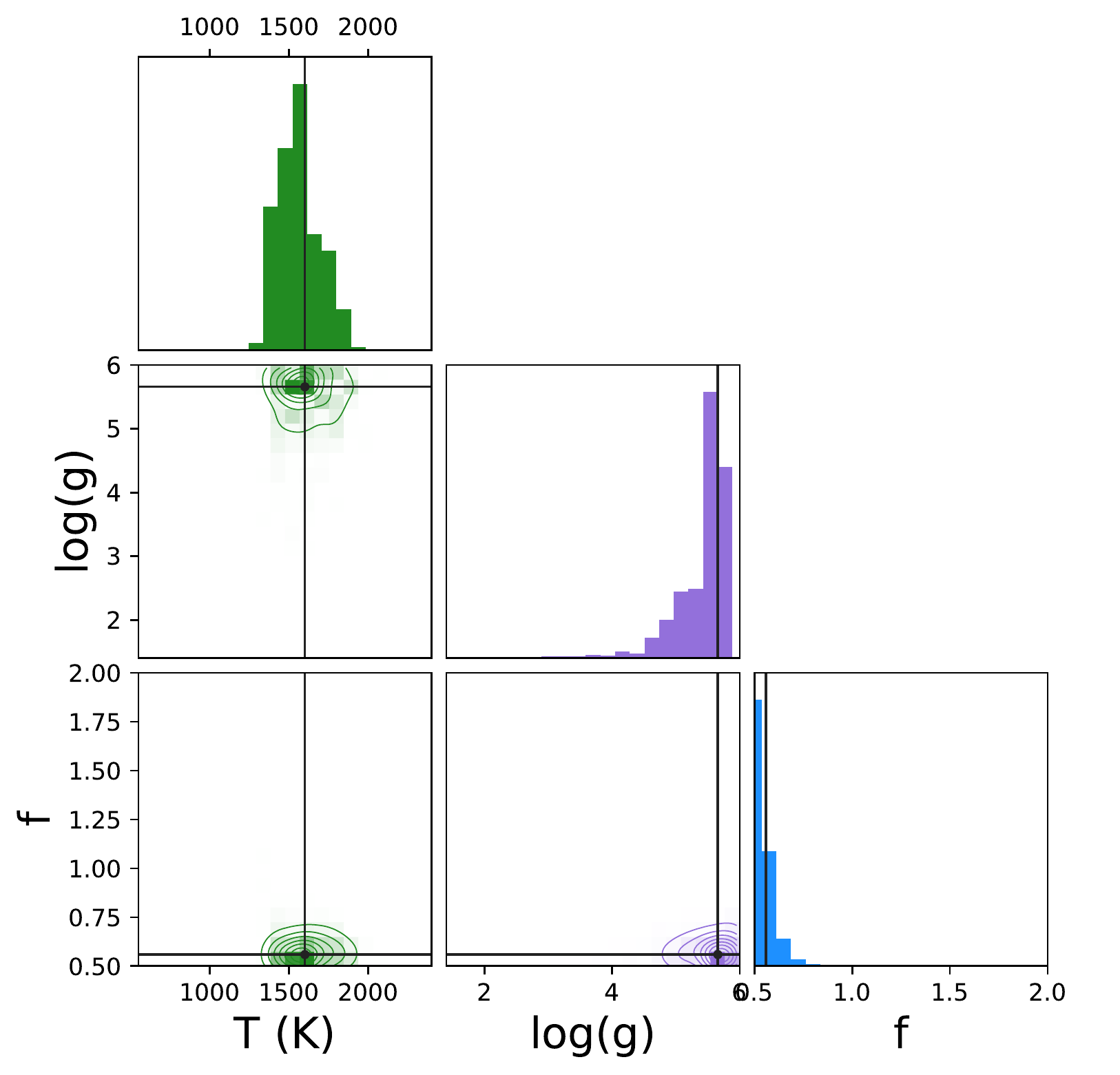}
\includegraphics[width=0.25\columnwidth,height=0.2\textheight]{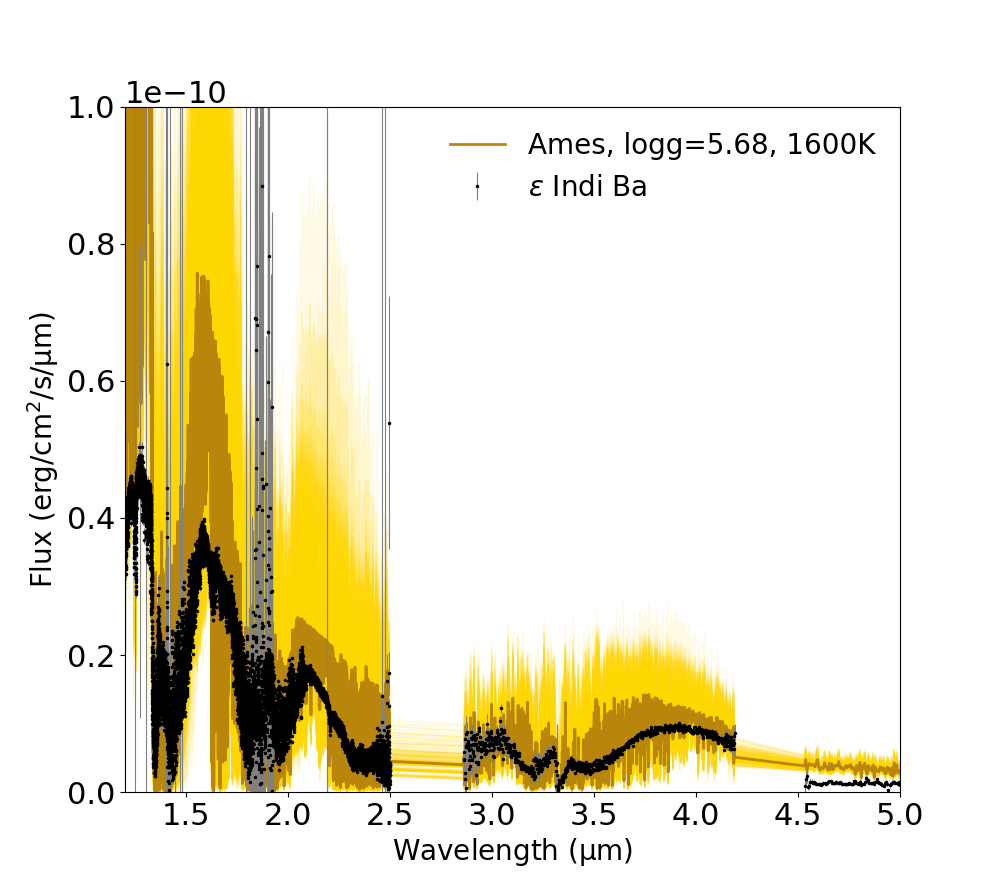}}
\subfloat[AMES-cond, no calibration factor]{\includegraphics[width=0.25\columnwidth,height=0.2\textheight]{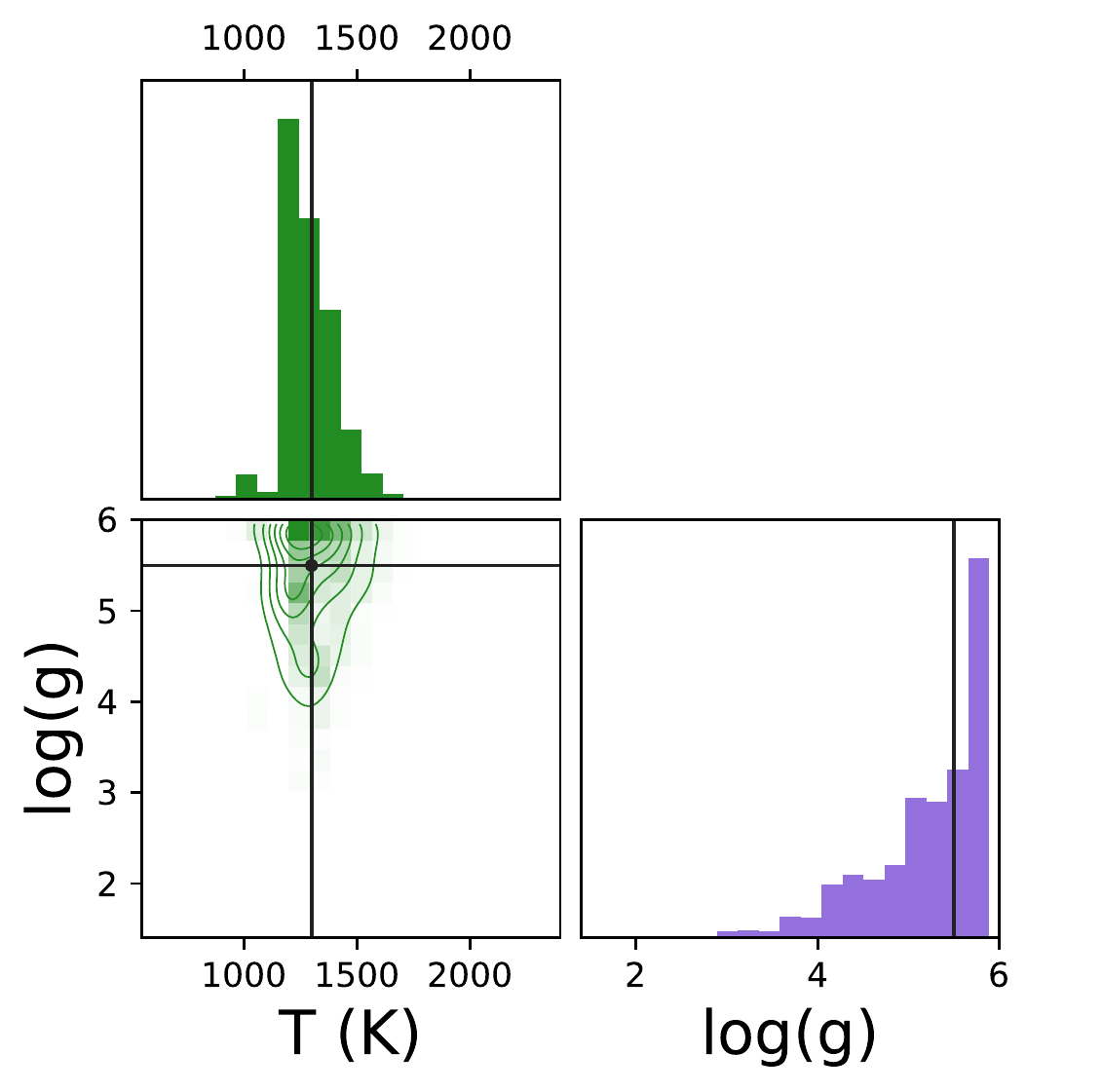}
\includegraphics[width=0.25\columnwidth,height=0.2\textheight]{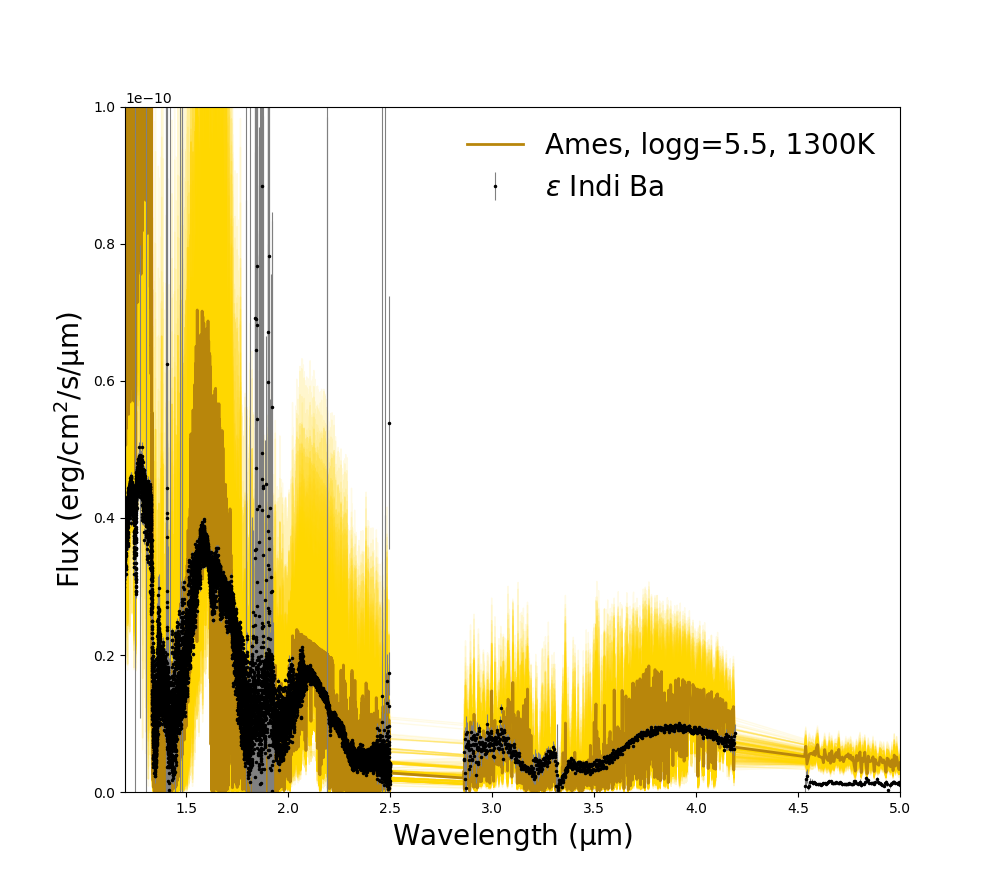}}
\end{center}
\vspace{-0.1in}
\caption{Posteriors for $\epsilon$ Indi Ba.}
\label{fig:epsa}
\end{figure*}

\begin{figure*}[h!]
\vspace{-0.1in}
\begin{center}
\subfloat[HELIOS, with calibration factor]{\includegraphics[width=0.25\columnwidth,height=0.2\textheight]{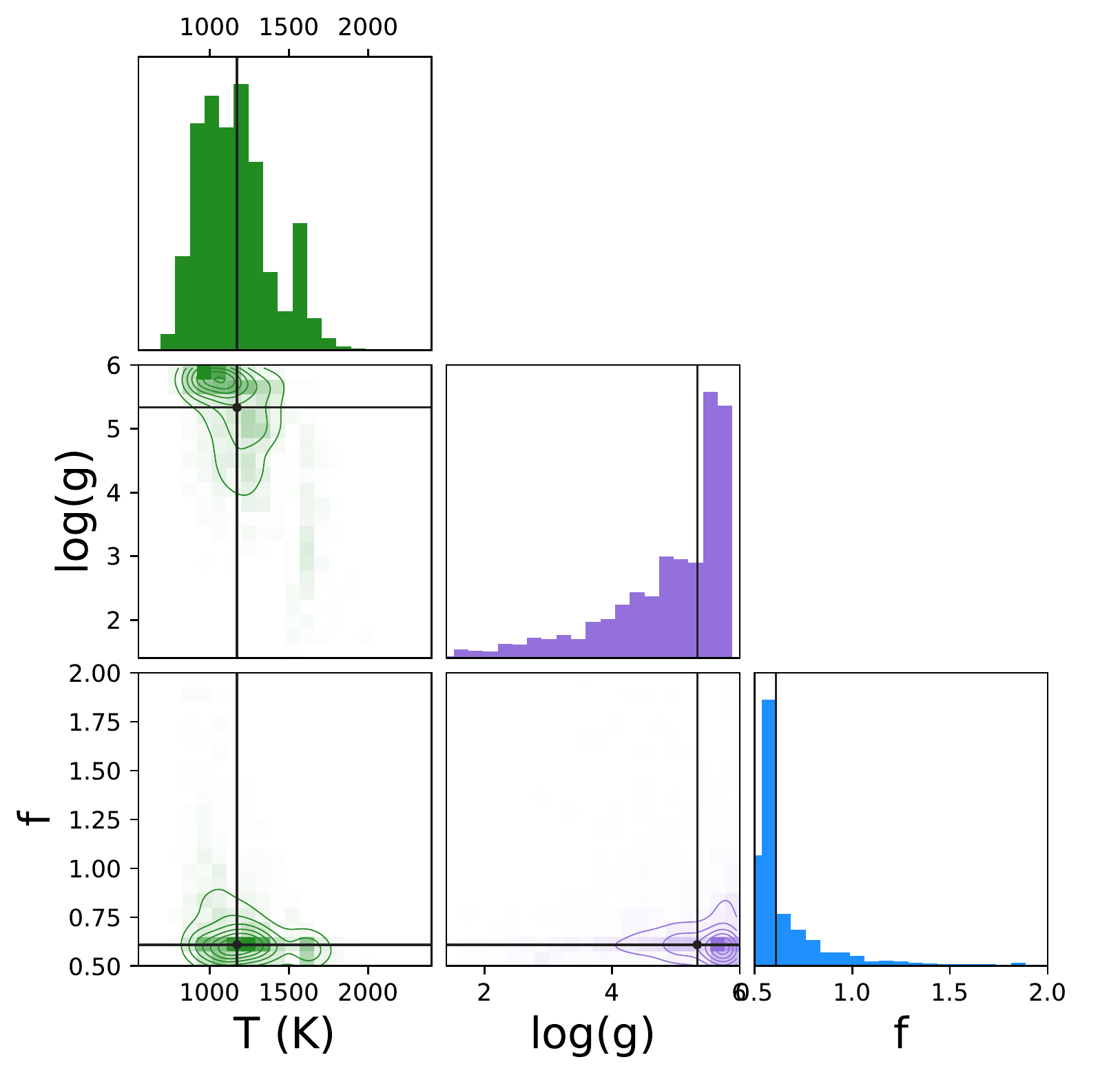}
\includegraphics[width=0.25\columnwidth,height=0.2\textheight]{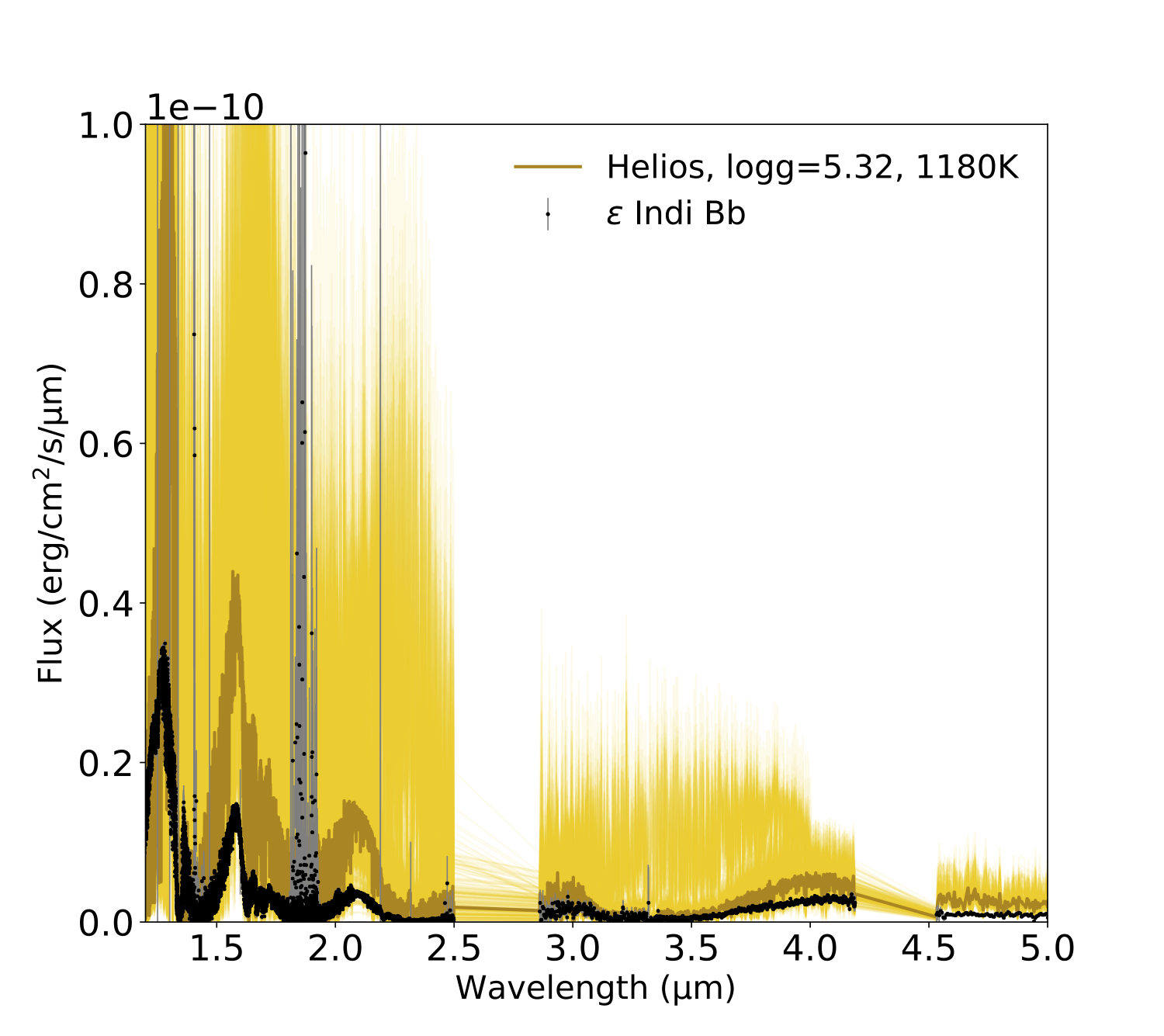}}
\subfloat[HELIOS, no calibration factor]{\includegraphics[width=0.25\columnwidth,height=0.2\textheight]{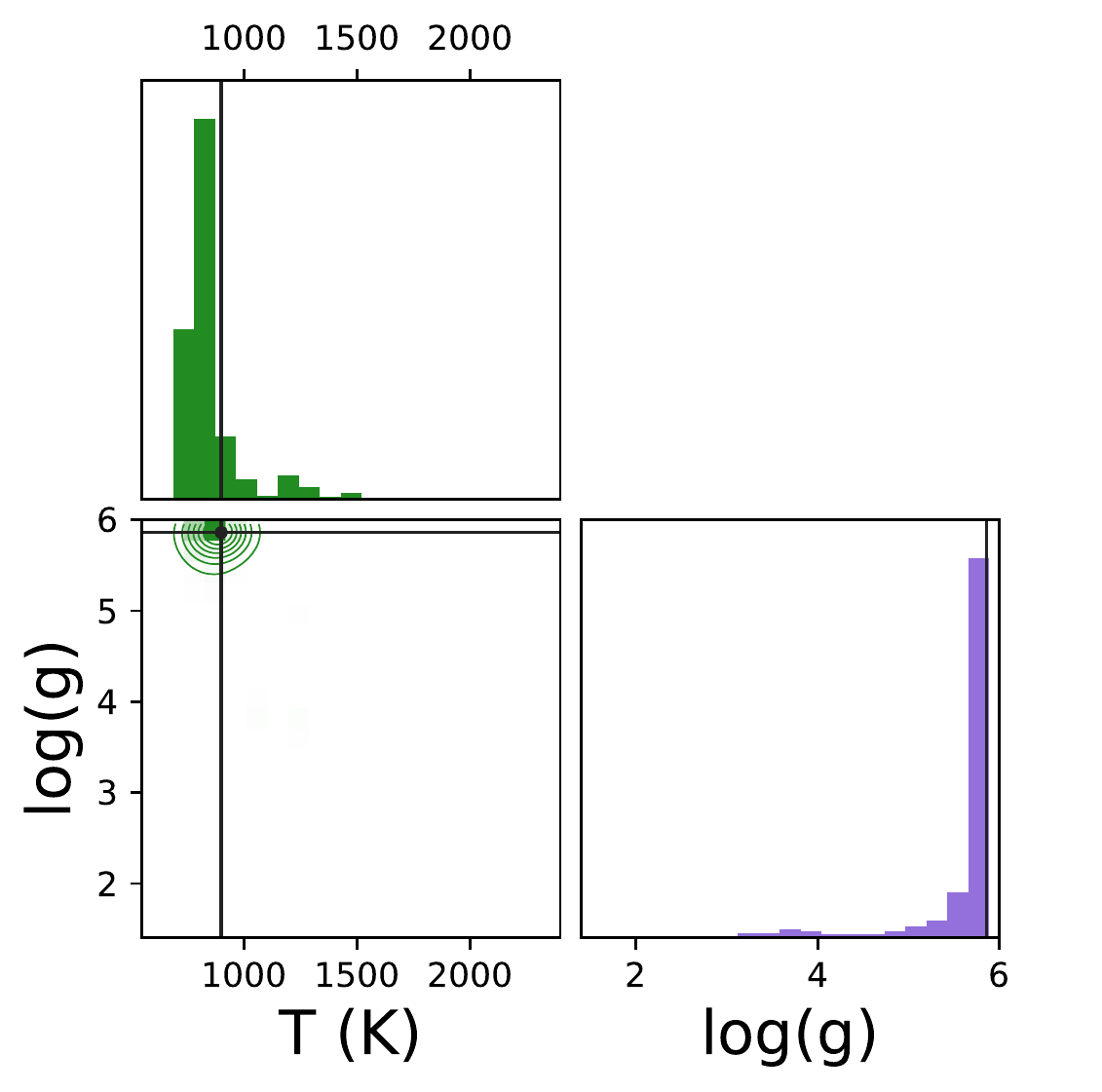}
\includegraphics[width=0.25\columnwidth,height=0.2\textheight]{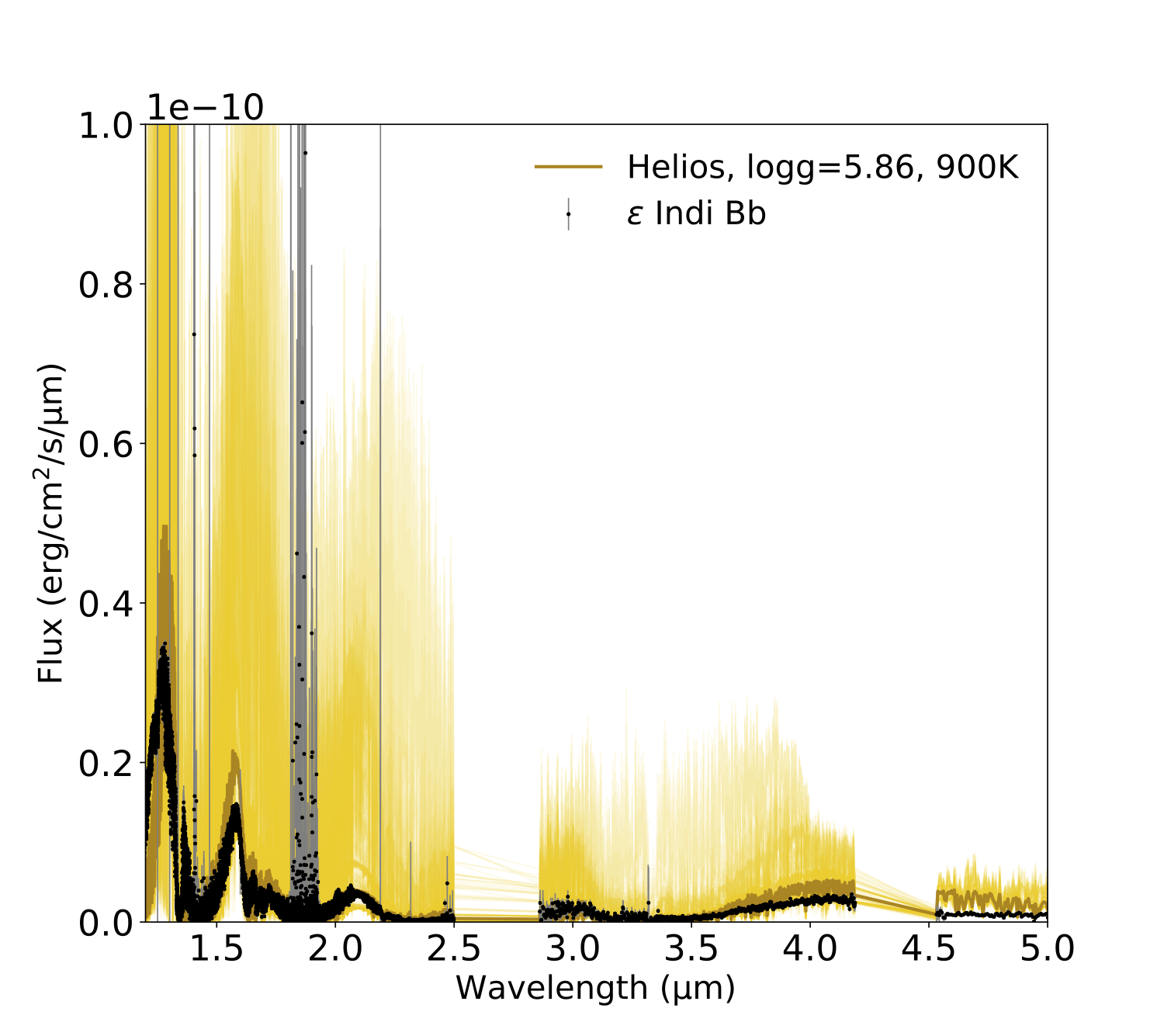}}

\subfloat[Sonora, with calibration factor]{\includegraphics[width=0.24\columnwidth,height=0.2\textheight]{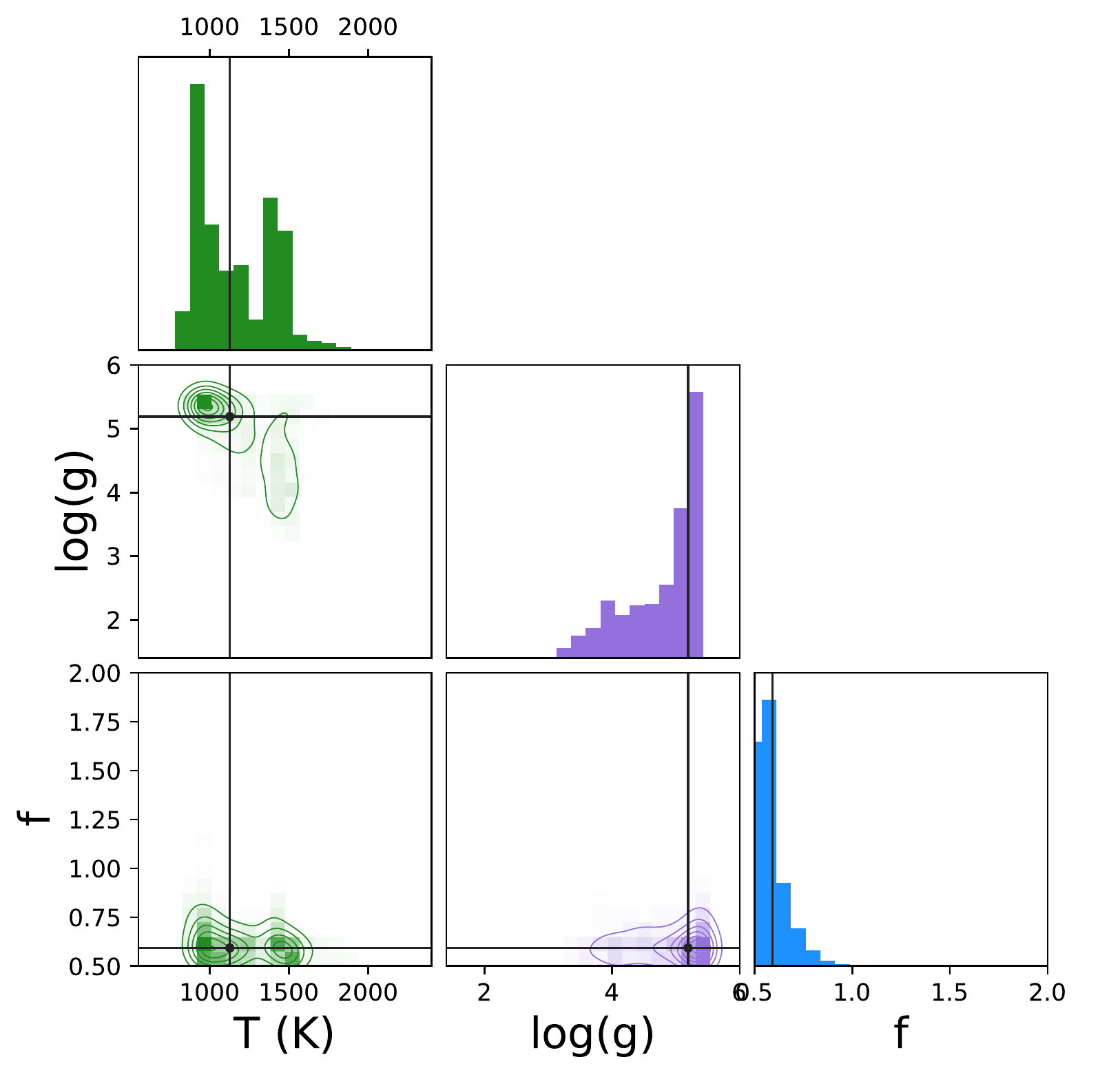}
\includegraphics[width=0.25\columnwidth,height=0.2\textheight]{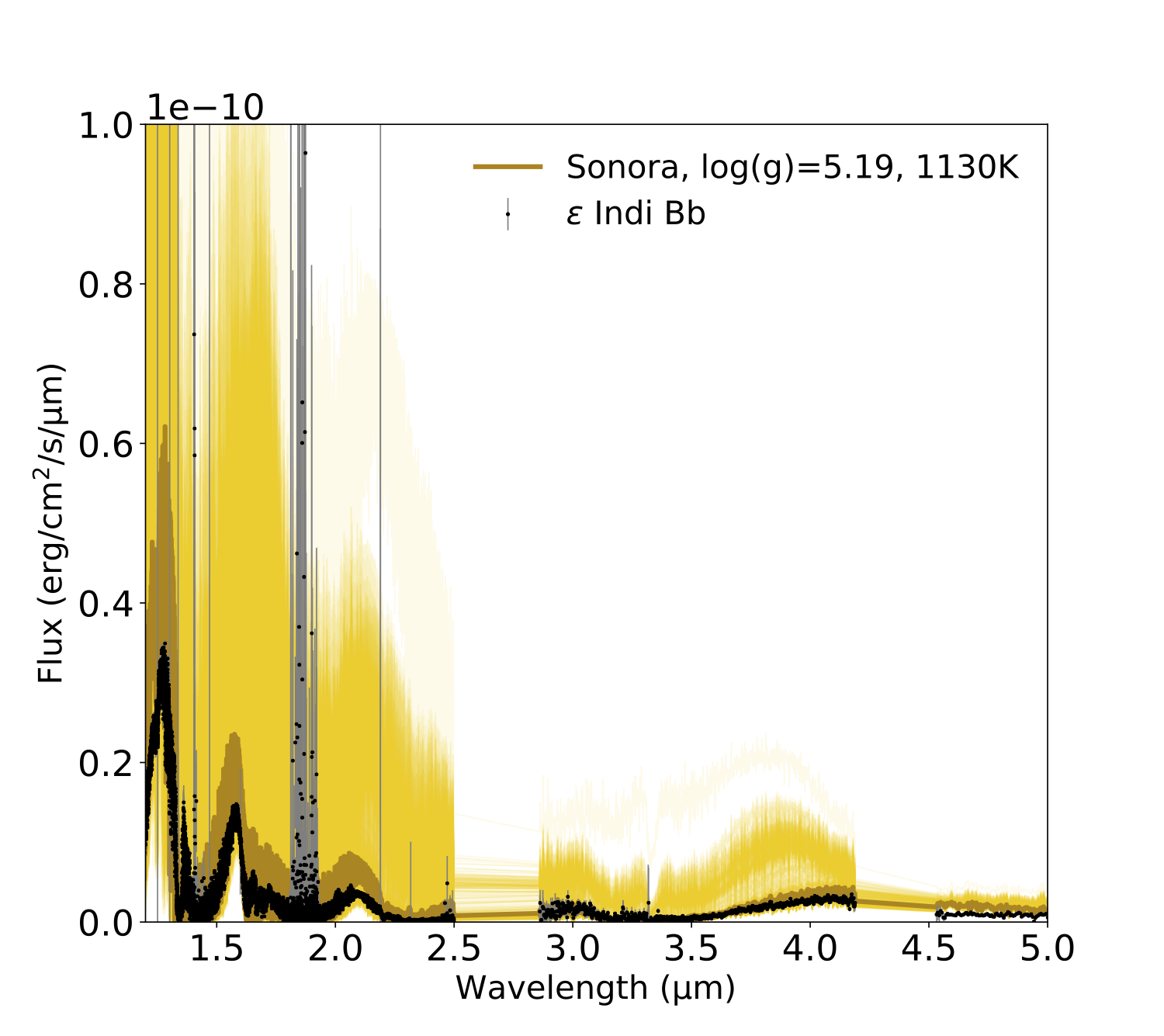}}
\subfloat[Sonora, no calibration factor]{\includegraphics[width=0.25\columnwidth,height=0.2\textheight]{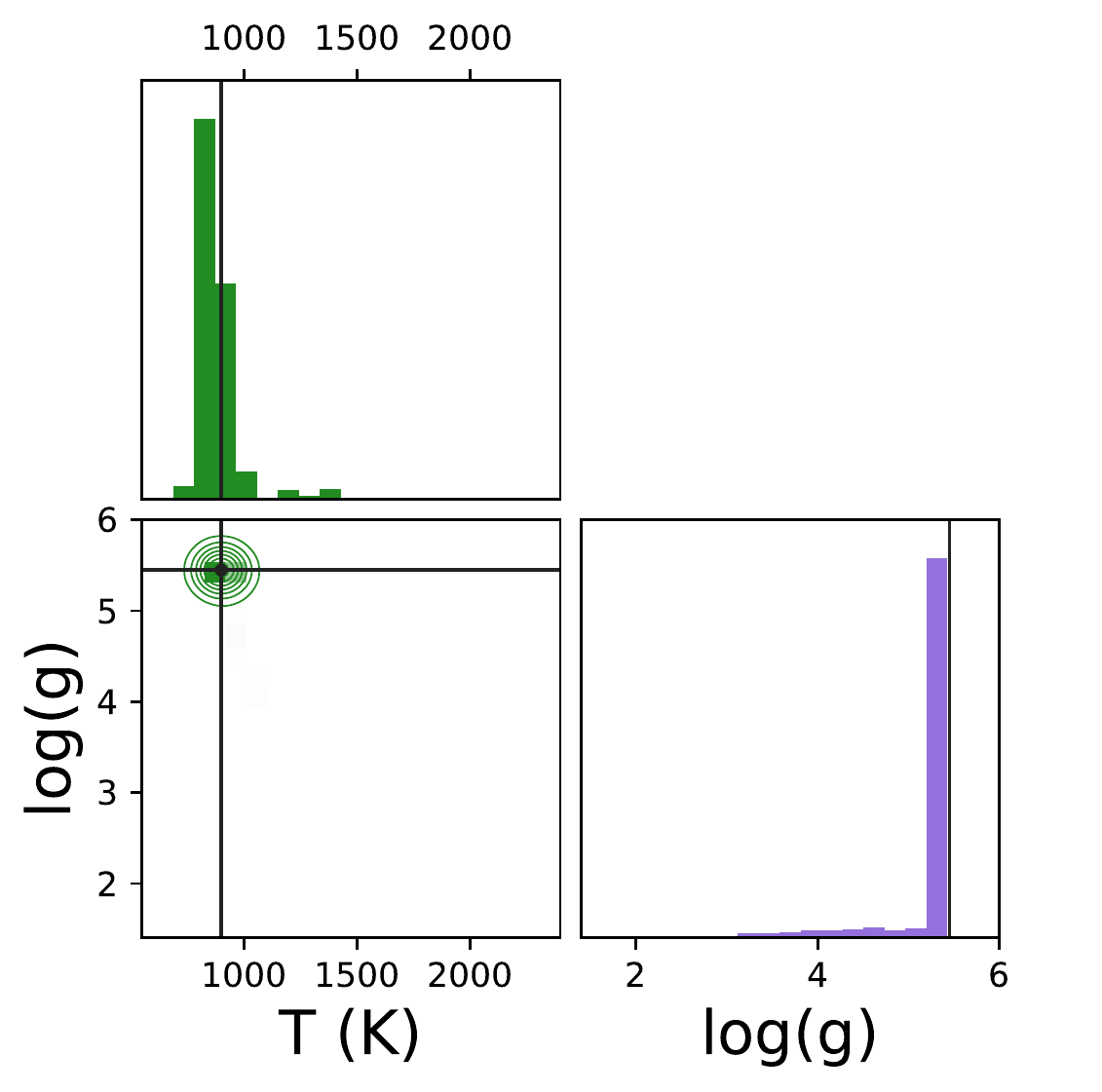}
\includegraphics[width=0.25\columnwidth,height=0.2\textheight]{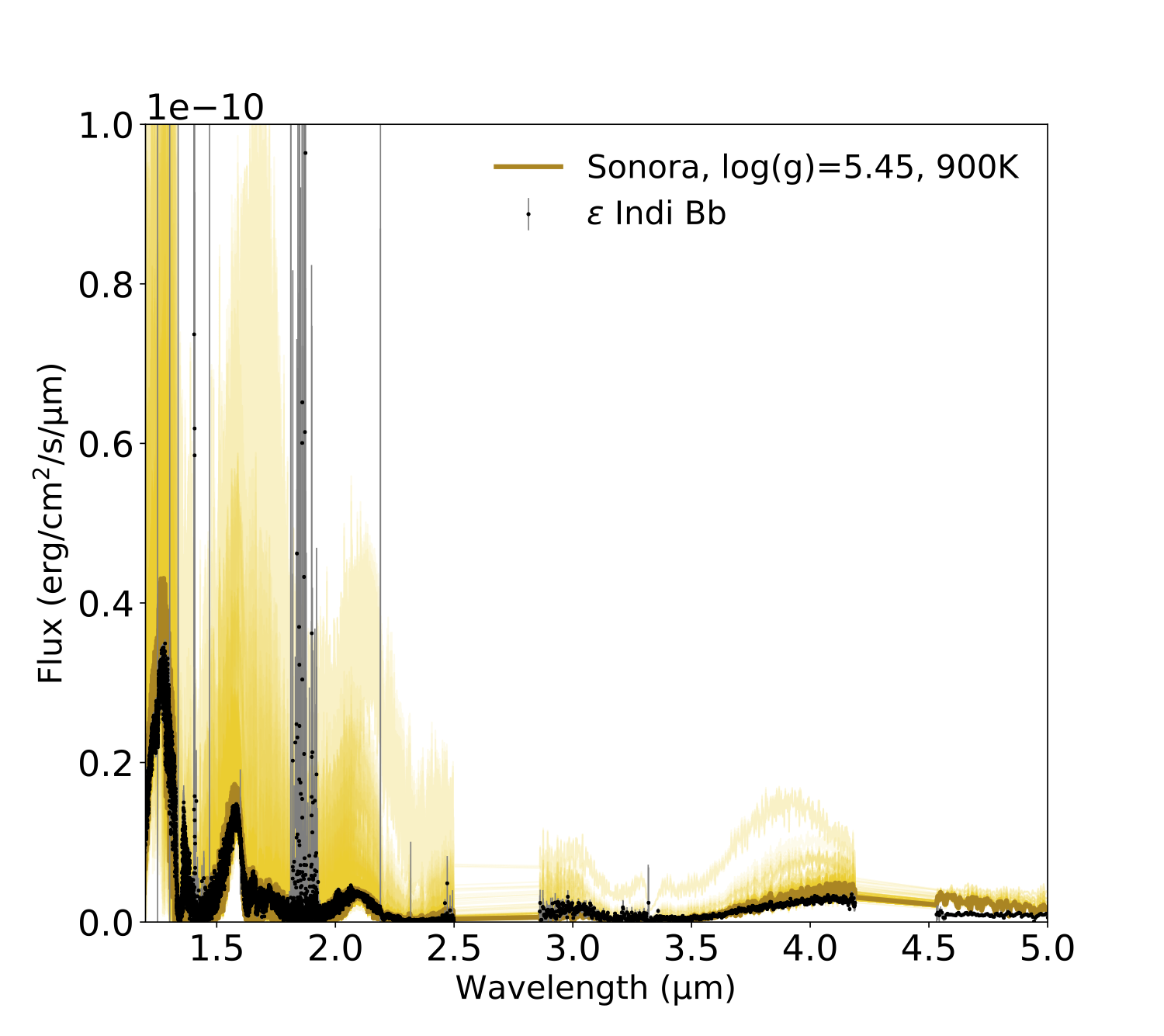}}

\subfloat[AMES-cond, with calibration factor]{\includegraphics[width=0.25\columnwidth,height=0.2\textheight]{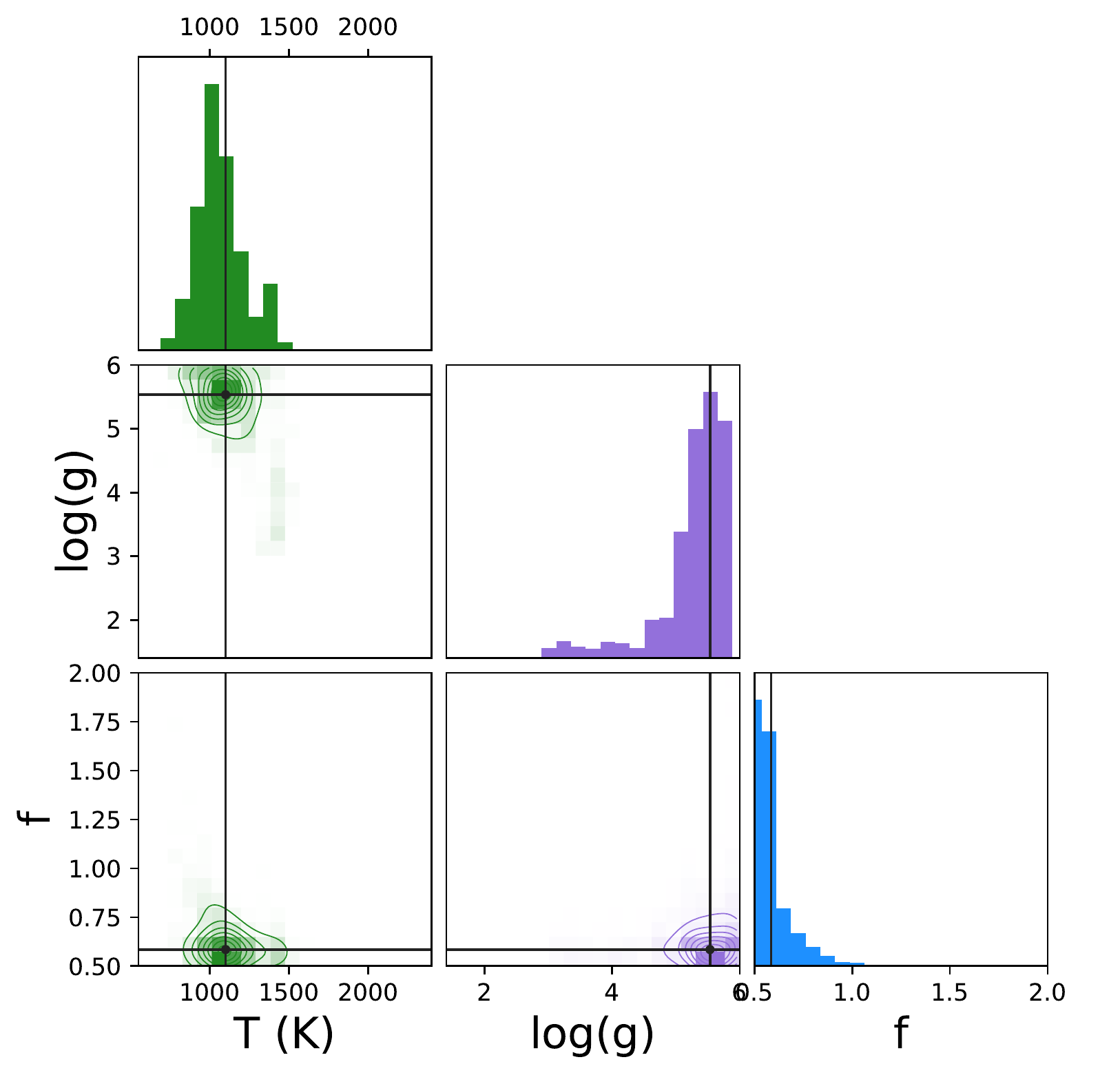}
\includegraphics[width=0.25\columnwidth,height=0.2\textheight]{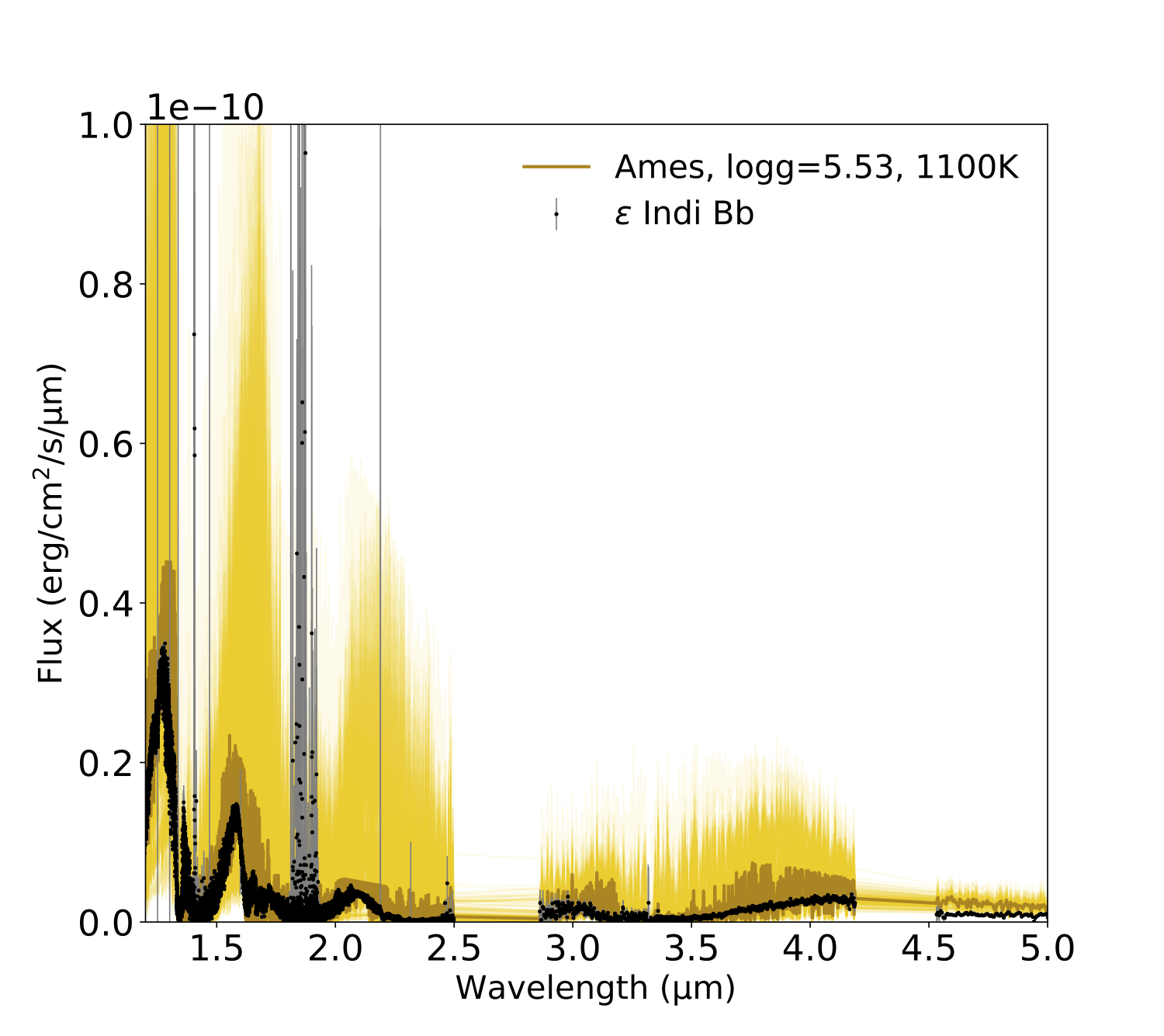}}
\subfloat[AMES-cond, no calibration factor]{\includegraphics[width=0.25\columnwidth,height=0.2\textheight]{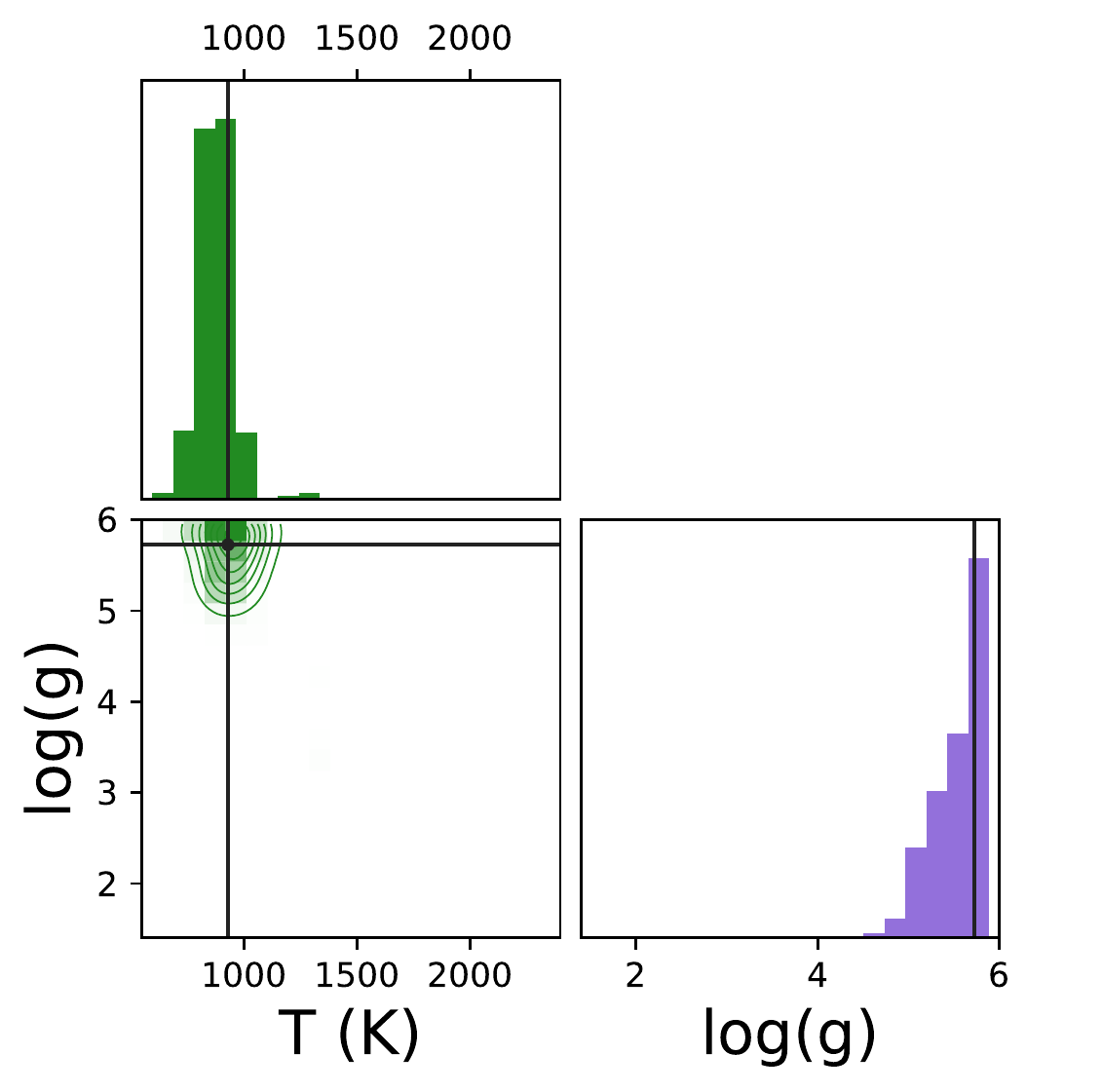}
\includegraphics[width=0.25\columnwidth,height=0.2\textheight]{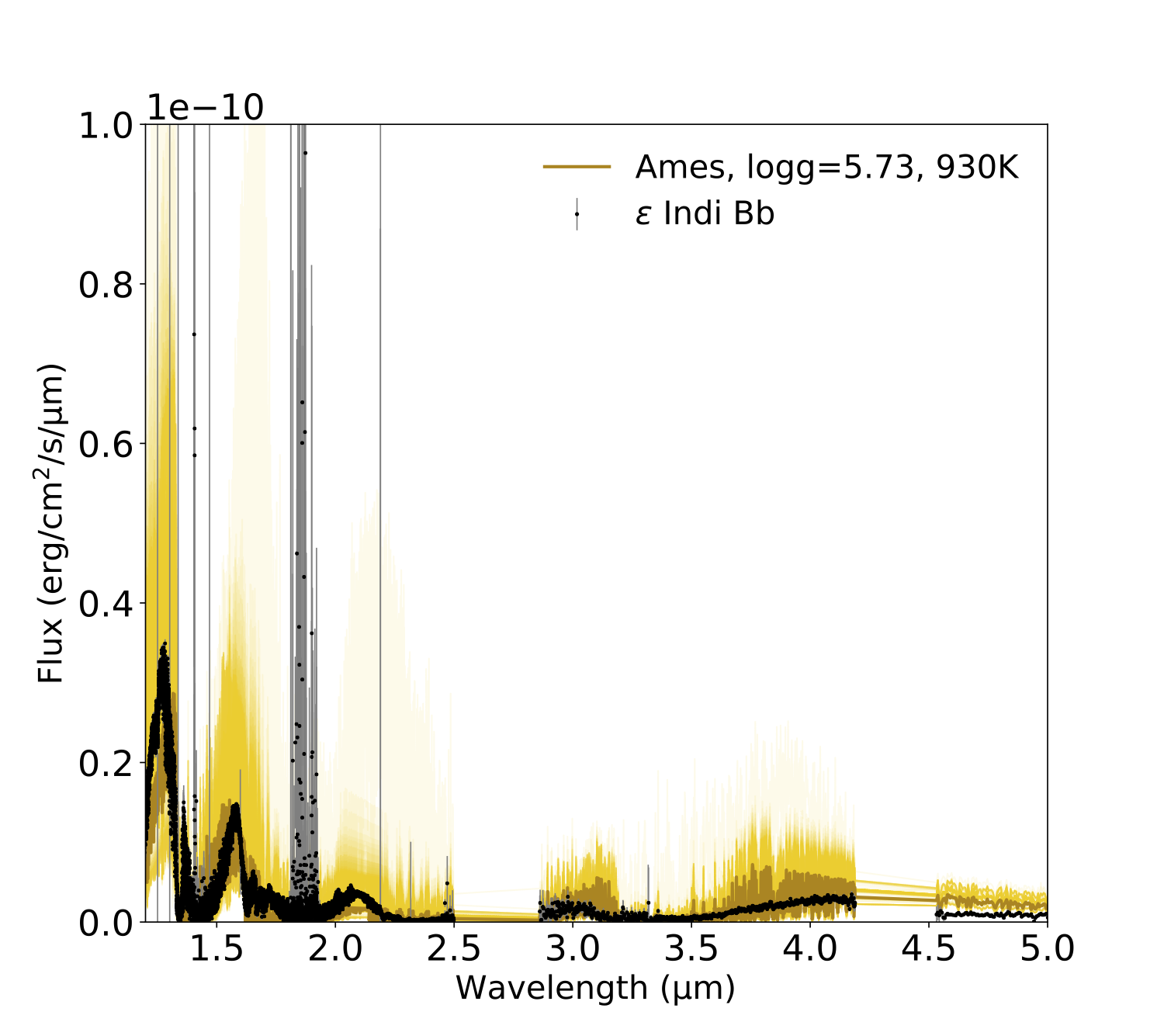}}
\end{center}
\vspace{-0.1in}
\caption{Posteriors for $\epsilon$ Indi Bb.}
\label{fig:epsb}
\end{figure*}

\end{document}